\documentclass[a4paper,11pt]{article}
\usepackage[skins]{tcolorbox}
\usepackage{mathtools,slashed}
\usepackage[T1]{fontenc}
\usepackage{slashed,verbatim,subfigure}
\usepackage[numbers,sort&compress]{natbib}
\usepackage{amsmath}
\usepackage{mathrsfs}
\usepackage{amsbsy}
\usepackage{amssymb}
\usepackage{appendix}
\usepackage{graphicx}
\usepackage[final]{pdfpages}
\usepackage{dcolumn}
\usepackage{bm}
\usepackage[colorlinks=true,linktocpage=true,
linkcolor=blue,citecolor=blue]{hyperref}
\usepackage[a4paper]{geometry}
\catcode`@11
\def\seceqaa{\@addtoreset{equation}{section}
	\def\theequation{A\arabic{equation}}}
\def\seceqbb{\@addtoreset{equation}{section}
	\def\theequation{B\arabic{equation}}}
\def\seceqcc{\@addtoreset{equation}{section}
	\def\theequation{C\arabic{equation}}}
\def\seceqdd{\@addtoreset{equation}{section}
	\def\theequation{D\arabic{equation}}}
\def\seceqee{\@addtoreset{equation}{section}
	\def\theequation{E\arabic{equation}}}
\catcode`@11
\newcommand{\be}{\begin{eqnarray}}
\newcommand{\ee}{\end{eqnarray}}
\begin{document}
\large
\title{Entanglement entropy and Page curve from the ${\cal M}$-theory dual of thermal QCD above $T_c$ at intermediate coupling}
\author{Gopal Yadav\footnote{email- gyadav@ph.iitr.ac.in}~~and~~Aalok Misra\footnote{email- aalok.misra@ph.iitr.ac.in}\vspace{0.1in}\\
Department of Physics,\\
Indian Institute of Technology Roorkee, Roorkee 247667, India}
\date{}
\maketitle
\begin{abstract}
Exploration of entanglement entropy (EE) and obtaining the Page curve in the context of eternal black holes associated with top-down non-conformal holographic thermal duals at intermediate coupling, has been entirely unexplored in the literature. We fill this gap by obtaining the Page curve of an eternal neutral black hole from a doubly holographic setup relevant to the ${\cal M}$-theory dual of thermal QCD-like theories at high temperatures (i.e. above $T_c$) and intermediate coupling (effected by inclusion of terms quartic in curvature in M theory). 
%On the gauge theory side we have thermal QCD at intermediate coupling which is a non-conformal theory. The gravity dual is coupled to, effectively, four dimensional thermal QCD bath. Despite the non-conformality, we find analogues of the "Hartman-Maldacena-like surface'' and "island surface''. 
Remarkably, excluding the higher derivative terms, the EE of the Hawking radiation from the on-shell Wald entanglement entropy  integral (for appropriate choices of constants of integration appearing in the embeddings) increases {\it almost} linearly due to dominance of entanglement entropy contribution from the Hartman-Maldacena(HM)-like surface $S_{\rm EE}^{\rm HM, \beta^0}, \beta\sim l_p^6$. This imparts a ``Swiss-Cheese'' structure to $S_{\rm EE}^{\rm HM, \beta^0}$ effecting  a ``Large N Scenario''(LNS).  Then after the Page time, the EE contribution from the Island Surface (IS) $S_{\rm EE}^{\rm IS, \beta^0}$ dominates and saturates the linear time growth of the entanglement entropy of Hawking radiation and leads to the Page curve. The same is also obtained via the areas of the Hartman-Maldacena-like/Island surfaces. Requiring the (IS) EE ($S_{\rm EE}^{\rm IS, \beta^0}$) per unit Hawking-Beckenstein entropy ($S_{\rm BH}$) post the Page time to be around 2, and positivity of the Page time, set respectively a lower and upper bound on the horizon radius $r_h$ (the non-extremality parameter). Further, with the inclusion of the ${\cal O}(R^4)$ terms in ${\cal M}$ theory, the fact the turning point associated with the HM-like surface is in the deep IR, requires a relationship between $l_p$ and $r_h$  along with a conjectural $\gamma\equiv e^{-{\cal O}(1)N^{1/3}}$-suppression (motivated by the aforementioned requirement $\frac{S_{\rm EE}^{\rm IS, \beta^0}}{S_{\rm BH}}\sim2$ up to leading order). We obtain a hierarchy with respect to $\gamma$ in $S_{\rm EE}^{\rm HM, \beta^0}, S_{\rm EE}^{\rm IS, \beta^0}$ and $S_{\rm EE}^{\rm HM, \beta}, S_{\rm EE}^{\rm HM, \beta}$ (at ${\cal O}(\beta)$).  This is due to the existence of massless graviton corresponding to a null mass eigenvalue of the Laplace-Beltrami equation in the internal space.
\end{abstract}
%\newpage
\tableofcontents

\section{Introduction and Motivation}
\label{introduction}
According to Hawking, black holes evolve from the pure state to a mixed state, i.e., Hawking radiation is thermal radiation. Evaporation of the black holes has no information about the initial state of the black hole which means information is lost in black hole evaporation process \cite{Hawking}. But it has to respect the fundamental principle of quantum mechanics which is unitarity. Entanglement entropy of the Hawking radiation starts decreasing after the Page time \cite{Page} of the evaporating black hole and for eternal black holes it reaches a constant value. Unitary evolution of black holes thus for has been understood via AdS/CFT correspondence \cite{AdS-CFT}. In gauge-gravity duality, on gravity dual side, we have black hole and on the gauge theory side we have conformal field theory which respects unitarity. Therefore, we can study the unitary evolution of the black holes on the CFT  side via gauge-gravity duality. A similar construction for QCD-like theories (equivalance class of theories which are IR confining, UV conformality and quarks transforms in the fundamental representation) at intermediate coupling was done in \cite{HD-MQGP} where ${\cal M}$-theory dual  ${\cal O}(R^4)$ corrections were considered. We are studying the black hole information paradox via gauge-gravity duality in ${\cal M}$-theory dual of \cite{HD-MQGP}. For a review on the black hole information paradox, see \cite{review-BH-IP,Harlow}.
\par
There is a prescription to calculate entanglement entropy in field theories which have a gravity dual. For conformal field theories it was given in \cite{RT} that entanglement entropy will be given by the area of the co-dimension two minimal surface in the bulk gravity dual. Generalisation of the Ryu-Takayanagi formula for non-conformal theories was given in \cite{KKM} and used made in \cite{MCTEQ} to discuss deconfinement phase transition in thermal QCD-like theories from entanglement entropy point of view. Further, if we consider higher derivative gravity theories then we can calculate entanglement entropy using \cite{Dong}. Authors in \cite{EW} included quantum corrections to the Ryu-Takayanagi formula at all order in $\hbar$ and entanglement entropy which is known as generalised entropy. At leading order in $\hbar$ in generalised entropy formula, we recover the Ryu-Takayanagi formula \cite{RT}. In Englehardt-Wall prescription one is required to extremize the generalised entropy, and surfaces which extremize the generalized entropy are known as Q(uantum) E(xtremal) S(urfaces). The island proposal was first introduced in \cite{AMMZ}, where the authors discussed that at late times islands come into the picture which contribute to the entanglement entropy. In this case, one is required to extremize the generalised entropy-like functional over all islands which include contribution from the island surfaces, and then minimize over all possible extrema. Islands which extremize the aforementioned generalised entropy-like functional are known as Q(uantum) E(xtremal) I(slands).
\par
In doubly holographic models we consider a black hole in the gravitational dual is coupled to an external CFT bath \cite{Island-HD}. For example, gravity (which contains the black hole) in $d$-dimensions is coupled to external bath in $d$-dimensions, where $d$-dimensional external bath has its own holographic dual in $(d+1)$-dimensions. In these kinds of models we consider two copies of the setup described earlier. We consider two kinds of surfaces. First is the Hartman-Maldacena-like surface \cite{Hartman-Maldacena}, which starts from the point where gravity is coupled to external bath, i.e., defect and crosses the black hole horizon and reaches up to the defect of thermofield double of doubly holographic models, and entanglement entropy contribution from this surface has linear time dependence which leads to information paradox at late times. Second, is the island surface, which start from the external bath and hit the Karch-Randall brane \cite{KR1,KR2}. Entanglement entropy contribution from the island surface is independent of time and dominates after the Page time. After combining the entanglement entropy contributions from both surfaces we obtain the Page curve. Doubly holographic model with gravitating bath was discussed in \cite{G-bath} and with non-gravitating baths models were discussed in \cite{PBD,Bath-WCFT,B-N-P-1,B-N-P-2}. Page curve calculation for the neutral Gauss-Bonnet black hole was done in \cite{NGB}\footnote{One of us (GY), thanks R.~X.~Miao to  bring their work to our attention}. 
\par
There is one more way via which information paradox was resolved. In this approach, initially we calculate the entanglement entropy contribution without the island surface using Cardy's formula \cite{Cardy-Formula} which depends on time and is responsible for the appearance of information paradox. Later, island surfaces emerge which contribute to the entanglement entropy and this contribution is independent of time which dominates after the Page time. Combining these two contributions we obtain the Page curve. Based on this approach some examples were discussed in \cite{Island-RNBH,NBH-HD,Yu-Ge,Island-SB,Omidi}. Since there is no known Cardy-like formula for non-conformal theories therefore we are following the doubly holographic models approach discussed earlier in our calculations. Islands and Page curves were studied in type IIB string theory in \cite{Island-IIB-1,Island-IIB-2,Island-IIB-3} and for flat space black holes, see \cite{flat-space-Page-curve}. Page curves of the Reissner-Nordstr\"om black hole in the presence of higher derivative terms was studied in \cite{charged-GB-BH}. 
\par
To study thermal QCD-like theories via gauge-gravity duality in the UV complete type IIB holographic dual, the authors started in \cite{metrics} where they constructed type IIB string dual of large-$N$ thermal QCD-like theories in the strong coupling limit. To study the finite coupling regime of thermal QCD-like theories authors constructed type IIA mirror of the aforementioned type IIB setup then uplifted it to the ${\cal M}$-theory in \cite{MQGP,NPB}. Applications of the same were studied in \cite{MQGP-app-1,MQGP-app-2,MQGP-app-3,MQGP-app-4}. Further, to have analytical control on the intermediate coupling regime of thermal QCD-like theories, the authors in \cite{HD-MQGP} incorporated HD terms on the gravity dual side. We are working with \cite{HD-MQGP} too in this paper to obtain the Page curve of a neutral black hole in the presence of higher derivative terms. The model \cite{HD-MQGP} has been reviewed in \cite{MChPT,MCTEQ} and the latter references are applications of the same. One of us (GY) studied the effect of rotation on the deconfinement temperature of thermal QCD-like theories at intermediate coupling from ${\cal M}$-theory  in \cite{Rotation-Tc-M-Theory}.

\par
In this paper we are calculating the Page curve of a neutral black hole where we have a non-conformal bath. In our case, on the gravity dual side we have ${\cal M}$-theory dual inclusive of higher derivative terms and on the gauge theory side we have thermal QCD-like theories at intermediate coupling which is a non-conformal theory. In this paper we are considering a doubly holographic setup similar to \cite{PBD,Bath-WCFT} where on the gauge theory side authors have a conformal theory but the story is different in our setup because we have a non-conformal theory. Similar to \cite{PBD,Bath-WCFT}, we consider two kinds of candidate surfaces: Hartman-Maldacena-like surface and island surface. Hartman-Maldacena-like surface is responsible for the linear time growth of the entanglement entropy of Hawking radiation \cite{Hartman-Maldacena}. After the Page time, the entanglement entropy contribution from the island surface dominates which is independent of time and we obtain the Page curve from ${\cal M}$-theory dual inclusive of ${\cal O}(R^4)$ corrections.
\par

We have organised the paper in the following way. In section \ref{setup} via two subsections \ref{M-theory-dual} and \ref{DHS},  we have described the holographic dual of thermal QCD-like theories from top-down approach in subsection \ref{M-theory-dual}  and construction of doubly holographic setup in ${\cal M}$-theory dual in subsection \ref{DHS}. Section \ref{ETW-sub} is devoted to discussion of end-of-the-world (ETW) ``brane'' embedding in the ${\cal M}$-theory dual. First, we have discussed that ETW-``brane" has constant embedding at the level of Einstein gravity in \ref{ETW-sub-sub-i} and then we have shown in \ref{ETW-sub-sub-ii} that when restricted to the aforementioned constant ETW-``brane'' embedding, there are no boundary terms generated from the most dominant $J_0$ terms at ${\cal O}(R^4)$ in the ``MQGP limit'' in the eleven dimensional supergravity action that would have covariant derivatives of the metric variation, and hence would have required suitable boundary terms to be constructed to cancel the same (much like the GHY boundary terms to cancel off similar terms in the EH action in Einstein gravity). Section \ref{Page-curve-WHD} is divided into three subsections. Subsections \ref{EE-HM-WHD} and \ref{EE-IS-WHD} involve calculations of the entanglement entropy of the Hawking radiation from the areas of Hartman-Maldacena-like and island surfaces, and in subsection \ref{Page-curve-WHD-terms} we have obtained the Page curve of eternal neutral black hole in the absence of higher derivative terms in eleven dimensional supergravity action. We do the Page curve calculation of the neutral black hole in the presence of higher derivative terms in section \ref{Page-curve-HD} via four subsections \ref{EE-HM-HD}, \ref{Swiss-Cheese-i}, \ref{EE-IS-HD} and \ref{Page-curve-plot-HD}. In subsections  \ref{EE-HM-HD} and \ref{EE-IS-HD} we obtain the entanglement entropies of the Hawking radiation for the Hartman-Maldacena-like and island surfaces in the presence of higher derivative terms, and then we obtain the Page curve in subsection \ref{Page-curve-plot-HD}. In subsection \ref{Swiss-Cheese-i} we discuss the ``Swiss-Cheese'' structure of entanglement entropy contribution from the Hartman-Maldacena-like surface.  Section \ref{massless_graviton} has a discussion on showing the existence of massless graviton by showing the existence of a null eigenvalue of the Laplace-Beltrami differential equation for the graviton wave function along the internal coordinates and hence  provide a physical reason for the exponential-in-$N$ suppression of entanglement entropies in  particular for the Island Surface.  We discuss our results and conclusion in section \ref{summary}.
\par
 Additionally there are four appendices. By dividing the appendix \ref{HM+IS-analyt+num} into two parts, we have computed angular integrals used in this paper in \ref{a-1} and turning point of the island surface in \ref{a-2}. In appendix \ref{HM} we have listed various $r$ dependent functions in entanglement entropy expression of the HM-like surface at ${\cal O}(\beta^0)$ and ${\cal O}(\beta)$, apart from that we have also computed EOM of the HM-like surface embedding in the same appendix. In appendix \ref{ISM} we have listed $r$ dependent functions appearing in the ${\cal O}(\beta)$ term of the entanglement entropy of island surface and then computed the derivatives of the Lagrangian with respect to embedding function and derivatives of the embedding function. In \ref{PT}, we have listed all the possible terms that we obtain in the holographic entanglement entropy for Hartman-Maldacena-like and island surfaces in \ref{PT-HM} and \ref{PT-IS}.

\section{Setup}
\label{setup}
In this section we will first summarise (in \ref{M-theory-dual}) the type IIB/IIA mirror dual  of thermal QCD-like theories (i.e., IR-confining, UV-conformal with quarks in the fundamental representation of flavor and color) and its no-braner ${\cal M}$-theory uplift worked out in \cite{metrics,MQGP,HD-MQGP}. We will then discuss (in \ref{DHS}) the doubly holographic setup in the ${\cal M}$-theory uplift. \par

\subsection{Top-Down UV-Complete Holographic Dual of Thermal QCD-Like Theories}
\label{M-theory-dual}
From the point of view of construction of a string/${\cal M}$-theory holographic dual truly close to realistic thermal QCD-like theories, one needs to consider finite gauge coupling ($g_s$) and finte number of colors ($N$). From the point of view of gauge-gravity duality, this entails looking at the strong-coupling limit of string theory - ${\cal M}$ theory - the same was dubbed as the "MQGP limit''  in \cite{MQGP} wherein $g_s\stackrel{<}{\sim}1, N_f \equiv {\cal O}(1), (N, M) \gg1$ such that $\frac{g_s M^2}{N}\ll1$. 

\begin{itemize}
\item {\bf Brane configuration of \cite{metrics}}:
Brane setup of \cite{metrics} has $N$ $D3$-branes, $M\ D5$-branes and $M\ \overline{D5}$-branes, $N_f\ D7$-branes $N_f\ \overline{D7}$-branes. Worldvolume coordinates are the aforementioned $D$-branes are summarised in Table 1. $M\ D5$-branes and $M\ \overline{D5}$-branes are wrapping the same vanishing two cycle  $S^2(\theta_1,\phi_1)$ but located at antipodal points ( average separation is ${\cal R}_{D5/\overline{D5}}$) of resolving $S^2_a(\theta_2,\phi_2)$. ${\cal R}_{D5/\overline{D5}}$ is the boundary between UV ($r>{\cal R}_{D5/\overline{D5}}$) and IR ($r<{\cal R}_{D5/\overline{D5}}$) on the gravity dual side in terms of radial coordinate. Flavor $D7$-branes were embedded holomorphically \cite{ouyang} in the resolved conifold geometry (\ref{Ouyang-definition}):
\begin{equation}
\label{Ouyang-definition}
\left(r^6 + 9 a^2 r^4\right)^{\frac{1}{4}} e^{\frac{i}{2}\left(\psi-\phi_1-\phi_2\right)}\sin\left(\frac{\theta_1}{2}\right)\sin\left(\frac{\theta_2}{2}\right) = \mu_{\rm Ouyang},
\end{equation}
where $\mu_{\rm Ouyang}$ is the Ouyang embedding parameter.  $N_f\ \overline{D7}$-branes are present only in the UV and UV-IR interpolating region whereas $N_f\ D7$-branes are present in the UV, UV-IR interpolating region and in the IR too. This provides a realization of chiral symmetry breaking in the setup and UV conformality in the theory which we will explain in the next bullet.
\begin{table}[h]
\begin{center}
\begin{tabular}{|c|c|c|}\hline
&&\\
S. No. & Branes & World Volume \\
&&\\ \hline
&&\\
1. & $N\ D3$ & $\mathbb{R}^{1,3}(t,x^{1,2,3}) \times \{r=0\}$ \\
&&\\  \hline
&&\\
2. & $M\ D5$ & $\mathbb{R}^{1,3}(t,x^{1,2,3}) \times \{r=0\} \times S^2(\theta_1,\phi_1) \times {\rm NP}_{S^2_a(\theta_2,\phi_2)}$ \\
&&\\  \hline
&&\\
3. & $M\ \overline{D5}$ & $\mathbb{R}^{1,3}(t,x^{1,2,3}) \times \{r=0\}  \times S^2(\theta_1,\phi_1) \times {\rm SP}_{S^2_a(\theta_2,\phi_2)}$ \\
&&\\  \hline
&&\\
4. & $N_f\ D7$ & $\mathbb{R}^{1,3}(t,x^{1,2,3}) \times \mathbb{R}_+(r\in[|\mu_{\rm Ouyang}|^{\frac{2}{3}},r_{\rm UV}])  \times S^3(\theta_1,\phi_1,\psi) \times {\rm NP}_{S^2_a(\theta_2,\phi_2)}$ \\
&&\\  \hline
&&\\
5. & $N_f\ \overline{D7}$ & $\mathbb{R}^{1,3}(t,x^{1,2,3}) \times \mathbb{R}_+(r\in[{\cal R}_{D5/\overline{D5}}-\epsilon,r_{\rm UV}]) \times S^3(\theta_1,\phi_1,\psi) \times {\rm SP}_{S^2_a(\theta_2,\phi_2)}$ \\
&&\\  \hline
\end{tabular}
\end{center}
\caption{Brane Configuration of \cite{metrics}}
\end{table}

\item
Based on earlier discussion, we can see that product color and flavor gauge groups are $SU(N+M)\times SU(N+M)$ and $SU(N_f)\times SU(N_f)$ when we are in the UV, i.e., $r>{\cal R}_{D5/\overline{D5}}$. When one is going from UV ($r>{\cal R}_{D5/\overline{D5}}$) to IR ($r<{\cal R}_{D5/\overline{D5}}$) via RG group flow then $SU(N_f)\times SU(N_f)$ changes into  $SU(N_f)$ (because there are no  $N_f\ \overline{D7}$-branes in the IR). This process can be interpreted as chiral symmetry breaking in the setup. Effect of transition from UV to IR at $r={\cal R}_{D5/\overline{D5}}$  on the product color gauge group is that it changes $SU(N+M)\times SU(N+M)$ to $SU(N+M)\times SU(N)$ (because no  $M \ \overline{D5}$-branes in the IR). This triggers Sieberg-like duality in the setup and if we perform repeated Sieberg-like dualities then in the IR we will be left with $SU(M)$ gauge theory where $M$ can be identified with number of colors in the theory and $M$ can be set to 3 \cite{Misra+Gale} in the "MQGP'' limit discussed later in this subsection.

\item Now let us see how we can say that the theory is non-conformal. This can be seen from the following RG-flow equations of the gauge couplings of the product gauge group:
\begin{eqnarray}
\label{nc}
\hskip -0.3in 4\pi^2 \left( \frac{1}{g^2_{SU(N+M)}}+\frac{1}{g^2_{SU(N)}}\right)e^\phi \sim \pi ; 
4\pi^2 \left( \frac{1}{g^2_{SU(N+M)}} - \frac{1}{g^2_{SU(N)}}\right)e^\phi \sim \frac{1}{2\pi\alpha^{'}} \int_{S^2} B_2.
\end{eqnarray}
If $\int_{S^2} B_2 =0$, then we will have a conformal theory and this is true in the UV in this setup because of presence of $M$ $\overline{D5}$-branes in the UV. In the IR, $\int_{S^2} B_2 \neq 0$, which implies non-conformality in the theory from (\ref{nc}). $N_f\ \overline{D7}$-branes cancels the net $N_f$ flavor $D7$-branes charges in the UV which corresponds to constancy of dilaton in the UV.

\item
{\bf String dual of thermal QCD-like theories \cite{metrics}}: On the gauge theory side we want to have finite temperature QCD. As we know that there are two phases of QCD: confining phase ($T<T_c$) and the deconfined phase ($T>T_c$). In the deconfined phase, finite temperature can be introduced by a black hole in the holographic dual side and in the confining phase, finite temperature can be introduced by a thermal background in the holographic dual side. In addition to finite temperature there is finite separation between the $M\ D5$-branes and $M\ \overline{D5}$, this corresponds to non-trivial resolution of the conifold geometry. Further, we need IR confinement in thermal QCD and it will be obtained on the gauge theory side by deforming the three cycle in the gravity dual side. From the aforementioned discussion it is clear that gravity dual of type IIB setup is a resolved warped deformed conifold. Further, we have fluxes (which contains the backreaction(also in the warp factor)) in the IR which are coming from $D3$-branes and the $D5$-branes.

 \item
{\bf Number of colors, i.e.,} ${\bf N_c=3}$: 
Number of colors $N_c$ can be identified with $M$ when Seiberg-like duality cascade ends in the IR \cite{Misra+Gale}, and $M$ can be set to 3 in the "MQGP'' limit. The idea is as follows: we can write  $N_c$ as the sum of effective number of $D3$-branes and $D5$-branes, i.e., $N_c = N_{\rm eff}(r) + M_{\rm eff}(r)$, where $N_{\rm eff}(r)$ and $M_{\rm eff}(r)$ are defined from $\tilde{F}_5$ and $\tilde{F}_3$:
\begin{eqnarray}
\label{F5tilde+M}
& & \hskip -0.3in \tilde{F}_5\equiv dC_4 + B_2\wedge F_3 = {\cal F}_5 + *{\cal F}_5;{\cal F}_5\equiv N_{\rm eff}\times{\rm Vol}({\rm Base\ of\ Resolved\ Warped\ Deformed\ Conifold}) ,\nonumber\\
&  & \hskip -0.3in  M_{\rm eff}(r) = \int_{S^3}\tilde{F}_3,
\end{eqnarray}
where $\tilde{F}_3 (\equiv F_3 - \tau H_3)\propto M_{\rm eff}(r)\equiv M\frac{1}{1 + e^{\alpha(r-{\cal R}_{D5/\overline{D5}})}}, \alpha\gg1$ with ${\cal R}_{D5/\overline{D5}}$ being the $D5-\overline{D5}$ separation along the blown-up $S^2$ which in turn is estimated to be $\sqrt{3}a$ in the type IIB dual with the same therefore acting as the outer boundary of the UV-IR interpolating region in same as well as its type IIA mirror gravity dual and its ${\cal M}$-theory uplift, and $S^3$ is dual to \\$\ e_\psi\wedge\left(\sin\theta_1 d\theta_1\wedge d\phi_1 - B_1\sin\theta_2\wedge d\phi_2\right)$, where $e_\psi\equiv d\psi + {\rm cos}~\theta_1~d\phi_1 + {\rm cos}~\theta_2~d\phi_2$) and $B_1$ is an `asymmetry factor' defined in \cite{metrics}. Since $N_{\rm eff} \in [N_{\rm UV},0_{\rm IR}]$ and $M_{\rm eff} \in [0_{\rm UV},M_{\rm IR}]$, therefore $N_c=M_{IR}$, where $M_{IR}$ implies that in the IR, number of colors is equal to $M$ and this can be set to 3 in the ``MQGP'' limit. Further we can also take $N_f=2(u/d)+1(s)$. Therefore we can see that $\frac{N_f}{N_c}$ is fixed in the IR (Veneziano-like limit \cite{Nitti_et_al}; but $N_f$ and $N_c$ remain finite themselves).

\item
{\bf The MQGP limit, Type IIA Strominger-Yau-Zaslow (SYZ) mirror  of \cite{metrics} and its  ${\cal M}$-theory  uplift at intemediate gauge coupling}:
\begin{itemize}
\item
The MQGP limit is defined as below \cite{MQGP, NPB}:
\begin{equation}
\label{MQGP_limit}
g_s\sim\frac{1}{{\cal O}(1)}, M, N_f \equiv {\cal O}(1),\ N \gg1,\ \frac{g_s M^2}{N}\ll1, \frac{\left(g_sM^2\right)g_sN_f}{N}\ll1.
\end{equation}
MQGP limit (\ref{MQGP_limit}) involves finite string coupling (intermediate gauge coupling on gauge theory side) and finite $M$ which is number of colors in the IR, i.e., $N_c=M$. Therefore ${\cal M}$ theory dual is holographic dual of thermal QCD-like theories at intermediate gauge coupling \cite{HD-MQGP}. 
\item
One can obtain the ${\cal M}$-theory uplift of type IIB setup by constructing the type IIA Strominger-Yau-Zaslow (SYZ) mirror (which is obtained by implementing triple $T$-duality along the three isometry directions)  setup of type IIB setup and then uplift the type IIA mirror to ${\cal M}$-theory. Three isometry directions produces a local special Lagrangian (sLag) $T^3$ $-$ which could be identified with the $T^2$-invariant sLag of \cite{M.Ionel and M.Min-OO (2008)} with a large base ${\cal B}(r,\theta_1,\theta_2)$ (of a $T^3(\phi_1,\phi_2,\psi)$-fibration over ${\cal B}(r,\theta_1,\theta_2)$) \cite{NPB,EPJC-2}\footnote{ In \cite{Misra+Gale} it was discussed in a footnote that to implement SYZ mirror symmetry idea is similar to \cite{SYZ-free-delocalization}, we need to replace the pair of $S^2$s by pair of $T^2$s from which we can get the correct $T$-duality coordinates. After getting the type IIA mirror, similar to the \cite{SYZ-free-delocalization} pertaining to $D5$-branes wrapping a vanishing $S^2$, uplift it to ${\cal M}$-theory with a bonafide $G_2$-structure that is free of the angular delocalization. One can descend back to type IIA which will therefore be free of delocalization now.} Three isometry directions are $x, y, z$ which are toroidal analogue of $\phi_1,\phi_2,\psi$ \footnote{As explained in \cite{Knauf-thesis}, the $T^3$-valued $(x, y, z)$ (used for effecting SYZ mirror in \cite{MQGP, NPB}) are defined via:
\begin {eqnarray*}
\label{xyz-definitions}
& & \phi_1 = \phi_{10} + \frac{x}{\sqrt{h_2}\left[h(r_0,\theta_{10,20})\right]^{\frac{1}{4}} \sin\theta_{10}\ r_0},\nonumber\\
& & \phi_2 = \phi_{20} + \frac{y}{ \sqrt{h_4}
\left[h( r_0,\theta_{10,20})\right]^{\frac{1}{4}}\sin\theta_{20}\ r_0}\nonumber\\
& & \psi = \psi_0 + \frac{z}{\sqrt{h_1} \left[h( r_0,\theta_{10,20})
\right]^{\frac{1}{4}}\ r_0},
\end {eqnarray*}
$h_{1,2,4}$ defined in \cite{metrics}, and one works up to linear order in $(x, y, z)$. In the IR, it can be shown \cite{theta0-theta} that $\theta_{10,20}$ can be promoted to global coordinates $\theta_{1,2}$ in all the results in the paper.}. %We know that if we apply $m$ $T$ dualities along the $Dp$-brane then we will get $D(p-m)$-brane and if we apply the $n$ $T$ duality normal to the worldvolume coordinates of the $Dp$-brane then we will obtain $D(p+n)$-brane. 
From Table {\bf 1} we see that $T$ duality along the $\psi$ direction converts the $N$ $D3$-branes, $M$ fractional $D3$-branes and $N_f$ flavor $D7$-branes into $N\ D4$-branes wrapping the $\psi$ circle, $M\ D4$-branes straddling a pair of orthogonal $NS5$-branes and $N_f$ flavor $D6$-branes. Now, from second and third $T$ duality along the $\phi_1$ and $\phi_2$ directions, we obtain $N$ $D6$-branes, $M$ $D6$-branes and $N_f$ $D6$-branes "wrapping" a non-compact three-cycle  $\Sigma^{(3)}(r, \theta_1, \phi_2$). Now if we uplift the type IIA mirror to ${\cal M}$-theory then it was found that $D6$-branes are converted into KK monopoles (variants of Taub-NUT spaces). Therefore we have no branes in the ${\cal M}$-theory and we will be left with ${\cal M}$-theory with a $G_2$-structure manifold - a no-braner uplift.

\end{itemize}

\item 
{\bf ${\cal M}$-theory uplift including ${\cal O}(R^4)$ and $G$-Structure Torsion Classes}
\begin{itemize}
\item
One of us (AM) along with V.~Yadav, worked out $O(l_p^6)$ corrections to the ${\cal M}$-theory ${\cal M}$-theory metric in the ``MQGP'' limit in \cite{HD-MQGP} and using the the aforementioned  ${\cal M}$-theory  metric, $G$-structure torsion classes were worked out for the six-, seven- and eight-folds of the ${\cal M}$-theory uplift and summarized in Table 2. The eleven-fold $M_{11}$ in the ${\cal M}$ theory uplift obtained in \cite{MQGP} turns out to be a warped product of $S^1(x^0)\times \mathbb{R}_{\rm conformal}$
and $M_7(r,\theta_{1,2},\phi_{1,2},\psi,x^{10})$, the latter being a cone over $M_6(\theta_{1,2},\phi_{1,2},\psi,x^{10})$ with $M_6(\theta_{1,},\phi_{1,2},\psi,x^{10})$ possessing the following nested fibration structure:
{\footnotesize
\begin{equation}
\hskip -0.4in
\label{M_6}
\begin{array}{cc}
&{\cal M}_6(\theta_{1,2},\phi_{1,2},\psi,x_{10})   \longleftarrow   S^1(x^{10}) \\
&\downarrow  \\
& \hskip 0.7in {\cal M}_5(\theta_{1,2},\phi_{1,2},\psi) \longleftarrow  {\cal M}_3(\phi_1,\phi_2,\psi)  \\
&\downarrow  \\
 &{\cal B}_2(\theta_1,\theta_2)
\end{array}.
\end{equation}
}
As shown in \cite{MQGP}, $p_1^2(M_{11}) = p_2(M_{11}) = 0$ up to ${\cal O}(\beta^0)$ where $p_a$ is the $a$-th Pontryagin class of $M_{11}$. This hence implied that $X_8=0$ in (\ref{D=11}) up to ${\cal O}(\beta^0)$.
\begin{table}[h]
\begin{center}
\begin{tabular}{|c|c|c|c|}\hline
S. No. & Manifold & $G$-Structure & Non-Trivial Torsion Classes \\ \hline
1. & $M_6=$\ {\rm non-K\"{a}hler\ conifold} & $SU(3)$ & $T^{\rm IIA}_{SU(3)} = W_1 \oplus W_2 \oplus W_3 \oplus W_4 \oplus W_5: W_4 \sim W_5$ \\ \hline
2. & $M_7 = S^1_{\cal M}\times_w M_6$ & $G_2$ & $T^{\cal M}_{G_2} = W_1 \oplus W_7 \oplus W_{14} \oplus W_{27}$ \\ 
&&& \hskip 0.3in $\downarrow r\sim (1+ \alpha_a)a,\ \alpha_a\sim 0.1-0.3$ \\
&&& $W_1={\cal O}\left(\frac{1}{N^{\alpha_{W_1}}}\right), \alpha_{W_1}>1,$ \\ 
&&&  {\rm therefore\ disregarded\ up to}\ ${\cal O}\left(\frac{1}{N}\right)$;\\
& & & $W_7\approx0$ \\ 
&&&\hskip -0.5in $\Rightarrow T^{\cal M}_{G_2} = W_{14} \oplus W_{27}$ \\ \hline
3. & $M_8 = S^1_t \times_w M_7$ & $SU(4)$ & $T^{\cal M}_{SU(4)} = W_2 \oplus W_3 \oplus W_5$ \\ \hline
4. & $M_8$ & $Spin(7)$ & $T^{\cal M}_{Spin(7)} = W_1 \oplus W_2$ \\ \hline
\end{tabular}
\end{center}
\caption{IR $G$-Structure Classification of Six-/Seven-/Eight-Folds in the type IIA/${\cal M}$-Theory Duals of Thermal QCD-Like Theories (at High Temperatures) and for $r\in$IR; $\times_w$ denotes a warped product}
\label{G-Structures-6+7+8-folds}
\end{table}
\end{itemize}
\end{itemize}

From the above discussions we find that the type IIB setup has the following properties: IR confinement, UV conformality, quarks transform in the fundamental representation of flavor and color groups and well defined in the confining and deconfined phases. Therefore it is an ideal holographic dual of thermal QCD-like theories.

\begin{figure}
\begin{center}
\includegraphics[width=0.8\textwidth]{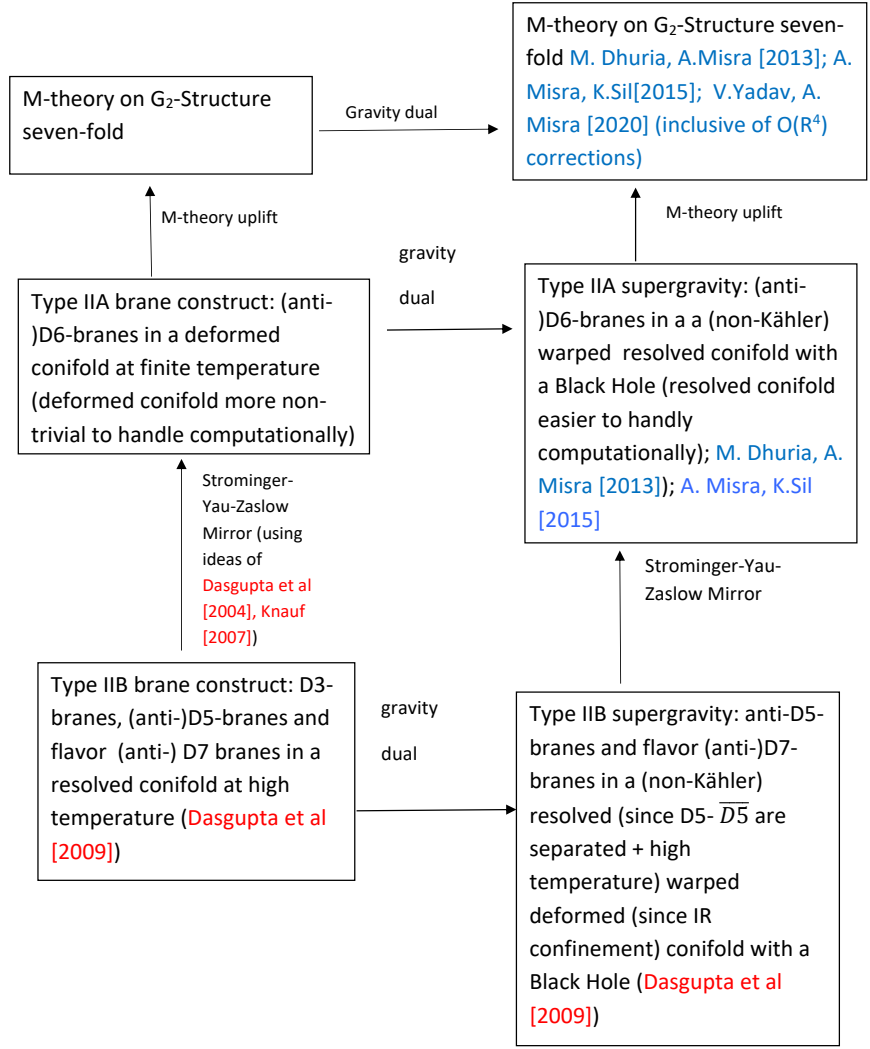}
\end{center}
\vskip -0.4in
\caption{Type IIB dual of large-$N$ thermal QCD like theories, its SYZ type IIA mirror and ${\cal M}$-theory uplift}
\label{Flowchart}
\end{figure}

The type IIB setup of \cite{metrics}, its Strominger-Yau-Zaslow type IIA mirror and its ${\cal M}$-theory uplift as constructed in \cite{MQGP}, \cite{NPB} as well as \cite{HD-MQGP} (${\cal M}$-theory uplift at intermediate `t Hooft coupling effected by inclusion of ${\cal O}(R^4)$ terms) are summarized in Fig. \ref{Flowchart}.

\subsection{Doubly Holographic Setup}
\label{DHS}
Let us first review the doubly holographic setup \cite{Island-HD}. There are three equivalent descriptions of doubly holographic models which are summarised below:
\begin{enumerate}
\item A bath which is a boundary CFT(BCFT) living at the boundary of $AdS_{d+1}$ whose boundary is (d-1) dimensional defect \cite{McAvity+Osborn,Cardy},  to collect the Hawking radiation .
 
\item Conformal field theory in $d$ dimensions is coupled to gravity on an ``End-of-The-World'' (ETW) brane in an asymptotically $AdS_d$ space ${\cal M}_d$ with a half-space bath CFT coupled to ${\cal M}_d$ via ``transparent boundary conditions'' at the defect (the intersection of the ETW-brane and the bath).

\item Einstein gravity in ${\cal M}_{d+1}$ asymptotic to $AdS_{d+1}$, the holographic dual of $d$-dimensional BCFT, containing ${\cal M}_d$ as an ETW brane \cite{RS,KR1,KR2}.
\end{enumerate}
Let $S({\cal R})$ be the von Neumann entropy of subregion ${\cal R}$ defined on constant time slice in description 1.  In the second description it will be given by the island-rule (${\cal I}$) \cite{AMMZ} as,
\begin{equation}
S({\cal R}) = {\rm min}_{\cal I}~{\rm ext}_{\cal I}~S_{gen}({\cal R} \cup {\cal I}),
\end{equation}
where generalised entropy functional is defined as \cite{EW}:
\begin{equation}
S_{gen}({\cal R} \cup {\cal I})= \frac{A(\partial {\cal I})}{4 G_N}+ S_{matter}({\cal R} \cup {\cal I}),
\end{equation}
where the first term in the above equation is the area of the boundary of the island surface and second term is the matter contribution from bath as well as island regions. In the third description, generalised entropy functional defined in semiclassical geometry can be computed classically using Ryu-Takayanagi formula in $(d+1)$ dimensions \cite{RT} from area of the co-dimension two minimal surface:
\begin{equation}
S_{gen}({\cal R} \cup {\cal I})= \frac{A(\gamma)}{4 G_N^{(d+1)}},
\end{equation}
where $\gamma$ is the co-dimension two surface in the ${\cal M}_{(d+1)}$ dimensional bulk space-time. Therefore one can calculate very easily entanglement entropy of the Hawking radiation in doubly holographic models using classical Ryu-Takayanagi formula in third description. \par
There is also a similar setup available in the literature as constructed in \cite{Bath-WCFT}. In \cite{Bath-WCFT}, authors constructed a similar setup as discussed earlier but in this setup bath is a warped conformal field theory(WCFT) \cite{WCFT} instead of CFT. This setup has also three equivalent descriptions which are summarised below:
\begin{enumerate}
\item Boundary warped conformal field theory(BWCFT) in two dimensions with one dimensional boundary.
\item Two dimensional JT gravity coupled to WCFT in two dimensions at the interface point via transparent boundary conditions.
\item Einstein gravity  on ${\cal M}_{3}$ asymptote to $AdS_{3}$ space containing ${\cal M}_2$ as a Planck brane \cite{AMMZ}.
\end{enumerate}

Similar to earlier discussion one can calculate the entanglement entropy of the Hawking radiation classically in the third description when one has warped conformal field theory in two dimensions as a bath using Ryu-Takayanagi formula \cite{RT}. Doubly holographic models have been studied in the literature mentioned in section \ref{introduction} and also in \cite{Ling+Liu+Xian,Omiya+Wei,RE-DH,Phase-BCFT,critical-islands,IITK1,IITK2,Liu et al,Li+Yang,JT-KR,Deng+An+Zhou,Geng}.\par

Motivated by these constructions we are generalising these kinds of setups to our case. After a double Wick-rotation, we have ${\rm QCD}_{2+1}$ at r=0, along $x^{1,2}$ and Wick-rotated $x^3$. One can think of a fluxed ETW-hypersurface or ``brane" $M_{10}=\mathbb{R}^2(x^{2,3}) \times_w M_8$ where $M_8$ is the (non-compact) $SU(4)/Spin(7)$-structure (\cite{HD-MQGP}, Table 2) eight-fold $M_8^{SU(4)/Spin(7)}(r,t,\theta_{1,2},\phi_{1,2},\psi,x^{10})$ at $x^1=0$ which has a black hole. The ETW-hypersurface can be interpreted as $\mathbb{R}^2(x^{2,3})\times_w M_8^{SU(4)/Spin(7)}(t,r,\theta_{1,2},\phi_{1,2},\psi,x^{10})$ containing black $M5$-brane  at $x^1=0$ with world volume $\Sigma^{(6)}=S^1(t)\times_w\mathbb{R}_{\geq0}(r)\times\Sigma^{(4)}$ where $\Sigma^{(4)}= n_1S^3(\theta_1,\phi_1,\psi) \times[0,1]_{\theta_2}+n_2S^2(\theta_1,\phi_1)\times S^2(\theta_2,x^{10})$ with $n_1$ determined by $\int_{S^3\times[0,1]}G_4$ and $n_2$ by $\int_{S^2\times S^2}G_4$; the ${\rm QCD}_{2+1}$ can be thought of as living on $M2$-branes with world volume $\Sigma^{(3)}(x^{1,2,3})$. ${\rm QCD}_{2+1}$ at $r=0$ would interact gravitationally via the pull-back of the ambient $M_{11} =\mathbb{R}^{1,2,3}\times_w M_8 = \mathbb{R}(x^1)\times_w M_{10}$ metric used to contract the non-abelian field strength in the gauge kinetic term obtained as part of the $M2$-brane($x^{1,2,3}$) world-volume action. This has some similarity with points 2. and 3. of the first paragraph of this section as regards the doubly holographic setup as described in \cite{Island-HD}. Let us be more specific and briefly describe the three equivalent descriptions alluded to towards the beginning of this sub-section. The doubly holographic setup constructed from the bottom-up approach (usually followed in the literature) as described at the beginning of this sub-section, has the following ${\cal M}$-theory description (the one we follow) of top-down double holography with $QCD_{2+1}$ bath  with the numbering matches the one used in the aforementioned three equivalent description:

\begin{enumerate}
\item {\bf Boundary-like Description:} $QCD_{2+1}$ (could be thought of as supported on an $M2$-brane with world-volume $\Sigma^{(3)}(x^{1, 2, 3})$, and) is living at the tip ($r=0$) of a non-compact seven-fold of $G_2$ structure which is a cone over a warped non-K\"{a}hler resolved conifold. The two-dimensional ``defect'' $\Sigma^{(2)}\cong\Sigma^{(3)}(x^{2, 3}; x^1=0)\cong \mathbb{R}^{2}(x^{2,3})$.  

\item {\bf Non-Conformal Bath-ETW Interaction Description:} Fluxed End-of-The-World (ETW)  hypersurface  $M_{10}\cong \mathbb{R}^2(x^{2,3})\times_w M_8^{SU(4)/Spin(7)}(t,r,\theta_{1,2},\phi_{1,2},\psi,x^{10})$ containing black $M5$-brane  at $x^1=0$ coupled to  $QCD_{2+1}$ bath living on $M2$ brane with world volume $\Sigma^{(3)}(x^{1, 2, 3})$ along the defect $\left.\Sigma^{(2)}\cong
M_{10}\cap \Sigma^{(3)}(x^{1, 2, 3})\right|_{x^1=0}$ via exchange of massless graviton.

\item {\bf Bulk Description:}  $QCD_{2+1}(x^{1, 2, 3})$ has holographic dual which is eleven dimensional ($M_{11}\cong \mathbb{R}(x^1)\times_w M_{10}\cong S^1(t)\times_w\mathbb{R}^3(x^{1, 2, 3})\times_w M_7^{G_2}(r,\theta_{1,2},\phi_{1,2},\psi,x^{10})$) ${\cal M}$-theory background (compactified on a seven-fold with $G_2$ structure) containing the fluxed ETW-hypersurface $M_{10} (x^1=0)$.
\end{enumerate}

\begin{figure}
\begin{center}
\includegraphics[width=1.05\textwidth]{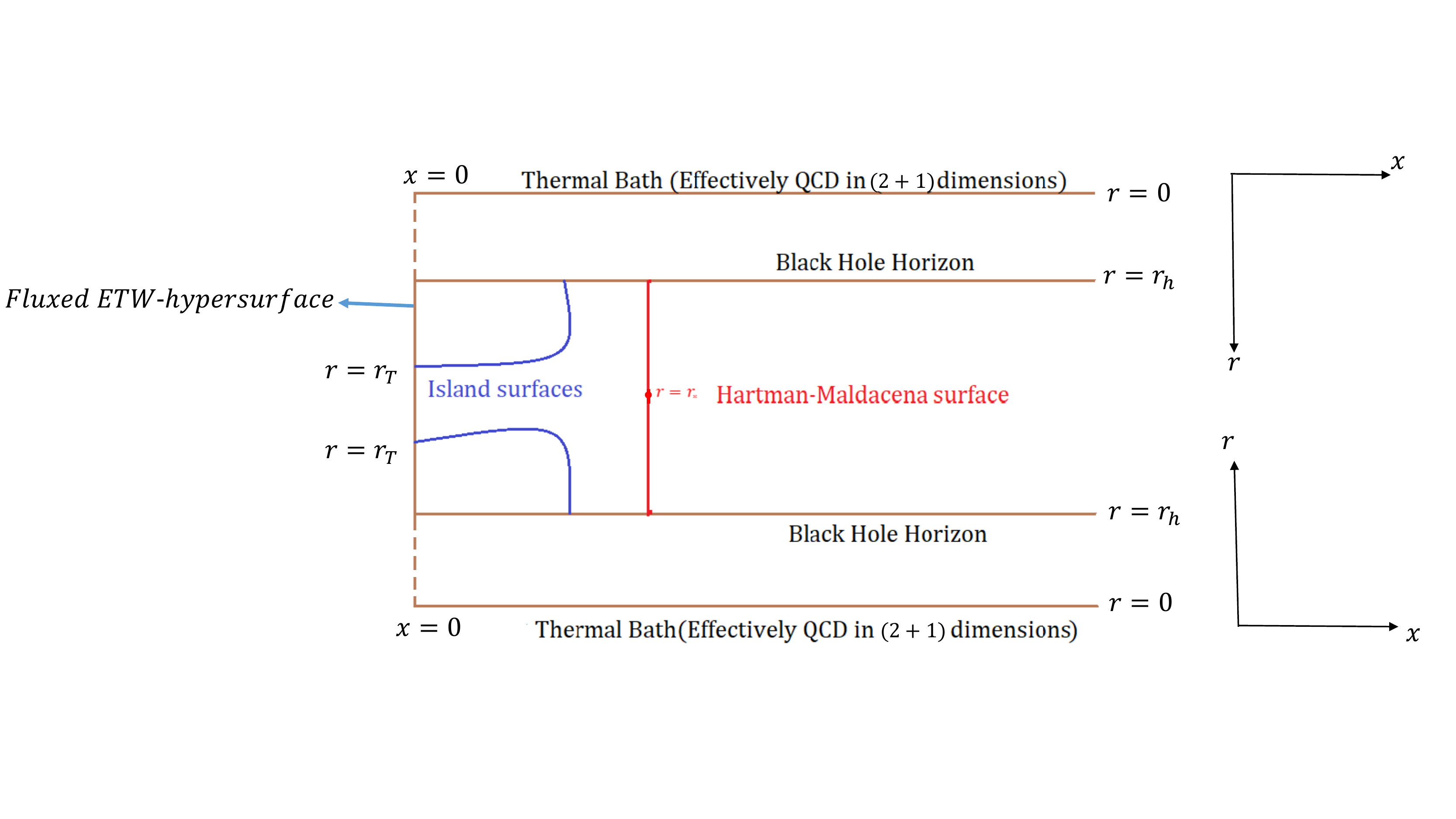}
\end{center}
\caption{Doubly holographic setup in ${\cal M}$-theory dual. ETW-black ``brane''  is coupled the thermal bath (at the tip of seven-fold of $G_2$-structure which is a cone over a warped non-K\"{a}hler resolved conifold), where  the Hawking radiation is collected, which effectively is thermal $QCD_{2+1}$ along $x^{1, 2, 3}$ after Wick-rotation along $x^3$ at $r=0$.   Blue curves correspond to the island surfaces and red curve corresponds to the Hartman-Maldacena-like surface.}
\label{Doubly Holographic Setup}
\end{figure}

The pictorial representation of the aforementioned ETW-brane(containing a black $M5$-brane)/$M2$-brane(supporting the non-conformal ``bath'' - $QCD_{2+1}$)-setup focusing only on the $x^1-r$-plane, along with the Hartmann-Maldacena-like surface and island surface, is given in Fig. \ref{Doubly Holographic Setup}. Unlike \cite{AMMZ}, \cite{Island-HD} and \cite{G-bath}, the (non-conformal) bath is at $r=0$ instead of the UV cutoff $r=r_{\rm UV}$\footnote{The cut-off, unlike most references in the literature is not at infinity but is such that $r_{\rm UV}\stackrel{<}{\sim}L\equiv\left(4\pi g_s N\right)^{1/4}$ thereby justifying dropping the ``1'' in the 10-dimensional warp factor $h$  in \cite{metrics}, \cite{MQGP}, \cite{HD-MQGP} [that appears in (\ref{TypeIIA-from-M-theory-Witten-prescription-T>Tc})], which otherwise would have been $1 + \frac{L^4}{r^4}\left[1 + {\cal O}\left(\frac{g_sM^2}{N}\right)\right]$.}. In the high temperature ($T>T_c$) ${\cal M}$-theory dual (\ref{TypeIIA-from-M-theory-Witten-prescription-T>Tc}), evidently $r\geq r_h$ (the metric component $g_{tt}^{\cal M}$ being proportional to a warp factor $e^B$ where $B\sim\log \left(1 - \frac{r_h^4}{r^4} + {\cal O}\left(\frac{g_sM^2}{N}\right)\right)$ from the solutions to the supergravity EOM implies that for $B\in\mathbb{R}, r\geq r_h$ in the MQGP limit (\ref{MQGP_limit})).\footnote{Further, as will be shown later, the area/entanglement entropy (inclusive of ${\cal O}(R^4)$ contributions) of the HM surface (see (\ref{AHM_i}), (\ref{Page-curve-beta0}), (\ref{SEE-HM-beta0-simp}), (\ref{SEE-HM-simp-beta0}) and (\ref{SEEHMbeta-iii})) as well as the IS (see 
(\ref{AIS-i}), (\ref{EE-IS-simp}), (\ref{on-shell-L-beta0}) and (\ref{SEEISbeta-iii})) are proportional to a positive power of $M$ - the number of fractional $D3$-branes in the parent type IIB dual \cite{metrics}. Now, the contribution to $r$ integral from $r\in$ UV, i.e., $r>\sqrt{3}a\sim \biggl[1 + {\cal O}\left(\frac{\sqrt{\beta}}{N}\right) + {\cal O}\left(\frac{g_sM^2}{N}\right)\biggr]r_h\sim\left(1 + \epsilon\right)r_h, 
|\epsilon|\ll1$ \cite{MChPT} will involve replacing $M$ by $M^{\rm UV}\equiv M_{\rm eff}(r\in{\rm UV}\Leftrightarrow r>\sqrt{3}a)$ - the UV-valued effective number of fractional $D3$-branes in the aforementioned parent type IIB dual - which is vanishing small (see the discussion beneath (\ref{F5tilde+M})) ensuring UV conformality. Therefore all integrals $\int^{r_{\rm UV}}_{r_*\ {\rm or}\ r_T}$ of integrands relevant to the HM/IS areas or entanglement entropies, will vanish as $r_{*, T}>r_h$ though being nearer to $r_h$ than $\sqrt{3}a$.}

 If $S({\cal R})$ be the von Neumann entropy of subregion ${\cal R}$ defined on constant time slice in description 1 then in the second description it will be given by the island-rule \cite{AMMZ} as:
\begin{equation}
S({\cal R}) = {\rm min}_{\cal I}~{\rm ext}_{\cal I}~S_{gen}({\cal R} \cup {\cal I}),
\end{equation}
where ${\cal I}$ is the island surface and generalised entropy functional is defined as \cite{EW}:
\begin{equation}
S_{gen}({\cal R} \cup {\cal I})= \frac{A(\partial {\cal I})}{4 G_N}+ S_{matter}({\cal R} \cup {\cal I}).
\end{equation}
As discussed earlier that in doubly holographic models we can calculate the generalised entropy functional defined in semiclassical geometry using classical Ryu-Takayanagi formula \cite{RT} in eleven dimensions in the third description as below:

\begin{equation}
\label{S-gen-WHD}
S_{gen}({\cal R} \cup {\cal I})= \frac{A(\gamma)}{4 G_N^{(11)}},
\end{equation}
where $A(\gamma)$ is the area of the minimal surface, $\gamma$, which is a co-dimension two surface in the eleven dimensional bulk space-time. By following \cite{PBD} we obtain the Page curve of a neutral  black hole using (\ref{S-gen-WHD}) without higher derivative terms in section \ref{Page-curve-WHD} from ${\cal M}$-theory dual. \par
 For higher derivative gravity theories we can calculate the entanglement entropy in the holographic dual theories from the results of \cite{Dong}. Since we are working with ${\cal M}$-theory dual therefore we can write the analogue of the generalised entropy functional defined in equation (\ref{S-gen-WHD}) using \cite{Dong} as below:
 \begin{equation}
\label{S-gen-HD}
S_{gen}({\cal R} \cup {\cal I})= S_{\rm gravity},
\end{equation}
where 
\begin{equation}
S_{\rm gravity} = \frac{S_{\rm EE}}{4 G_N^{(11)}},
\end{equation}
where $S_{\rm EE}$ can be calculated using \cite{Dong}. We will use equation (\ref{S-gen-HD}) to obtain the Page curve from ${\cal M}$-theory dual in the presence of higher derivative terms in section \ref{Page-curve-HD}. The idea is that we need to calculate equation (\ref{S-gen-HD}) for the Hartman-Maldacena-like surface which may produce a linear time growth of entanglement entropy of the Hawking radiation and for the island surface which dominates after the Page time. Combining these two contributions will generate the analogue of the Page curve.

\section{ETW-``brane'' Embedding}
\label{ETW-sub}
In the doubly-holographic approach, End of The World (ETW)-``brane" embedding is an important issue related to the existence of islands. The issue is well understood at the level of Einstein gravity at LO (i.e. Einstein-Hilbert(EH) action and the boundary Gibbons-Hawking-York surface terms), but not much is known for higher derivative gravity, e.g. $D=11$ supergravity at ${\cal O}(R^4)$;  in top-down models such as the one considered by us, the ETW-"brane'' would be a fluxed hypersurface ${\cal W}$ \footnote{One of us (GY) would like to thank C.~F.~Uhlemann for a brief clarification on this point.}.  We will now address this issue in this section as two subsections - in the first we will obtain the ETW embedding at the level of EH+GHY action, and in the second we will show that in the MQGP limit (\ref{MQGP_limit}), the aforementioned LO embedding receives no corrections. \par
The ${\cal N}=1, D=11$ supergravity action without higher derivative terms is given by:
\begin{eqnarray}
\label{D=11}
& & \hskip -0.8inS = \frac{1}{2\kappa_{11}^2}\int_{M_{11}} \sqrt{-G^{\cal M}}\left[  {\cal R} *_{11}\wedge 1 - \frac{1}{2}G_4\wedge *_{11}G_4 -
\frac{1}{6}C\wedge G\wedge G\right]+\frac{1}{\kappa_{11}^2}\int_{r=r_{\rm UV}} d^{10}x \sqrt{h} K \nonumber\\
& & +\frac{1}{\kappa_{11}^2}\int_{\cal W} d^{10}x \sqrt{\gamma} ({\cal K}-9T).
\end{eqnarray}
$\gamma$ , ${\cal K}$ and $T$ in (\ref{D=11}) are the induced metric, trace of extrinsic curvature (${\cal K}_{mn}$) and tension of the ETW brane (${\cal W}$: $\partial M_{11} = \left\{r=r_{\rm UV}\right\}\cup {\cal W}$). Metric for the ${\cal M}$-theory dual when $T>T_c$ on gauge theory side is given by:
\begin{eqnarray}
\label{TypeIIA-from-M-theory-Witten-prescription-T>Tc}
\hskip -0.1in ds_{11}^2 & = & e^{-\frac{2\phi^{\rm IIA}}{3}}\Biggl[\frac{1}{\sqrt{h(r,\theta_{1,2})}}\left(-g(r) dt^2 + \left(dx^1\right)^2 +  \left(dx^2\right)^2 +\left(dx^3\right)^2 \right)
\nonumber\\
& & \hskip -0.1in+ \sqrt{h(r,\theta_{1,2})}\left(\frac{dr^2}{g(r)} + ds^2_{\rm IIA}(r,\theta_{1,2},\phi_{1,2},\psi)\right)
\Biggr] + e^{\frac{4\phi^{\rm IIA}}{3}}\left(dx^{11} + A_{\rm IIA}^{F_1^{\rm IIB} + F_3^{\rm IIB} + F_5^{\rm IIB}}\right)^2,
\end{eqnarray}
where $A_{\rm IIA}^{F^{\rm IIB}_{i=1,3,5}}$ are the type IIA RR 1-forms obtained from the triple T/SYZ-dual of the type IIB $F_{1,3,5}^{\rm IIB}$ fluxes in the type IIB holographic dual of \cite{metrics}, and $g(r) = 1 - \frac{r_h^4}{r^4}$.  Near the $\psi=2n\pi, n=0, 1, 2$-coordinate patch, we can write the metric (\ref{TypeIIA-from-M-theory-Witten-prescription-T>Tc}) in the following form (explicit form of the metric in terms of the parameters of the model can be read off from the appendix {\bf A} of \cite{MCTEQ}):
\begin{equation}
\label{symbolic-metric}
ds^2=\alpha(r)[-g(r) dt^2+\sigma(r) dr^2+dx_{\mu}dx^{\mu}]+g_{mn}dx^mdx^n ,
\end{equation}
where $x_\mu(\mu=1,2,3)$ represent spatial coordinates, $r$ is the radial coordinate and $x^m(m=5,6,7,8,9,10)$ represent six angular coordinates $(\theta_{1,2},\phi_{1,2},\psi,x^{10})$ in the conifold geometry. The volume of the compact six-fold can be obtained by the following equation:
\begin{equation}
\label{V_int}
{\cal V}_{\rm int}=\int \prod_m dx^m \sqrt{-g} .
\end{equation}
On comparing equations (\ref{TypeIIA-from-M-theory-Witten-prescription-T>Tc}) and (\ref{symbolic-metric}), we obtain:
\begin{eqnarray}
\label{alpha+sigma}
& &
\alpha(r, \theta_{1,2})=\frac{e^{-\frac{2\phi^{\rm IIA}}{3}}}{\sqrt{h(r,\theta_{1,2})}},\nonumber \\
& &  \sigma(r, \theta_{1,2})=\frac{h(r,\theta_{1,2})}{g(r)},
\end{eqnarray}
where $\phi^{\rm IIA}$ is the type IIA dilaton profile and can be defined as $G_{x_{10}x_{10}}^{\cal M}=e^{\frac{4\phi^{\rm IIA}}{3}}$. For ${\cal M}$-theory dual, the expressions for $\alpha(r, \theta_{1,2}), \sigma(r,\theta_{1,2}), H(r,\theta_{1,2})\equiv {\cal V}_{\rm int}^2(r,\theta_{1,2})\alpha(r,\theta_{1,2})^3$ can be easily read off from (\ref{TypeIIA-from-M-theory-Witten-prescription-T>Tc}) and (\ref{alpha+sigma}).

The ETW-''brane'' ${\cal W}: AdS_4^\infty\times_w M_6$ with $G_4$ fluxes threading a homologous sum of four-cycles $S^3\times[0,1]$ and $S^2\times S^2$ in $M_6=M_5(\theta_{1,2}, \phi_{1,2},\psi)\times S^1(x^{10})\hookrightarrow M^{SU(4)/Spin(7)}_8(t,r,\theta_{1,2}, \phi_{1,2},\psi,x^{10})$ (in the large-$N$ MQGP limit) implies: $\partial M_{11}=\left\{r=r_{\rm UV}\right\}\cup {\cal W}$. Assuming the End of The World(ETW)-"Brane'' embedding,
\begin{equation}
\label{ETW_i}
x^1 = x^1(r),
\end{equation}
and substituting the same into (\ref{symbolic-metric}) yields:
\begin{equation}
\label{ETW_ii}
ds^2=\alpha(r)\left[-g(r) dt^2+\left(\sigma(r)+ \left(\frac{dx^1}{dr}\right)^2\right) dr^2+\sum_{\mu=2}^3dx_{\mu}dx^{\mu}\right]+g_{mn}dx^mdx^n. 
\end{equation}

\subsection{At ${\cal O}(\beta^0)$}
\label{ETW-sub-sub-i}
Assuming $T$ to be the tension of the ETW-``brane", and if ${\cal K} $ and ${\cal K}_{\tilde{m}\tilde{n}}, \tilde{m}, \tilde{n}=r, t, \mu, m$ are respectively the extrinsic curvature scalar and tensor on the end-of-the-world (ETW)-``brane", then \cite{Takayanagi-ETW},
\begin{equation}
\label{ETW_iii}
{\cal K}_{mn} - {\cal K} h_{mn} = - 9 T h_{mn},
\end{equation} 
where $h_{mn}$ is the induced metric on the end-of-the-world (ETW)-``brane". Let us look at the $rr$-component of (\ref{ETW_iii}) in the IR. One can argue that in the IR, the embedding function $x^1(r)=x(r)$ always appears as $x'(r),  x''(r)$ in (\ref{ETW_iii}),  and writing $a = \left(b + {\cal O}\left(\frac{g_sM^2}{N}\right)\right)r_h$ \cite{EPJC-2}, 
the terms LO and NLO in $N$ and
{\footnotesize
\begin{eqnarray}
\label{ETW_iv}
  {\cal K}\left(N, g_s; r_h\right) & \sim & -\kappa_K\frac{\left(3 b^2-1\right) \sqrt{ {r_h}} ( \log N -9 \log ( {r_h}))}{\sqrt[4]{ {g_s}} \sqrt[4]{N}
     \sqrt{r- {r_h}} \left(\left(3 b^2-1\right)  \log N +9 \log ( {r_h})\right) \sqrt[3]{  {N_f}
   ( \log N -3 \log ( {r_h}))}}\nonumber\\
%   & &  + {\cal O}\left(\sqrt{r-r_h}\right),\nonumber\\
\hskip -0.1in  {\cal K}_{rr}\left(N, M, N_f, g_s; r_h; x'(r);x''(r)\right) & \sim & -\kappa_{K_{rr}}^{(1)}\frac{\sqrt{{g_s} N} \Sigma(r_h; N, N_f)}{r_h (r- {r_h})^2} +\kappa_{K_{rr}}^{(2)} \frac{{g_s}^3 M^2 {N_f} \log ({r_h}) (\log N -12 \log ({r_h})) \Sigma(r_h; N, N_f)}{ {r_h}^2 \sqrt{{g_s} N}
   (r-{r_h})^2}\nonumber\\
   & & +\kappa_{K_{rr}}^{(3)}\frac{{r_h} x'(r) \Sigma(r_h; N, N_f)}{ \sqrt{{g_s} N}} +\kappa_{K_{rr}}^{(4)}\frac{{r_h}^2 x''(r)\Sigma(r_h; N, N_f)}{
   \sqrt{{g_s} N}},\nonumber\\
 h_{rr}\left(N, M, N_f, g_s; r_h; x'(r)\right) & \sim & -2\kappa_{K_{rr}}^{(1)}\frac{  \sqrt{g_s N} \Sigma(r_h; N, N_f)}{r_h (r- {r_h})}
  -\kappa_{K_{rr}}^{(3)} \frac{{r_h}^2 x'(r) \Sigma(r_h; N, N_f)}{ \sqrt{{g_s} N}} 
  \nonumber\\
    & & -2 \kappa_{K_{rr}}^{(2)}\frac{3 {g_s}^3 M^2 {N_f} \log
   ({r_h}) (\log N -12 \log ({r_h})) \Sigma(r_h; N, N_f)}{ {r_h} \sqrt{{g_s} N} (r-{r_h})},
\end{eqnarray}
}
wherein $\Sigma(r_h; N, N_f)\equiv \left(6  \log N    {N_f}-3   {N_f} \log \left(9 a^2
    {r_h}^4+ {r_h}^6\right)\right)^{2/3}$ the numerical pre-factors are respectively collected in $\kappa_{K}, \kappa_{K_{rr}^{(i=1,2,3,4)}}$. As $b=\frac{1}{\sqrt{3}}+\epsilon$, (e.g., in the $\psi=2n\pi, n=0, 1, 2$-coordinate patches, $|\epsilon|\sim r_h^2\left(|\log r_h|\right)^{9/2}N^{-9/10-\alpha_b}, \alpha_b>0$ \cite{HD-MQGP}), one can approximate  (\ref{ETW_iii}) by $K_{mn} \sim -9 T h_{mn}$. Looking then at $m=n=r$ component of the same, one notes that unlike  ${\cal K}_{rr}\left(N, M, N_f, g_s; r_h; x'(r);x''(r)\right)$, $ h_{rr}\left(N, M, N_f, g_s; r_h; x'(r)\right)$ does not have an $x''(r)$ term. Therefore,
\begin{equation}
\label{x''=0}
x''(r)=0.
\end{equation}
Further, we note that the LO-in-$N$ terms in the IR ($r={\cal O}(1)r_h$ of) ${\cal K}_{rr}\left(N, M, N_f, g_s; r_h; x'(r); x''(r)\right)$ are proportional to $h_{rr}\left(N, M, N_f, g_s; r_h;x'(r)\right)$ with a proportionality constant -2 resulting in the ETW-``brane" tension 
\begin{equation}
\label{T_ETW}
T\sim\frac{1}{r_h}.
\end{equation}
But, at NLO in $N$, the coefficients of the same are proportional to each other but with a proportionality constant that is -1. These two can be reconciled by $x'(r)=0$, i.e.,
%The LHS of (\ref{ETW_iii}) then turns out to be:
%\begin{eqnarray}
%\label{ETW_v}
%-\frac{\sqrt{ {g_s} N} \left(6  \log N    {N_f}-3   {N_f} \log \left(9 a^2
   % {r_h}^4+ {r_h}^6\right)\right)^{2/3}}{ (r- {r_h})^3},
%\end{eqnarray}
%and similarly, the RHS of (\ref{ETW_iii}) turns out to be:
%\begin{eqnarray}
%\label{ETW_vi}
%-\frac{T \sqrt{ {g_s} N} \left(6  \log N    {N_f}-3   {N_f} \log \left(9 a^2
%    {r_h}^4+ {r_h}^6\right)\right)^{2/3}}{ {r_h}^2}.
%\end{eqnarray}
\begin{equation}
\label{ETW_vii}
x^1(r)={\rm constant},
\end{equation}
and we take the constant to be zero.

\subsection{No Boundary Terms Generated at ${\cal O}(R^4)$}
\label{ETW-sub-sub-ii}
Eleven dimensional supergravity action including ${\cal O}(R^4)$ terms is given by:
{\footnotesize
\begin{eqnarray}
\label{D=11_O(l_p^6)}
& & \hskip -0.8inS = \frac{1}{2\kappa_{11}^2}\int_{M_{11}} \sqrt{-G^{\cal M}}\left[  {\cal R} *_{11}1 - \frac{1}{2}G_4\wedge *_{11}G_4 -
\frac{1}{6}C\wedge G\wedge G\right] + \frac{1}{\kappa_{11}^2}\int_{r=r_{\rm UV}} d^{10}x \sqrt{h} K \nonumber\\
& & \hskip -0.8in+ \frac{1}{(2\pi)^43^22^{13}}\left(\frac{2\pi^2}{\kappa_{11}^2}\right)^{\frac{1}{3}}\int_{M_{11}} d^{11}x\sqrt{-G^{\cal M}}\left( J_0 - \frac{1}{2}E_8\right) + \left(\frac{2\pi^2}{\kappa_{11}^2}\right)\int C_3\wedge X_8+\frac{1}{\kappa_{11}^2}\int_{\cal W} d^{10}x \sqrt{\gamma} ({\cal K}-9T),\nonumber\\
& & 
\end{eqnarray}
}
where:
\begin{eqnarray}
\label{J0+E8-definitions}
& & \hskip -0.8inJ_0  =3\cdot 2^8 (R^{HMNK}R_{PMNQ}{R_H}^{RSP}{R^Q}_{RSK}+
{1\over 2} R^{HKMN}R_{PQMN}{R_H}^{RSP}{R^Q}_{RSK})\nonumber\\
& & \hskip -0.8inE_8  ={ 1\over 3!} \epsilon^{ABCM_1 N_1 \dots M_4 N_4}
\epsilon_{ABCM_1' N_1' \dots M_4' N_4' }{R^{M_1'N_1'}}_{M_1 N_1} \dots
{R^{M_4' N_4'}}_{M_4 N_4},\nonumber\\
& & \hskip -0.8in\kappa_{11}^2 = \frac{(2\pi)^8 l_p^{9}}{2};
\end{eqnarray}
$\kappa_{11}^2$ being related to the eleven-dimensional Newtonian coupling constant, and $G=dC$ with $C$ being the ${\cal {M}}$-theory three-form potential with the four-form $G$ being the associated four-form field strength. Givent the hierarchy $|t_8^2G^2R^3|<|E_8|<|J_0|$ \cite{HD-MQGP} in the MQGP limit (\ref{MQGP_limit}), we will only work with $J_0$; $\gamma$ , ${\cal K}$ and $T$ in (\ref{D=11_O(l_p^6)}) are defined earlier.

From \cite{HD-MQGP}, 
\begin{eqnarray}
\label{ETW_OR4_i}
& & \delta J_0 \sim -\delta g_{\tilde{M}\tilde{N}}\Biggl[g^{M\tilde{M}}R^{H\tilde{N}NK} + g^{N\tilde{N}}R^{HM\tilde{M}K} + g^{K\tilde{M}}R^{HMN\tilde{N}}\Biggr]\tilde{\chi}_{HMNK}\nonumber\\
& & -\frac{1}{2}\Biggl[g^{H\tilde{H}}\left(D_{[N_1}\Biggl(D_{K_1]}\delta g_{M_1\tilde{H}} + D_{M_1}\delta g_{|K_1]\tilde{H}} - D_{\tilde{H}}\delta g_{|K_1]M_1}\Biggr)\right)\chi_H^{M_1N_1K_1}\Biggr],
\end{eqnarray}
where,
\begin{eqnarray}
\label{ETW_OR4_ii}
& & \tilde{\chi}_{HMNK} \equiv R_{PMNQ} R_H^{\ RSP}R^Q_{\ RSK} - \frac{1}{2}R_{PQKN}R_H^{\ RSP}R^Q_{\ RSM};\nonumber\\
& & \chi_H^{M_1N_1K_1} \equiv R_{P\ \ \ \ Q}^{\ M_1N_1}R_H^{\ RSP} R^{Q\ \ \ K_1}_{\ RS} - \frac{1}{2}R_{PQ}^{\ \ M_1N_1}R_H^{\ RSP}R^{Q\ \ \ K_1}_{\ RS}.
\end{eqnarray}
We thus see that a typical boundary term involving covariant derivatives of the metric variations that one would need to cancel out by an appropriate boundary term (using Stokes theorem) is:
\begin{eqnarray}
\label{ETW_OR4_iii}
& & \int_{{\cal W}}D_{[K_1|}\delta g_{M_1\tilde{H}} \chi^{\tilde{H}M_1[N_1K_1]}d\Sigma_{|N_1]} = \int_{{\cal W}}D_{[K_1|}\delta g_{M_1\tilde{H}} \chi^{\tilde{H}M_1[N_1K_1]}\eta_{|N_1]}\sqrt{-h}d^{10}y.
\end{eqnarray}
The (dual to the) unit normal vector to ${\cal W}$ given by:
\begin{equation}
\label{ETW_OR4_iv}
\eta_M = \left(\eta_r, \eta_x, \eta_{M\neq r, x}\right) = \frac{\left(-\frac{dx^1(r)}{dr}\sqrt{G^{rr}_{\cal M}}, \sqrt{G^{x^1x^1}_{\cal M}},{\bf 0}\right)}{\sqrt{\left(G^{xx}_{\cal M}\right)^2 + \left(G^{rr}_{\cal M}\right)^2\left(\frac{dx^1(r)}{dr}\right)^2}}. 
\end{equation}
Therefore,
\begin{equation}
\label{ETW_OR4_v}
\left.\eta_M\right|_{x^1(r)={\rm constant}} = \left(0, \sqrt{G^{xx}_{\cal M}},{\bf 0}\right).
\end{equation} 
What we will now show that the ETW embedding (\ref{ETW_vii}) continues to hold even with the inclusion of ${\cal O}(l_p^6)$ corrections. From (\ref{ETW_OR4_v}), one sees that one will be required to considera $\chi^{HM[x^1 N]}$. Using that the only linearly independent non-vanishing Riemann curvature tensor for the metric (\ref{symbolic-metric}) with one index along $x^1$ is $R_{x^1tx^1t}$, and  (\ref{ETW_OR4_ii}), one obtains:
\begin{eqnarray}
\label{ETW_OR4_vi}
& & \chi^{tx^1tx^1} = \frac{1}{2}\left(g_{x^1x^1}^{\cal M}\right)^2\left(R_t^{\ x^1x^1t}\right)^2R^{tx^1x^1t}.
\end{eqnarray}  
\footnote{Using
\begin {eqnarray*}
\label{ETW_OR4_vii}
& & R_t^{x^1x^1t}\sim -\frac{ \left(9 a^2+{r_h}^2\right)}{ {N_f} {r_h}^2 \left(6 a^2+{r_h}^2\right)
   (\log N -3 \log ({r_h})) \sqrt[3]{\frac{12 \pi }{{g_s}}+3 \log N  {N_f}-9 {N_f} \log
   ({r_h})}};\nonumber\\
   & & R^{tx^1tx^1} \sim \frac{ \left(9 a^2 ({g_s} \log N  {N_f}+4 \pi )-3 {g_s} {N_f} \left(9
   a^2+{r_h}^2\right) \log ({r_h})+{g_s} \log N  {N_f} {r_h}^2\right)}{ \sqrt{{g_s}}
   {N_f}^3 {r_h}^3 \left(6 a^2+{r_h}^2\right) (r-{r_h}) (\log N -3 \log ({r_h}))^3}.
\end {eqnarray*}
We therefore see that up to LO in $N$ and $\log N, |\log r_h|$, and in the IR (near $r=r_h$)
\begin {eqnarray*}
\label{ETW_OR4_viii}
& & \left.\beta \chi^{tx^1tx^1}\right|_{x^1={\rm constant}}\sim \frac{\beta}{N}\frac{1}{\sqrt{g_s}N_f^{25/3}r_h^3(r-r_h)\left(\log N - 3 |\log r_h|\right)^{10/3}}.
\end {eqnarray*}}
We therefore see from (\ref{ETW_OR4_iii})($\chi^{ttx^1t}=\chi^{x^1x^1x^1t}=0$):
\begin{eqnarray}
\label{ETE_OR4_ix}
& &  \left.\int_{{\cal W}}D _{[K_1|}\delta g^{\cal M}_{M_1\tilde{H}} \chi^{\tilde{H}M_1[N_1K_1]}d\Sigma_{|N_1]}\right|_{x^1={\rm constant}} \sim \int_{{\cal W}}D_{t}\delta g^{\cal M}_{x^1t}
\chi^{tx^1[x^1t]}\eta_{x^1}\sqrt{-h}d^{10}y.
\end{eqnarray}
Now, $g^{\cal M}_{x^1t}=0$. We therefore see that restricted to $x^1=$constant, there is no surface term generated if one chooses $\delta g^{\cal M}_{x^1t}=0$ - we will assume to do so in the rest of the paper.

\section{Page Curve without Higher Derivative Terms }
\label{Page-curve-WHD}
To obtain the page curve we are required to calculate the entanglement entropy contribution from Hartman-Maldacena-like surface and island surface and then we have to see behaviour of these entropies as function of time. \par In this section we are going to perform this analysis in a ${\cal M}$-theory uplift of type IIB set up \cite{metrics} without inclusion of higher derivative terms in eleven dimensional supergravity action i.e. up to ${\cal O}(\beta^0)$ term (\ref{D=11}), which is holographic dual of thermal QCD-like theories in four dimensions at finite coupling. We are following \cite{PBD} to obtain the Page curve of an eternal black hole by computing areas of the Hartman-Maldacena-like and Island surfaces.
 We are considering the doubly holographic setup as discussed in section \ref{setup}. We are collecting the radiation of black-hole in a bath (which is effectively a thermal QCD-like theories in 2+1D). In this setup we have two kinds of extremal surfaces- Hartman-Maldacena-like surface and island surface. In subsection \ref{EE-HM-WHD} we will calculate entanglement entropy contribution from Hartman-Maldacena-like surface and in subsection \ref{EE-IS-WHD} we will calculate entanglement entropy contribution from island surface. We will obtain the Page curve of an eternal neutral black hole in the absence of higher derivative terms in subsection \ref{Page-curve-WHD-terms} by combining the entanglement entropies of Hartman-Maldacena-like and island surfaces.

\subsection{Entanglement Entropy Contribution from Hartman-Maldacena-like Surface}
\label{EE-HM-WHD}
In this subsection we are going to compute the entanglement entropy of the Hartman-Maldacena-like surface by computing area of the co-dimension two surface in the bulk similar to \cite{PBD}. \par
To compute the time dependent entanglement entropy of Hartman-Maldacena-like surface, we consider induced metric on constant $x^1$ slice and can be obtained from equation (\ref{symbolic-metric}) as below:
\begin{equation}
\label{metric-HM-t(r)}
ds^2|_{x^1=x_R}=\alpha(r)\Biggl[\left(-g(r)  \dot{t}(r)^2 +\sigma(r)\right)dr^2+(dx^2)^2+(dx^3)^2\Biggr]+g_{mn}dx^mdx^n.
\end{equation}
\\
Area density functional of Hartman-Maldacena-like Surface can be obtained from equation (\ref{metric-HM-t(r)}) as:
\begin{equation}
\label{Area-HM}
{\cal A}_{HM}(t_b)=\frac{A}{{\cal V}_2}=\int dt \sqrt{\left(-g(r)H(r) +\sigma(r)H(r) \dot{r}(t)^2\right)} \equiv \int dt {\cal L},
\end{equation}
where $H(r)={\cal V}_{int}^2\alpha(r)^3$ and ${\cal V}_2=\int \int dx^2 dx^3$ . Due to absence of explicit $t$ dependence in the Lagrangian, we have constant of motion E (which is the energy of the minimal surface):
\begin{equation}
E = \frac{\partial {\cal L}}{\partial \dot{r}(t)} \dot{r}(t) -{\cal L},
\end{equation}
where $\dot{r}(t)=\frac{dr(t)}{dt}$. Therefore using equation (\ref{Area-HM}) we can simplify the above equation as,
\begin{equation}
E = \frac{g(r)H(r)}{\sqrt{\left(-g(r)H(r) +\sigma(r)H(r) \dot{r}(t)^2\right)}}.
\end{equation}
On solving the above equation for $\dot{r}(t)$, we get:
\begin{equation}
\label{r-dot}
\dot{r}(t)=\pm \sqrt{\left(\frac{g(r)}{\sigma(r)}\left(1+\frac{g(r)H(r)}{E^2}\right)  \right)}
\end{equation}
If there is a surface at $r=r^*$ at which,
\begin{equation}
\label{rdot-star}
\dot{r}(t)|_{r=r^*}=0
\end{equation}
Then equation (\ref{r-dot}) will be simplified as,
\begin{equation}
\label{r_*}
\frac{g(r^*)}{\sigma(r^*)}\left(1+\frac{g(r^*)H(r^*)}{E^2}\right)=0.
\end{equation}
From the above equation we get:
\begin{equation}
E^2=-g(r^*)H(r^*),
\end{equation}
$r^*$ is the maximum value of $r$ for a surface with energy $E$. In the full geometry Hartman-Maldacena-like surface can be viewed as a surface which starts at $x^1=x_{\cal R}$ reaches up to $r^*$ and then goes to its thermofield double partner. Equation (\ref{Area-HM}) can be rewritten as:
\begin{equation}
\label{AHM-sim}
{\cal A}_{HM}(t_b)=\int dr \sqrt{-\frac{g(r)H(r)}{\dot{r}(t)^2}+\sigma(r)H(r)}.
\end{equation}
From equations (\ref{r-dot}) and (\ref{AHM-sim}), we get:
\begin{equation}
\label{AHM-simplified}
{\cal A}_{HM}(t_b)=2 \int_{r_h}^{r^*} dr {\frac{\sqrt{\sigma(r)g(r)}H(r)}{E\sqrt{\left(1+\frac{g(r)H(r)}{E^2}\right)}}}.
\end{equation}
Let us define time using the following integral:
\begin{equation}
\label{tdiff}
\int_{r_h}^{r^*} \frac{dr}{\dot{r}(t)}=t_*-t_b,
\end{equation}
where $t_*=t(r^*)$ and $t_b$ is the boundary time at $r=r_h$.
From equations (\ref{r-dot}) and  (\ref{tdiff}) we obtain the boundary time in terms of energy as:
\begin{equation}
\label{t_b}
t_b=- P \int_{r_h}^{r^*} \frac{dr}{\sqrt{\frac{g(r)}{\sigma(r)}\left(1+\frac{g(r)H(r)}{E^2}\right)}}.
\end{equation}
Therefore entanglement entropy of the Hartman-Maldacena-like surface is:
\begin{equation}
\label{SHM}
{\cal S}_{HM}(t_b) = \frac{{\cal A}_{HM}(t_b)}{4 G_N^{(11)}}.
\end{equation}
\subsubsection{Hartman-Maldacena-like Surface Analytics/Numerics }
The area of the Hartman-Maldacena-like surface after integrating out the angular coordinates and therefore incorporating a $(2\pi)^4$ arising from integration w.r.t. $\phi_{1,2}, \psi, x^{10}$, and is thus given as:
\begin{eqnarray}
\label{AHM_i}
& &  \hskip -0.3in A_{\rm HM} \sim (2\pi)^4 \int_{r_h}^{r^*} dr \Biggl({\pi ^9 E M^2 \sqrt[10]{N} \log ^2(2) (\log (64)-1)^2 N_f^4 g_s^{9/4} \log ^2(r) (\log (N)-3 \log (r))^4 }\nonumber\\
& & \times \left(N_f g_s \left(r^2 (2 \log (N)-18 \log (r)+3)-2 r_h^2 (\log (N)-54 r \log (r))\right)+8 \pi 
   r_h^2\right){}^2 \Biggr).
\end{eqnarray}

Assuming $\left|\log r_h\right|\gg\log N$ \cite{Bulk-Viscosity-McGill-IIT-Roorkee} in (\ref{AHM_ii}), one obtains:
\begin{equation}
\label{AHM-i}
A_{\rm HM}\sim {\cal O}(1)\times10^5M^2 \sqrt[10]{N} N_f^6 g_s^{17/4} r_h^4 \left(r_*-r_h\right) \log ^4\left(r_h\right) \left(\log (N)-3 \log \left(r_h\right)\right)^4.
\end{equation}
Now, $t_b$ is given by the principal value of the following integral:
\begin{eqnarray}
\label{tb_i}
& & \hskip -0.5in t_b \sim{\cal P}\int_{r_h}^{r_*} dr \Biggl(\frac{ E^2\sqrt{N} r^2 \sqrt{g_s} \left(N_f g_s \left(r^2 (2 \log (N)-18 \log (r)+3)-2 r_h^2 (\log (N)-54 r \log (r))\right)+8 \pi 
   r_h^2\right){}^2}{\left(r^4-r_h^4\right) \left(N_f g_s \left(\left(2 r^2-6 a^2\right) \log (N)+3 r \left(108 a^2 \log (r)+r-6 r \log (r)\right)\right)+24 \pi  a^2\right)^2}\Biggr)\nonumber\\
   & & \sim \lim_{\epsilon_1\rightarrow0}\left[4 \pi ^{33/2} E^2 \sqrt{{g_s}} \sqrt{N} \left(\frac{\log ({r_h}-r)}{4 {r_h}}-\frac{\log (r+{r_h})}{4
   {r_h}}+\frac{\tan ^{-1}\left(\frac{r}{{r_h}}\right)}{2 {r_h}}\right)\right]_{r=r_h+\epsilon_1}^{r=r_*}\nonumber\\
   & & -\frac{ E^2 \sqrt{{g_s}} \sqrt{N} (\log (-\epsilon )-\log ({r_h}-{r_*}))}{{r_h}}+\frac{\pi
   ^{33/2} E^2 \sqrt{{g_s}} \sqrt{N} ({r_*}-{r_h})}{2
   {r_h}^2} + {\cal O}\left(({r_*}-{r_h})^2\right).
\end{eqnarray}
The Principal value requires: $r_* = r_h + \epsilon_1$. Writing $a=\left(\frac{1}{\sqrt{3}}+\epsilon\right)r_h$, upto leading order in $\epsilon$, one obtains:
\begin{eqnarray}
\label{tb}
& & t_b= \frac{ E^2\pi^{33/2} \sqrt{N} \sqrt{g_s} \left(r_*-r_h\right)}{2 r_h^2}.
\end{eqnarray}
Writing $\epsilon_1=\tilde{\epsilon}_1r_h$, one sees that for (i) $\tilde{\epsilon}_1\sim0.5$, one will have to include terms up to ${\cal O}(\tilde{\epsilon}_1^2)$ in $t_b$, (ii) $\tilde{\epsilon}_1\sim\frac{1}{\sqrt{2}}$, one will have to include terms up to ${\cal O}(\tilde{\epsilon}_1^3)$ in $t_b$, (iii) $\tilde{\epsilon}_1\sim1$, one will have to include terms up to ${\cal O}(\tilde{\epsilon}_1^4)$ in $t_b$, and (iv) $\tilde{\epsilon}_1\sim\sqrt{5}$, one will have to include terms up to ${\cal O}(\tilde{\epsilon}_1^5)$ in $t_b$. We see that up to the Page time, $\tilde{\epsilon}_1\ll1$ and therefore one is justified in retaining terms only linear in $\tilde{\epsilon}_1$ as in (\ref{tb}).

From equations (\ref{AHM-i}) and (\ref{tb}), one obtains the entanglement entropy corresponding to Hartman-Maldacena-like surface as:
{
\begin{eqnarray}
\label{Page-curve-beta0}
& & S_{\rm HM}^{\beta^0} = \frac{A_{\rm HM}}{4G_N^{(11)}}\sim \frac{{\cal O}(1)\times10^{-4}M^2  N_f^6 g_s^{15/4} r_h^6 \log ^4\left(r_h\right) \left(\log (N)-3 \log \left(r_h\right)\right){}^4}{E^2 G_N^{(11)} N^{2/5}}t_b.
\end{eqnarray}
}
We therefore obtain a linear growth of the entanglement entropy corresponding to Hartman-Maldacena-like surface. 

\subsection{Entanglement Entropy Contribution from Island Surface}
\label{EE-IS-WHD}
In this subsection we compute the entanglement entropy corresponding to the Island surface by computing area of the co-dimension two surface in the bulk by following \cite{PBD}. Embedding for the Island surface is $x(r)$.\par
To compute the entanglement entropy of island surface we consider constant $t$ slice. Therefore using equation  (\ref{symbolic-metric}), we can write induced metric of the island surface in the following form:
\begin{equation}
\label{induced-metric-IS}
ds^2|_{constt- time}=\alpha(r)\Biggl[\left(\sigma(r) +\dot{x}(r)^2\right)dr^2+(dx^2)^2+(dx^3)^2\Biggr]+g_{mn}dx^mdx^n,
\end{equation}
here we have represented $x^1(r)$ by $x(r)$ and $\dot{x}(r) \equiv \frac{dx(r)}{dr}$. Now we can calculate area density functional of the island surface as given below:
\begin{equation}
\label{AIS}
{\cal A}_{IS}=\frac{A}{{\cal V}_2} = \int dr \sqrt{\left( H(r) \sigma(r)+ H(r)\dot{x}(r)^2\right)} \equiv \int dr {\cal L}  ,
\end{equation}
where,
\begin{equation}
\label{H(r)}
H(r)= {\cal V}_{int}^2\alpha(r)^3 .
\end{equation} 
Since $x(r)$ is cyclic coordinate therefore conjugate momentum corresponding to $x(r)$ i.e. $p_{x(r)}=\frac{\partial {\cal L}}{\partial \dot{x}(r)}$ is constant of motion:
\begin{equation}
\label{px}
p_{x(r)}={\cal C},
\end{equation}
where ${\cal C}$ is a constant. Using equation (\ref{AIS}) we simplify the equation (\ref{px}) as
\begin{equation}
\label{dx/dr}
\frac{H(r)\dot{x}(r)}{{\cal L}} = {\cal C}.
\end{equation}
On solving the above equation for $\dot{x}(r)$, we obtained:
\begin{equation}
\label{dotx}
\dot{x}(r)=\pm {\cal C}\sqrt{\frac{\sigma(r)}{H(r)-{\cal C}^2 \sigma(r)}}
\end{equation}
Let us assume that there is a turn around point at $r=r_T$ at which island surface satisfies following condition:
\begin{equation}
\label{dr/dx}
\left(\frac{dr}{dx}\right)_{r=r_T} =0.
\end{equation}
From equations (\ref{dotx}) and (\ref{dr/dx}), constant ${\cal C}$ can be obtained which is given below:
\begin{equation}
{\cal C}=\pm \sqrt{\frac{H(r_T)}{\sigma(r_T)}}.
\end{equation}
Using the aforementioned value of ${\cal C}$, equation (\ref{dx/dr}) simplifies to the following form:
\begin{equation}
\label{dx/dr-1}
\frac{dx(r)}{dr}=\pm \sqrt{\frac{H(r_T) \sigma(r_T)}{H(r)\sigma(r_T)-H(r_T)\sigma(r)}} .
\end{equation}
Now from equations (\ref{AIS}) and (\ref{dx/dr-1}) area density functional of the island surface simplifies as,
\begin{equation}
\label{AIS-integral}
{\cal A}_{IS}=2 \int_{r_h}^{r_T} \sqrt{H(r) \sigma(r) + \frac{H(r) H(r_T)\sigma(r)}{H(r)\sigma(r_T)-H(r_T)\sigma(r)}}.
\end{equation}
Therfore entanglement entropy contribution from the island surface is:
\begin{equation}
\label{EE-IS-WHD-TERMS}
{\cal S}_{IS} = \frac{{\cal A}_{IS}}{4 G_N^{(11)}}
\end{equation}
From the above equation it is clear that entanglement entropy contribution from island surface is independent of time. Therefore entropy contribution to the Hawking-radiation from island surface will be constant all the time.

\subsubsection{Island Surface Analytics}
The entanglement entropy upon evaluation of (\ref{AIS-integral}) yields:
\begin{eqnarray}
\label{AIS-i}
& &
{\cal A}_{\cal IS}=\int_{r_h}^{r_T} \Biggl(\frac{2 \sqrt{2} \pi  M N^{3/10} r N_f^2 g_s^{11/8} \log (r) (\log (N)-3 \log (r))^2}{\sqrt{r^4-r_h^4}} \nonumber\\
& & \times \left(N_f g_s \left(r^2 (2 \log (N)-18 \log (r)+3)-2 r_h^2 (\log (N)-54 r \log (r))\right)+8 \pi 
   r_h^2\right)\Biggr),
\end{eqnarray}
which hence obtains:
%{\scriptsize
%\begin{eqnarray}
%& &
%{\cal A}_{\cal IS}= \Biggl[\sqrt{2} \pi  M N^{3/10} N_f^3 g_s^{19/8} \log \left(r_h\right) \left(\log (N)-3 \log \left(r_h\right)\right)^2 \Biggl(216 r_h \tilde{r}_T  \log \left(r_h\right) \,
 %  _2F_1\left(-\frac{1}{4},\frac{1}{2};\frac{3}{4};\frac{1}{\tilde{r}_T^4}\right)\nonumber\\
%  & & -r_h  \left( \log (N) \left(2 \log \left(\sqrt{1-\frac{1}{\tilde{r}_T^4}}+1\right)-\log
 %  \left(\frac{1}{\tilde{r}_T^4}\right)\right)  +\sqrt{\tilde{r}_T^4-1} \left(18 \log \left(r_h\right)-2 \log (N)-3\right)+73 \sqrt{\pi} r_h \log \left(r_h\right)\right)\Biggr) \Biggr],\nonumber\\
 %  & &
%\end{eqnarray}
%}
%where $\tilde{r}_T\equiv\frac{r_T}{r_h} > 1$. Therefore entanglement entropy from the island surface (\ref{EE-IS-WHD-TERMS}) turns out to be:
{\scriptsize
\begin{eqnarray}
\label{EE-IS-simp}
& & {\cal S}_{IS} \sim \frac{1}{4 G_N^{(11)}} \Biggl[  M N^{3/10} N_f^3 g_s^{19/8} \log \left(r_h\right) \left(\log (N)-3 \log \left(r_h\right)\right){}^2  \Biggl(216 r_h  \tilde{r}_T  \log \left(r_h\right) \,
   _2F_1\left(-\frac{1}{4},\frac{1}{2};\frac{3}{4};\frac{1}{\tilde{r}_T^4}\right)\nonumber\\
  & & -r_h  \left(\log (N) \left(2 \log \left(\sqrt{1-\frac{1}{\tilde{r}_T^4}}+1\right)-\log
   \left(\frac{1}{\tilde{r}_T^4}\right)\right)  +\sqrt{\tilde{r}_T^4-1} \left(18 \log \left(r_h\right)-2 \log (N)-3\right)+73 \sqrt{\pi} r_h \log \left(r_h\right)\right)\Biggr) \Biggr],\nonumber\\
   & &
\end{eqnarray}
}
\noindent where $\tilde{r}_T\equiv\frac{r_T}{r_h} > 1$; $r_T$ is estimated in (\ref{r_T-i}) - (\ref{r_T-iii}) (and the text containing the same).

\subsection{Page curve at ${\cal O}(\beta^0)$ from Areas of Hartman-Maldacena-like and Island Surfaces}
\label{Page-curve-WHD-terms}
In this subsection we will obtain the Page curve of an eternal neutral black hole in the absence of higher derivative terms in the eleven dimensional supergravity action, explicitly by using the results of previous two subsections. \par
\begin{figure}
\begin{center}
\includegraphics[width=0.60\textwidth]{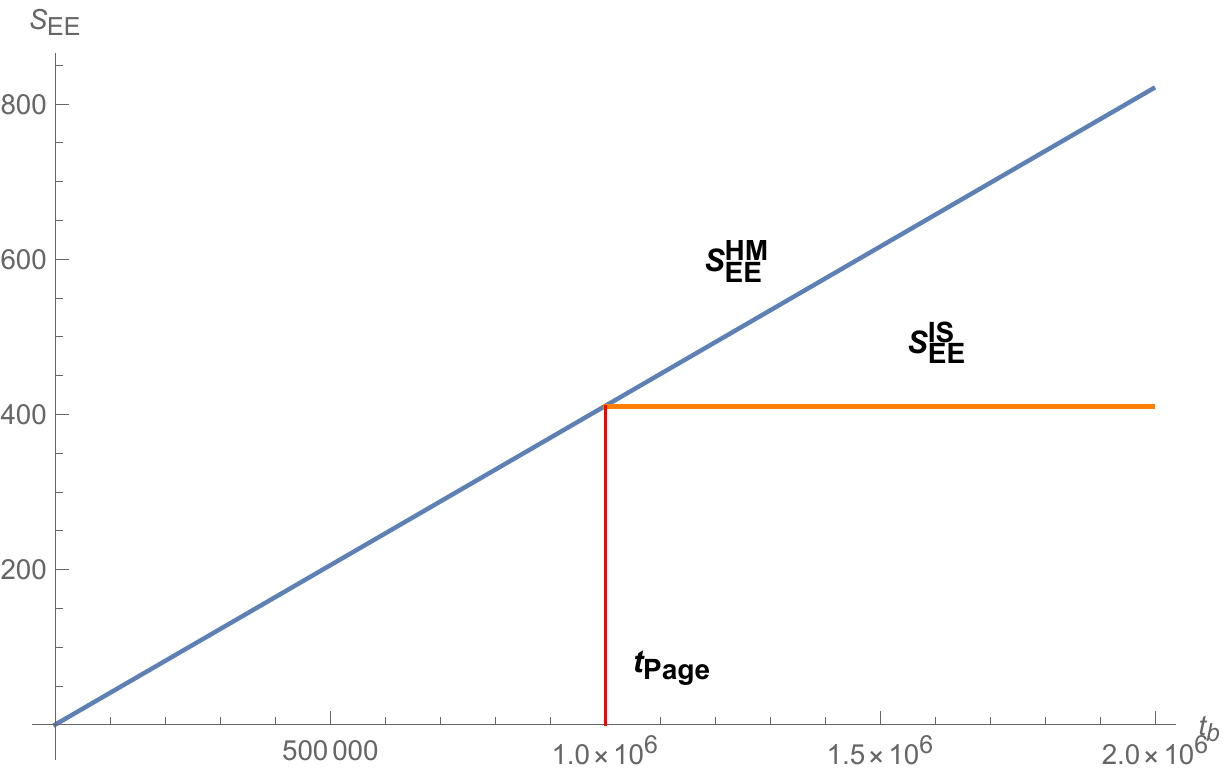}
\end{center}
\caption{Page curve up to ${\cal O}(\beta^0)$ of an eternal black hole from doubly holographic setup in ${\cal M}$-theory dual. Blue line in the graph corresponds to the entanglement entropy contribution from the Hartman-Maldacena-like surface's area, and orange line corresponds to the entanglement entropy contribution from the island surface's area; $E$ in (\ref{Page-curve-beta0}) is set to be $1.1$ to get the same order of magnitude of $t_{\rm Page}$ as in Fig.\ref{Page-time-vs-SBH-plot}.}
\label{Page-Curve-Areas}
\end{figure}

From (\ref{Page-curve-beta0}), for the aforementioned values of $g_s, M, N_f, N, r_h$, $S_{\rm EE}^{\rm HM}=\frac{5\times10^{-4}t_b}{E^2}$, and from equation (\ref{EE-IS-simp}) we see that entanglement entropy contribution of the island surface is independent of time, and for the aforementioned values, is given by 421. This contribution dominates after the Page time. As we saw in subsection \ref{EE-HM-WHD} that entanglement entropy of the Hartman-Maldacena-like surface has linear time dependence therefore when we combine both the contributions then we will find that initially entanglement entropy of the Hawking radiation will increase with boundary time then after the Page time it is the island surface contribution that dominates therefore entanglement entropy will stop increasing and reach a constant value. Therefore we will obtain the Page curve of an eternal black hole in the absence of higher derivative terms from a doubly holographic setup where gravity dual is ${\cal M}$-theory. The same is plotted in Fig. \ref{Page-Curve-Areas}. The Page time is given by: $t_{\rm Page}=8.5\times10^5 E^2$. To obtain the same $t_{\rm Page}$ after inclusion of ${\cal O}(\beta)$ "anomaly terms'' as will be obtained in \ref{Page-curve-plot-HD}, $E=1.1$. With this $t_{\rm Page}$, one sees that $\tilde{\epsilon}_1\sim 10^{-5}\ll1$, as stated beneath (\ref{tb}).
 We have also plotted the Page curve in section \ref{Page-curve-HD} in figure \ref{Page-curve-plot} where we have done our calculations in the presence of higher derivative terms.

\section{Page Curve with Higher Derivative Terms}
\label{Page-curve-HD}
In this section we are going to obtain the Page curve of a neutral black hole in the presence of higher derivative terms which are quartic in Riemann curvature tensor. We have divided this section into four subsections. In subsection \ref{EE-HM-HD}, we will compute the entanglement entropy of the HM-like surface, in subsection \ref{Swiss-Cheese-i}, we will discuss the the "Swiss-Cheese'' structure of the same, in subsection \ref{EE-IS-HD}, we will compute the entanglement entropy of the island surface and then finally in subsection \ref{Page-curve-plot-HD}, we will obtain the Page curve of an eternal black hole using the results obtained in previous subsections.
\par
We will perform the same analysis as we did in previous section. In the presence of higher derivative terms we will compute entanglement entropy of Hartman-Maldacena-like surface and island surface using \cite{Dong}.
 \par
Holographic entanglement entropy in general higher derivative gravity theories can be calculated using the following formula in $AdS_{d+2}/CFT_{d+1}$ correspondence \cite{Dong}:
\begin{equation}
\label{HD-Entropy}
S_{EE}=\int d^dy \sqrt{-g}\Biggl[\frac{\partial {\cal L}}{\partial R_{z\bar{z}z\bar{z}}}+\sum_{\alpha}\left(\frac{\partial^2 {\cal L}}{\partial R_{zizj}\partial R_{\bar{z}m\bar{z}l}}\right)_{\alpha} \frac{8 K_{zij}K_{\bar{z}ml}}{(q_{\alpha}+1)}\Biggr],
\end{equation}
where $g$ is the determinant of the induced metric on co-dimension two surface, $(a,b)$ are along the normal directions, $(i,j,k,l)$ are along tangential directions, $K_{zij}=\frac{1}{2}\partial_z G_{ij}$ and its trace is defined as $K_z=K_{zij}G^{ij}$. First term in the above equation is Wald entanglement entropy and to calculate the second term one is required to go through the following steps:
\begin{itemize}
\item Label each term as $\alpha$ after differentiating twice the Lagrangian with respect to Riemann tensor.
\item Do the following transformations of the specific components of Riemann tensors in each term.
\begin{eqnarray}
\label{riemann-transformations}
R_{abij}=r_{abij}+ g^{ml}\left(K_{ajm}K_{bil}-K_{aim}K_{bjl}\right)\nonumber\\
& & \hskip -3.2in R_{aibj}=r_{aibj}+g^{ml} K_{ajm} K_{bil}- Q_{abij}\nonumber\\
& & \hskip -3.2in R_{ijml}=r_{ijml}+g^{ab}\left(K_{ail}K_{bjm}-K_{aij}K_{bml} \right),
\end{eqnarray}
where $Q_{abij}=\partial_a K_{bij}$.
\item In the $\alpha$th term, let number of $Q_{aaij}$ and $Q_{bbij}$ be $\gamma$ and number of $K_{aij}$, $R_{abci}$ and $R_{aijk}$ be $\delta$. Then $q_\alpha$ for the $\alpha$th term is defined as: 
\begin{equation}
\label{q-alpha}
q_{\alpha}=\gamma + \frac{\delta}{2}.
\end{equation}
\item By using equation (\ref{riemann-transformations}) one can then obtain values of $r_{abij},r_{aibj},r_{ijml}$ and then substitute back in equation (\ref{HD-Entropy}). By doing so we will obtain the entanglement entropy expression in terms of original Riemann tensors i.e. in terms of $R_{abij},R_{aibj}$ etc. 
\end{itemize}
Let us outline that how we can calculate the holographic entanglement entropy (\ref{HD-Entropy}) in general higher derivative gravity theories:
\begin{itemize}
\item If holographic dual is $(d+1)$ dimensional gravitational background then obtain the induced metric for the co-dimension two surface i.e. in $(d-1)$ dimensions in terms of the embedding function.

\item Calculate (\ref{HD-Entropy}) for the aforementioned induced metric. By doing so we will obtain the holographic entanglement entropy in terms of embedding function and its derivatives.

\item Work out the equation of motion of the embedding function and find its solution.

\item Substituting the solution obtained in previous step into the action will give the holographic entanglement entropy in higher derivative gravity theories.
\end{itemize}
\par

{\bf Contribution from $J_0$ term:}

We obtain the four terms from Wald entanglement entropy formula, i.e.,
$\frac{\partial J_0}{\partial R_{z\bar{z}z\bar{z}}}$ for both the extremal surfaces. which are listed in appendix \ref{PT-HM} for Hartman-Maldacena-like surface and in \ref{PT-IS} for the island surface.
\begin{eqnarray}
\label{Wald-J0-c-r}
\frac{\partial J_0}{\partial R_{z\bar{z}z\bar{z}}}=-4 x^2 \left(\frac{\partial J_0}{\partial R_{txtx}}\right).
\end{eqnarray}

To calculate the second term in equation (\ref{HD-Entropy}), we are required to calculate the four kinds of differentiations for Hartman-Maldacena-like and island surfaces as given below:
{\scriptsize
\begin{eqnarray}
\label{second-der-J0}
& &
\left(\frac{\partial^2 {J_{0}}}{\partial R_{zizj}\partial R_{\bar{z}m\bar{z}l}}\right) K_{zij}K_{\bar{z}ml}=\left(\frac{\partial^2 {J_0}}{\partial R_{xixj}\partial R_{xmxl}}+x^4\frac{\partial^2 {J_0}}{\partial R_{titj}\partial R_{tmtl}} -2 x^2 \frac{\partial^2 {J_0}}{\partial R_{titj}\partial R_{xmxl}}+4 x^2 \frac{\partial^2 {J_0}}{\partial R_{tixj}\partial R_{xmtl}} \right)\nonumber\\
& & \hskip 2in \times \left(\frac{1}{x^2}K_{tij}K_{tml}+K_{xij}K_{xml} -\frac{i}{x} K_{tij} K_{xml} +\frac{i}{x} K_{xij} K_{tml}\right),
\end{eqnarray}
}
 Coefficient appearing in the numerator and denominators of the above equation is $x=x_R$ for Hartman-Maldacena-like surface and $x=x(r)$ for the island surface. We have calculated and listed all four kinds of differentiations appearing in equation (\ref{second-der-J0}) for Hartman-Maldacena-like surface in appendix \ref{PT-HM} and for the island surface in appendix \ref{PT-IS}.

\subsection{Entanglement Entropy Contribution from Hartman-Maldacena-like Surface}
\label{EE-HM-HD}
In this subsection we will be calculating the entanglement entropy corresponding to the Hartman-Maldacena-like surface using Dong's formula (\ref{HD-Entropy}). \par
In ${\cal M}$-theory dual, Hartman-Maldacena-like surface is a co-dimension two surface which is located at $x^1=x_R$. Therefore using equation (\ref{HD-Entropy}) we can write expression for the entanglement entropy for Hartman-Maldacena-like surface in the following form:
\begin{equation}
\label{HD-Entropy-HM}
S_{EE}=\int dr dx^2 dx^3  d\theta_1 d\theta_2dxdydz dx^{10} \sqrt{-g}\Biggl[\frac{\partial {\cal L}}{\partial R_{z\bar{z}z\bar{z}}}+\sum_{\alpha}\left(\frac{\partial^2 {\cal L}}{\partial R_{zizj}\partial R_{\bar{z}m\bar{z}l}}\right)_{\alpha} \frac{8 K_{zij}K_{\bar{z}ml}}{(q_{\alpha}+1)}\Biggr],
\end{equation}
where $g$ is the determinant of the induced metric (\ref{metric-HM-t(r)}) on the co-dimension two surface. \\
{\bf Contribution from ${\cal O}(\beta^0)$ action:}
For the metric (\ref{metric-HM-t(r)}) first term in equation (\ref{HD-Entropy-HM}) simplifies to the following form for the ${\cal O}(\beta^0)$ term in action (\ref{D=11_O(l_p^6)}):
\begin{eqnarray}
\label{Lag-wald-beta0}
\int d{\cal V}_9 \sqrt{-g}\left(\frac{\partial R}{\partial R_{z\bar{z}z\bar{z}}}\right)=\int d{\cal V}_9 {\cal L}_{0}^{\rm HM} \sim \int dr\left(  x_R^2 \lambda (r) \sqrt{\alpha (r) \left(\sigma (r)-\left(1-\frac{r_h^4}{r^4}\right) t'(r)^2\right)}\right),
\end{eqnarray}
where $x_R$ is constant and $\alpha(r), \sigma(r), \lambda(r)$ are given in (\ref{alpha-sigma-lambda}).

{\bf Contribution from ${\cal O}(\beta)$ term:}
At ${\cal O}(\beta)$, one is required to calculate two type of terms, first one is Wald entanglement entropy term and the second one is the anomaly term.

\noindent {\bf Wald entanglement entropy term, i.e., $\left(\frac{\partial J_0}{\partial R_{z\bar{z}z\bar{z}}}\right)$:}\\
For Hartman-Maldacena-like surface, we obtain:
\begin{eqnarray}
\frac{\partial J_0}{\partial R_{z\bar{z}z\bar{z}}}=-4 x_R^2 \frac{\partial J_0}{\partial R_{txtx}}.
\end{eqnarray}
Out of four terms that we have listed for $\frac{\partial J_0}{\partial R_{txtx}}$  in appendix \ref{PT-HM}, most dominant terms that contribute to the Lagrangian in the large $N$ limit are given below (which have been calculated for the metric (\ref{metric-HM-t(r)})). 
{\footnotesize
\begin{eqnarray}
\label{Lag-wald-HM}
\int d{\cal V}_9 \sqrt{-g}\frac{\partial J_0}{\partial R_{z\bar{z}z\bar{z}}}=\int d{\cal V}_9 {\cal L}_{\cal W}^{\rm HM}=\int dr \left(-4 x_R^2 (\lambda_1(r)+\lambda_2(r)) \sqrt{\alpha (r) \left(\sigma (r)-\left(1-\frac{r_h^4}{r^4}\right) t'(r)^2\right)}\right),
\end{eqnarray}
}
where $\lambda_{1,2}(r)$ are given in (\ref{lambda1-2}).

\noindent{\bf Anomaly Term:} 
Using equation (\ref{second-der-J0}) and appendix \ref{PT-HM}, we find that following are the possible most dominant terms in the large-$N$ limit that contribute the Lagrangian from second term in equation (\ref{HD-Entropy-HM}):
\begin{eqnarray}
\label{Lag-anomaly-HM}
& & \int d{\cal V}_9 \ {\cal L}_{\cal A}^{\rm HM}= \int d{\cal V}_9 \sqrt{-g} \left(\frac{\partial^2 J_0}{\partial R_{zizj}\partial R_{\bar{z}m\bar{z}l}} K_{zij}K_{\bar{z}ml}\right) =\int d{\cal V}_9 \sqrt{-g} \left(\frac{\partial^2 J_0}{\partial R_{zizj}\partial R_{\bar{z}m\bar{z}l}} \frac{K_{tij}K_{tml}}{x_R^2}\right), \nonumber\\ 
& & \int d{\cal V}_9 \  {\cal L}_{\cal A}^{\rm HM}=\int dr \frac{1}{x_R^2} \Biggl(Z(r) {\cal L}_1+ x_R^4 W(r){\cal L}_2 -2 x_R^2 U(r){\cal L}_3 +4 x_R^2 \left(U(r)+ 2 V(r)\right){\cal L}_4 \Biggr),
\end{eqnarray}
where ${\cal L}_{1,3}$ are given in (\ref{L-terms-HM}), $Z(r), W(r), U(r), V(r)$ are given in (\ref{Z-W-U-V}), and use has been made of (\ref{dJ0overdR+d2J0overdR2}). 

From equations (\ref{Lag-wald-beta0}), (\ref{Lag-wald-HM}), (\ref{Lag-anomaly-HM}) and (\ref{L-terms-HM}), total holographic entanglement entropy corresponding to Hartman-Maldacena-like surface can be written as:
\begin{equation}
\label{Lag-total-HM}
S_{\rm EE}^{\rm total,\ {\rm HM}}=\int dr {\cal L}_{\rm {Total}}^{\rm {HM}}=\int dr \left({\cal L}_{0}^{\rm HM}+ {\cal L}_{\cal W}^{\rm HM}+ {\cal L}_{\cal A}^{\rm HM}\right)
\end{equation}
Using (\ref{DLag-tprime}) - (\ref{EOM-HM}), and writing $t(r) = t_0(r) + \beta t_1(r)$, the EOM up to ${\cal O}(\beta^0)$ is:
\begin{eqnarray}
\label{EOM-beta0}
& & N^{3/10} (r-r_h)^{5/2} p_1\left(r_h\right) {t_0}''(r)+\frac{5}{2} N^{3/10} (r-r_h)^{3/2} p_1\left(r_h\right) {t_0}'(r)=0. \nonumber\\ 
\end{eqnarray}
The solution to (\ref{EOM-beta0}) will be given by:
\begin{eqnarray}
\label{solution-t0}
& & {t_0}(r) = c_2-\frac{2 c_1}{3 \left(r-r_h\right){}^{3/2}}. 
\end{eqnarray}
Substituting (\ref{solution-t0}) into (\ref{EOM-HM}), one obtains:
{\footnotesize
\begin{eqnarray}
\label{solution-t1}
& & {t_1}(r)=c_3+\frac{2 r_h^{3/2} \sqrt{\kappa _{\sigma }} \left(11 {r_h}^7 \left(p_8^{\beta }\left(r_h\right)+p_9^{\beta }\left(r_h\right)\right)-3 \left(r-r_h\right){}^7 \left(p_8^{\beta
   }\left(r_h\right)+p_9^{\beta }\left(r_h\right)\right)-11 c_1{}^3 c_2 p_1\left(r_h\right)\right)}{33 c_1{}^3 M N_f^{5/3} g_s^{7/3} \left(r-r_h\right){}^{3/2} \sqrt{\kappa _{\alpha }} \kappa _{\lambda }
   \log \left(r_h\right) \left(\log (N)-9 \log \left(r_h\right)\right) \left(\log (N)-3 \log \left(r_h\right)\right){}^{5/6}}.
\end{eqnarray}
}
Retaining the term up to leading order in $N$ we can write embedding function $t(r)$ as:
\begin{eqnarray}
\label{t(r)}
t(r)=c_2-\frac{2 c_1}{3 \left(r-r_h\right){}^{3/2}}+\beta  \left(c_3-\frac{2 c_2}{3 \left(r-r_h\right){}^{3/2}}\right),
\end{eqnarray}
where it is understood that $t(r=r_h)\equiv t_b$ is evaluated as $t\left(r_h\biggl[1+\frac{1}{N^{n_{t_b}}}\biggr]\right), n_{t_b}\equiv{\cal O}(1)$, in the large-$N$ limit. Since (in ${\cal R}_{D5/\overline{D5}}=1$-units) $\frac{c_1}{r_h^{3/2}} \sim c_2$, i.e., $c_1 \sim r_h^{\alpha} c_2$ implying $\alpha = 3/2$, i.e. $l_p^6 \sim r_h^{3/2}$ or $l_p \sim r_h^{1/4}$ i.e.,
\begin{equation}
\label{lp-rh-relation} 
l_p \sim \left(g_s^{\frac{4}{3}}\alpha'\ ^2 \left(\frac{r_h}{{\cal R}_{D5/\overline{D5}}}\right)\right)^{\frac{1}{4}}.
\end{equation}
 
\begin{figure}
\begin{center}
\includegraphics[width=0.70\textwidth]{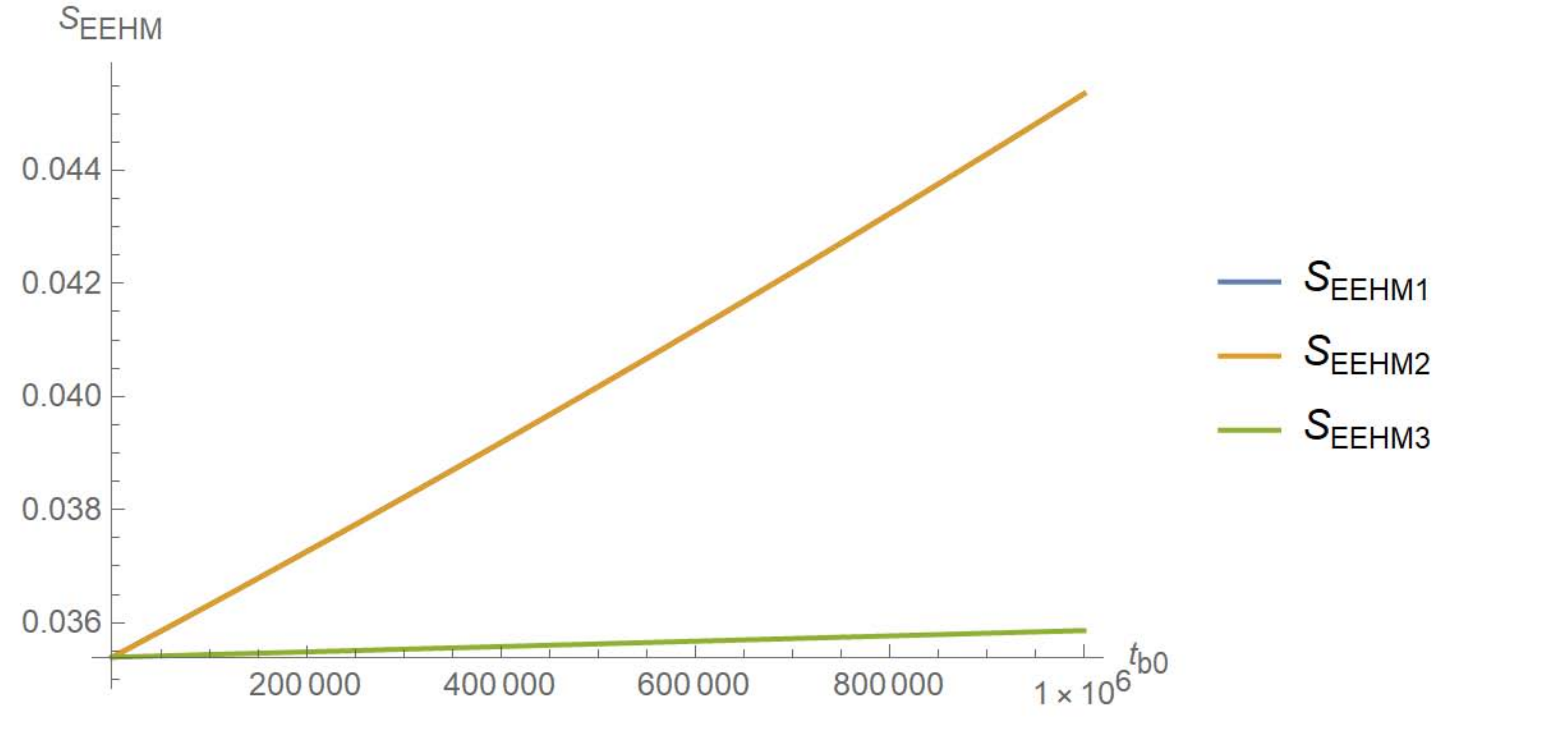}
\end{center}
\caption{}
\label{SEEHM-vs-tb0-curve-plot}
\end{figure}

To determine the turning point i.e. $r_*$ of the Hartman-Maldacena-like surface we need to impose the following condition:
\begin{equation}
\label{r*-HM}
\left(\frac{1}{t'(r)}\right)_{r=r_*}=\frac{\left(r_*-r_h\right){}^{5/2}}{c_1}-\frac{\beta  c_2 \left(r_*-r_h\right){}^{5/2}}{c_1{}^2}=0,
\end{equation}
Since $\frac{c_1}{r_h^{3/2}} \sim c_2$, therefore (\ref{r*-HM}) can be approximated by the following equation for the purpose of estimating $r_*$:
\begin{equation}
\label{rstar-equation}
(r_*-r_h)^{3/2}-\beta  \kappa_{r_*}^\beta=0,
\end{equation}
whose solution is
\begin{equation}
\label{solution-rstar}
r_*=r_h+\beta ^{2/3} {\left(\kappa_{r_*}^\beta\right)}^{2/3}
\end{equation}
Now we can calculate the entanglement entropy of the Hawking radiation for the Hartman-Maldacena-like surface as follow:
 \begin{eqnarray}
 \label{SEE-HM-beta0}
 S_{\rm EE}^{\beta^0, \rm HM}= \int_{r_h}^{r_*} dr \Biggl(\lambda (r) \sqrt{\alpha (r) \left(\sigma (r)-\left(1-\frac{r_h^4}{r^4}\right) t_0'(r){}^2\right)}\Biggr),
 \end{eqnarray}
using equations (\ref{alpha-sigma-lambda}) and (\ref{solution-t0}) above equation can be rewritten as
{\footnotesize
\begin{eqnarray}
\label{SEE-HM-beta0-simp}
& &
S_{\rm EE}^{\beta^0, \rm HM}\sim \int_{r_h}^{r_*}\left(\frac{\sqrt{\left(r-r_h\right) r_h} \log \left(r_h\right) \left(\log (N)-9 \log \left(r_h\right)\right) \left(\log (N)-3 \log \left(r_h\right)\right){}^{5/6}}{r_h^3 \sqrt{r_h^2}}\right)\nonumber\\
& & \times \Biggl(-\frac{2^{5/6} M N^{13/10} \log (2) (\log (64)-1) N_f^{5/3} g_s^{10/3} \sqrt{\kappa _{\alpha }} \kappa _{\lambda } \sqrt{\kappa _{\sigma }}}{81\ 3^{2/3} \pi ^{11/12}}\Biggr)\nonumber\\
& & \sim \frac{2 \left(r_*-r_h\right){}^{3/2} \log \left(r_h\right) \left(\log (N)-9 \log \left(r_h\right)\right) \left(\log (N)-3 \log \left(r_h\right)\right){}^{5/6}}{3 r_h^{7/2}}\nonumber\\
& & \times \Biggl(-\frac{2^{5/6} M N^{13/10} \log (2) (\log (64)-1) N_f^{5/3} g_s^{10/3} \sqrt{\kappa _{\alpha }} \kappa _{\lambda } \sqrt{\kappa _{\sigma }}}{81\ 3^{2/3} \pi ^{11/12}}\Biggr) \nonumber\\
& &  \sim \frac{2 \left(\beta ^{2/3} {\left(\kappa^\beta_{r_*}\right)}^{2/3}\right)^{3/2} \log \left(r_h\right) \left(\log (N)-9 \log \left(r_h\right)\right) \left(\log (N)-3 \log \left(r_h\right)\right){}^{5/6}}{3 r_h^{7/2}}\nonumber\\
& & \times \Biggl(-\frac{2^{5/6} M N^{13/10} \log (2) (\log (64)-1) N_f^{5/3} g_s^{10/3} \sqrt{\kappa _{\alpha }} \kappa _{\lambda } \sqrt{\kappa _{\sigma }}}{81\ 3^{2/3} \pi ^{11/12}}\Biggr)
 \nonumber\\
& & \sim -\frac{2\ 2^{5/6} \beta  {g_s}^{10/3} \sqrt{\kappa_\alpha} \kappa^\beta_{r_*} \kappa_{\lambda} \sqrt{\kappa_{\sigma}} M N^{13/10} \log (2) (\log (64)-1) N_f^{5/3} \log \left(r_h\right)
   \left(\log (N)-9 \log \left(r_h\right)\right) \left(\log (N)-3 \log \left(r_h\right)\right){}^{5/6}}{243\ 3^{2/3} \pi ^{11/12} r_h^{7/2}} \nonumber\\
\end{eqnarray}
}
If $t_b=t(r=r_h+\epsilon)$ where $\epsilon = \frac{r_h}{N^{n_{t_b}}}, n_{t_b}\equiv{\cal O}(1)$ then
\begin{equation}
\label{tb0}
t_{b_0}=c_2-\frac{2 c_1 N^{3 n_{t_b}/2}}{3 r_h^{3/2}},
\end{equation}
we obtain the $r_h$ from the above equation as below:
\begin{equation}
\label{rh-tb-relation}
r_h=\frac{\left(\frac{2}{3}\right)^{2/3}}{\left(\frac{(c_2-{t_{b_0}}) \
N^{-\frac{3 n_{t_b}}{2}}}{c_1}\right)^{2/3}}
\end{equation}
It will be argued when discussing the computation of the entanglement entropy corresponding to the Island Surface, that $\beta$ turns out to be related to black hole horizon radius (see the discussion around (\ref{scalings})). If $\left|\log(r_h)\right| \gg \log(N)$ then equation (\ref{SEE-HM-beta0-simp}), using (\ref{scalings}), can be approximated by
\begin{eqnarray}
\label{SEE-HM-simp-beta0}
S_{\rm EE}^{\beta^0, {\rm HM}} \sim e^{-\frac{3\kappa_{l_p}N^{\frac{1}{3}}}{2}}\frac{M N^{13/10} N_f^{5/3}\left(\left|\log(r_h)\right|\right)^{17/6}}{r_h^2}.
\end{eqnarray}
Since $\frac{1}{r_h^2} \sim \left(\frac{c_2}{c_1}-\frac{t_{b_0}}{c_1}\right)^{4/3} N^{-2 n_{t_b}} = \left(\frac{c_2}{c_1}\right){}^{2/3} \left(1-\frac{t_{b_0}}{c_2}\right){}^{4/3} N^{-2 n_{t_b}}$. For $t_{b_0} \ll c_2$, \\
 $\frac{1}{r_h^2} \sim \left(\frac{c_2}{c_1}\right){}^{4/3} \left(1-\frac{4 t_{b_0}}{3 c_2}\right) N^{-2 n_{t_b}}$.
Therefore,
\begin{eqnarray}
\label{SEE-HM-tb0}
& & S_{\rm EE}^{\beta^0, {\rm HM}} \sim  e^{-\frac{3\kappa_{l_p}N^{\frac{1}{3}}}{2}}M N^{13/10} N_f^{5/3}\left(1-\frac{4 t_{b_0}}{3 c_2}\right) \left(\frac{2}{3}\log \left(\frac{c_2-t_{b_0}}{c_1}\right) -n_{t_b} \log(N)\right)^{\frac{17}{6}}\nonumber\\
& & \approx e^{-\frac{3\kappa_{l_p}N^{\frac{1}{3}}}{2}}M N^{13/10} N_f^{5/3}\left(1-\frac{4 t_{b_0}}{3 c_2}\right)\left(\frac{2}{3}\log \left(\frac{c_2}{c_1}\right) -n_{t_b} \log(N)\right)^{\frac{17}{6}},\nonumber\\
\end{eqnarray}
when $c_1,c_2 <0$. The Fig. \ref{SEEHM-vs-tb0-curve-plot} for  $N=10^{3.3}, M=N_f=3, g_s=0.1, c_1=-10^3, c_2=-10^8$, $n_{t_b}=1$ obtained from (\ref{SEEHM-tb0}), shows the linearization assumed in obtaining the last line of (\ref{SEE-HM-tb0}), can be approximately justified. In it,
\begin{eqnarray}
\label{SEEHM-tb0}
& & S_{\rm EEHM1}\sim 51.024 \left(\frac{t_{{b_0}}}{7.5\times10^7}+1\right) \left(\frac{2}{3} \log
   \left(\frac{t_{{b_0}}+10^8}{10^3}\right)-7.6\right)^{\frac{17}{6}},\nonumber\\
& & S_{\rm EEHM2}\sim51.024 \left(\frac{t_{{b_0}}}{10^8}+1\right){}^{4/3} \left(\frac{2}{3} \log
   \left(\frac{t_{{b_0}}+10^8}{10^3}\right)-7.6\right)^{\frac{17}{6}},\nonumber\\
& &  S_{\rm EEHM3}\sim 0.0354 \left(\frac{t_{{b_0}}}{7.5\times10^7}+1\right).
\end{eqnarray}
From equation (\ref{SEE-HM-tb0}) we see that entanglement entropy of the Hawking radiation coming from the Hartman-Maldacena-like surface has linear time dependence. Therefore it is increasing with time and diverges at late times, i.e., when $t_{b_0} \rightarrow \infty$. 

\subsection{"Swiss-Cheese'' Structure of $S_{\rm EE}^{\beta^0, {\rm HM}}$ in a Large-$N$-Scenario}
\label{Swiss-Cheese-i}
In this subsection we discuss the "Swiss-Cheese'' structure of entanglement entropy of the Hawking radiation corresponding to the Hartman-Maldacena-like surface at $O(\beta^0)$.\par
\begin{figure}
\begin{center}
\includegraphics[width=0.70\textwidth]{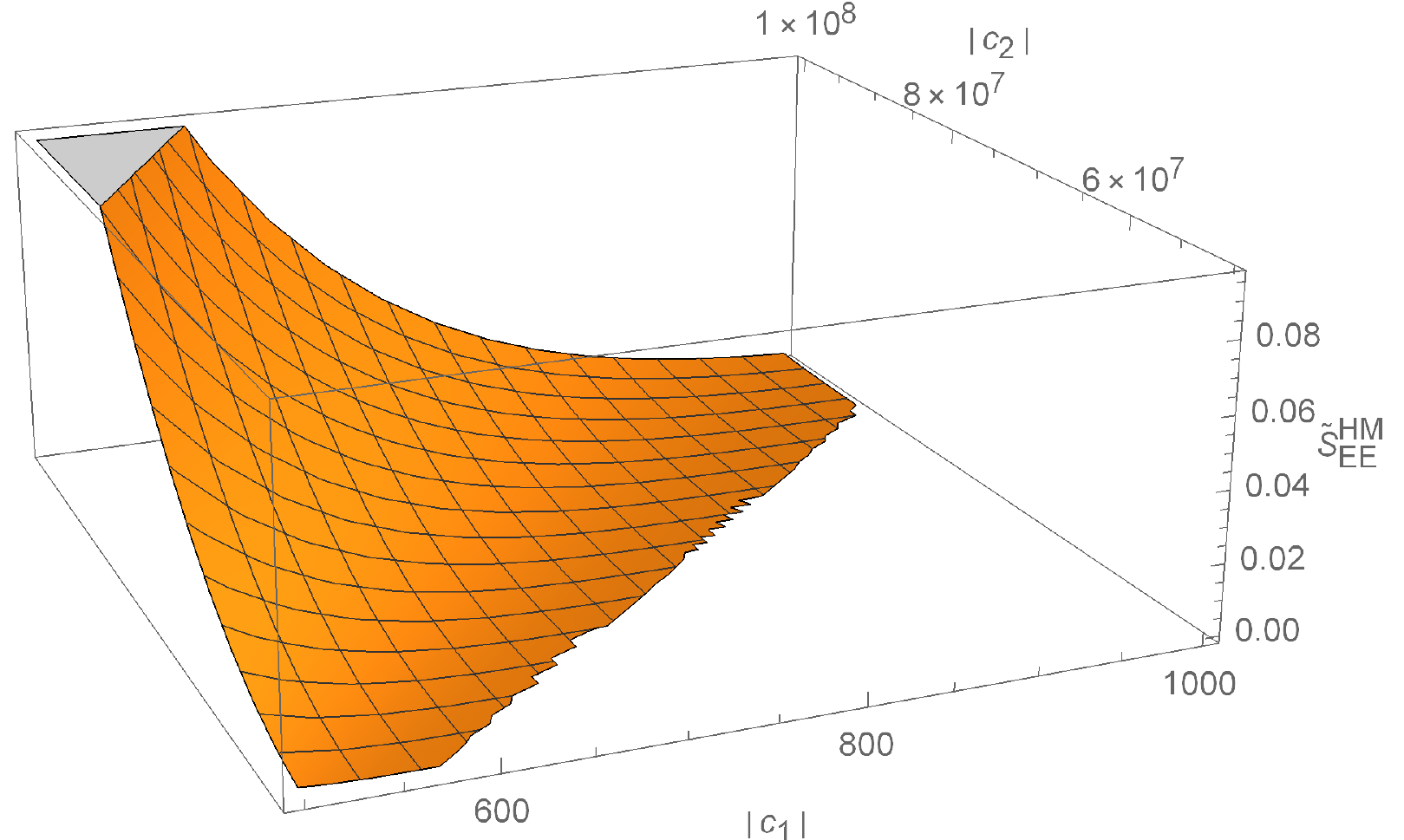}
\end{center}
\caption{Plot of $\tilde{S}_{\rm EE}^{\rm HM}$ as a bi-function of $\left(|c_1|, |c_2|\right)$ at $t=t_{\rm Page}=10^6$}
\label{Swiss-Cheese}
\end{figure}

The expression for $S_{\rm EE}^{\beta^0, {\rm HM}}$ in (\ref{SEE-HM-tb0}) for the values of constants of intergation $c_{1,2}$ used beneath the same, suggests the following hierarchy:
\begin{equation}
\label{Swiss-Cheese-c12}
|c_2|\sim e^{\kappa_{c_2}|c_1|^{1/3}},\ |c_1|\sim N.
\end{equation}
Further, assuming $c_{1,2}<0$, (\ref{SEE-HM-tb0}) can be rewritten as:
\begin{eqnarray}
\label{SHMEEbeta0-modc12}
& & S_{\rm EE}^{\beta^0, {\rm HM}} \sim  e^{-\frac{3\kappa_{l_p}N^{\frac{1}{3}}}{2}}M N^{13/10} N_f^{5/3}\left(1+\frac{4 t_{b_0}}{3 |c_2|}\right) \left(\frac{2}{3}\log \left(\frac{|c_2|+t_{b_0}}{|c_1|}\right) -n_{t_b} \log(N)\right)^{\frac{17}{6}}.
\end{eqnarray}
We thus see,
\begin{eqnarray}
\label{dSEEHbeta0overdmodc2-i}
& & \frac{\partial S_{\rm EE}^{\beta^0, {\rm HM}}}{\partial |c_2|} \sim  e^{-\frac{3\kappa_{l_p}N^{\frac{1}{3}}}{2}}M N^{13/10} N_f^{5/3}\Biggl[-\frac{4 t_{b_0}}{3 |c_2|^2}\left(\frac{2}{3}\log \left(\frac{|c_2|+t_{b_0}}{|c_1|}\right) -n_{t_b} \log(N)\right)^{\frac{17}{6}}\nonumber\\
& & + \frac{17}{6}\frac{2}{3|c_2|}\left(1+\frac{4 t_{b_0}}{3 |c_2|}\right)\left(\frac{2}{3}\log \left(\frac{|c_2|+t_{b_0}}{|c_1|}\right) -n_{t_b} \log(N)\right)^{\frac{11}{6}} \Biggr].
\end{eqnarray}
For $N=10^{3.3}, M=N_f=3, g_s=0.1, c_1=-10^3, c_2=-10^8$, $n_{t_b}=1$ as in the previous sub-section, as $t_b\leq t_{\rm Page}\sim10^6$,
\begin{eqnarray}
\label{dSEEHbeta0overdmodc2-ii}
& & \frac{\partial S_{\rm EE}^{\beta^0, {\rm HM}}}{\partial |c_2|} \sim e^{-\frac{3\kappa_{l_p}N^{\frac{1}{3}}}{2}}M N^{13/10} N_f^{5/3}\Biggl[-\leq 10^{6-16}\times{\cal O}(1)+{\cal O}(1)\times {\cal O}(1)\times 10^{-8} \Biggr]\approx  {\cal O}(1)\times 10^{-8}\nonumber\\
& & \sim e^{-\frac{3\kappa_{l_p}N^{\frac{1}{3}}}{2}}M N^{13/10} N_f^{5/3}{\cal O}(1)\times 10^{-8} >0.
\end{eqnarray}
Similarly,
\begin{eqnarray}
\label{dSEEHbeta0overdmodc1-i}
& & \frac{\partial S_{\rm EE}^{\beta^0, {\rm HM}}}{\partial |c_1|} \sim  -e^{-\frac{3\kappa_{l_p}N^{\frac{1}{3}}}{2}}M N^{13/10} N_f^{5/3}\Biggl[\left(1+\frac{4 t_{b_0}}{3 |c_2|}\right)\frac{1}{|c_1|} \Biggr]<0.
\end{eqnarray}
We also note that in the $|c_2|\gg|c_1|$-limit and assuming $|c_1|\sim N$,
\begin{eqnarray}
\label{SHMEEbeta0-largemodc2}
& & S_{\rm EE}^{\beta^0, {\rm HM}} \sim  e^{-\frac{3\kappa_{l_p}N^{\frac{1}{3}}}{2}}M N^{13/10} N_f^{5/3}\left(\log |c_2|\right)^{11/6}\left(12\log |c_2|-(34 + 51 n_{t_b})\log |c_1|\right),
\end{eqnarray}
which mimics  a Swiss-Cheese volume written out in terms of a single "large divisor'' volume $\log |c_2|$ and a single "small divisor'' volume $\log |c_1|$ (with 12 and $34+51n_{t_b}$ encoding information about some version of "classical intersection numbers'' of these "divisors'').  Alternatively, defining $S_{\rm EE}^{\beta^0, {\rm HM}} \equiv e^{-\frac{3\kappa_{l_p}N^{\frac{1}{3}}}{2}}M N^{13/10} N_f^{5/3}\tilde{S}_{\rm EE}^{\rm HM}$, one can think of (\ref{SHMEEbeta0-largemodc2}) as an open Swiss-Cheese surface (in the same sense as (\ref{dSEEHbeta0overdmodc1-i}) and (\ref{SHMEEbeta0-largemodc2}), i.e., $\tilde{S}_{\rm EE}^{\rm HM}$ decreases as $|c_1|$ increases and $\tilde{S}_{\rm EE}^{\rm HM}$ increases as $|c_2|$ increases) in $\mathbb{R}_{\geq0}\left(\tilde{S}_{\rm EE}^{\rm HM}\right)\times\mathbb{R}^2_{+}\left(|c_1|, |c_2|\right)$ - see Fig. \ref{Swiss-Cheese}. 

 Therefore, (\ref{Swiss-Cheese-c12}) along with (\ref{dSEEHbeta0overdmodc2-ii}), (\ref{dSEEHbeta0overdmodc1-i}) and (\ref{SHMEEbeta0-largemodc2}), suggests very curiously a "Swiss-Cheese'' structure of $S_{\rm EE}^{\beta^0, {\rm HM}}$ in a Large $N$ Scenario (reminiscent of the "Large Volume Scenario'' in moduli stabilizations in string theory \cite{LVS,Swiss-Cheese-LVS}). 

\subsection{Entanglement Entropy Contribution from Island Surface}
\label{EE-IS-HD}
In this subsection we are calculating the entanglement entropy corresponding to the Island surface using Dong's formula (\ref{HD-Entropy}). \par
Similar to Hartman-Maldacena-like surface, island surface is also a co-dimension two surface and located at constant time slice. Therefore we can write the expression for the holographic entanglement entropy similar to Hartman-Maldacena-like surface in the following form:
\begin{equation}
\label{HD-Entropy-IS}
S_{EE}=\int dr dx^2 dx^3  d\theta_1 d\theta_2dxdydz dx^{10} \sqrt{-g}\Biggl[\frac{\partial {\cal L}}{\partial R_{z\bar{z}z\bar{z}}}+\sum_{\alpha}\left(\frac{\partial^2 {\cal L}}{\partial R_{zizj}\partial R_{\bar{z}m\bar{z}l}}\right)_{\alpha} \frac{8 K_{zij}K_{\bar{z}ml}}{(q_{\alpha}+1)}\Biggr],
\end{equation}
where all the quantities in the above equation are same as equation (\ref{HD-Entropy-HM}).   \\
{\bf Contribution from ${\cal O}(\beta^0)$ term:}
Holographic entanglement entropy contribution to the Lagrangian (\ref{D=11_O(l_p^6)}) from the induced metric (\ref{induced-metric-IS}) for the island surface at ${\cal O}(\beta^0)$ turns out to be:
\begin{eqnarray}
\label{Lag-beta0-IS}
\int d{\cal V}_9 \sqrt{-g}\left(\frac{\partial R}{\partial R_{z\bar{z}z\bar{z}}}\right) =\int d{\cal V}_9 {\cal L}_{0}^{\rm IS}= \int dr\left( 4 \lambda_5(r) x(r)^2 \sqrt{\alpha (r) \left(\sigma (r)+x'(r)^2\right)}\right),
\end{eqnarray}
where, 
\begin{eqnarray}
\label{lambda5}
& &
\lambda_5(r)=\kappa_{\lambda_5}\frac{ M N^{17/10} N_f^{4/3} g_s^{5/2} r_h^2 \log (N) \log (r) \sqrt[3]{\log (N)-3 \log (r)}}{r^4 \alpha _{\theta _1}^3 \alpha _{\theta _2}^2} \nonumber\\
& & \sim \frac{M N^{21/20} \log (2) (\log (64)-1) N_f^{4/3} g_s^{13/4} r_h^2 \kappa _{\lambda _5} \log (N) \log (r) \sqrt[3]{\log (N)-3 \log (r)}}{r^4},
\end{eqnarray}
where $\kappa_{\lambda_5}$ is the numerical factor. Angular integrations in equations (\ref{lambda5}), (\ref{lambda3-4}) and (\ref{Z-W-U-V-1}) have been performed using (\ref{angular-integrations-2}) and (\ref{angular-integrations-1}).

{\bf Contribution from ${\cal O}(\beta)$ term:}
At ${\cal O}(\beta)$ we are required to calculate the two terms to obtain the holographic entanglement entropy (\ref{HD-Entropy-IS}). First term is coming from $\left(\frac{\partial J_0}{\partial R_{z\bar{z}z\bar{z}}}\right)$ and is given below:
\begin{equation}
\frac{\partial J_0}{\partial R_{z\bar{z}z\bar{z}}} = - 4 x(r)^2\frac{\partial J_0}{\partial R_{txtx}},
\end{equation}
from appendix \ref{PT-IS} we find that following terms are the most dominant terms in the large-$N$ limit:
\begin{eqnarray}
\label{Lag-Wald-IS}
\int d{\cal V}_9 \sqrt{-g}\frac{\partial J_0}{\partial R_{z\bar{z}z\bar{z}}}=\int d{\cal V}_9 {\cal L}_{\cal W}^{\rm IS}=\int dr \left(4 x(r)^2 (\lambda_3(r)+\lambda_4(r)) \sqrt{\alpha (r) \left(\sigma (r)+x'(r)^2\right)}\right),\nonumber\\
\end{eqnarray}
where $\lambda_{3,4}$ are given in (\ref{lambda3-4}).
 We obtain the second term in equation (\ref{HD-Entropy-IS}) using (\ref{second-der-J0}) and appendix \ref{PT-IS}. For the island surface, most dominant terms in the large-$N$ limit that contribute to the Lagrangian are given below:
\begin{eqnarray}
\label{Lag-anomaly-IS}
& & \hskip -0.5in  \int d{\cal V}_9 {\cal L}_{\cal A}^{\rm IS} =\int d{\cal V}_9 \sqrt{-g} \left( \frac{\partial^2 J_0}{\partial R_{zizj}\partial R_{\bar{z}m\bar{z}l}} K_{zij}K_{\bar{z}ml}\right) =\int d{\cal V}_9 \sqrt{-g} \left( \frac{\partial^2 J_0}{\partial R_{zizj}\partial R_{\bar{z}m\bar{z}l}} K_{xij}K_{xml}\right), \nonumber\\ 
& &  =\int dr \left(Z_1(r) {\cal L}_1+ x(r)^4 W_1(r){\cal L}_2 -2 x(r)^2 U_1(r){\cal L}_3 +4 x(r)^2 \left(U_1(r)+ 2 V_1(r)\right){\cal L}_4 \right),
\end{eqnarray}
where,
\begin{eqnarray}
\label{L1234}
 \hskip -1in {\cal L}_1= {\cal L}_2=\frac{\sqrt{\alpha (r) \left(\sigma (r)+x'(r)^2\right)} \left(\alpha '(r) \left(\sigma (r)+x'(r)^2\right)+\alpha (r) \left(\sigma '(r)+2 x'(r) x''(r)\right)\right)^2}{x'(r)^2 \left(\sigma (r)+x'(r)^2\right)^4};\nonumber\\
   & &  \hskip -4.7in {\cal L}_3= {\cal L}_4=\frac{\sqrt{\alpha (r) \left(\sigma (r)+x'(r)^2\right)}}{x'(r)^2},
\end{eqnarray}
and various $r$ dependent functions appearing in equation (\ref{Lag-anomaly-IS}) are given in (\ref{Z-W-U-V-1}).
 From equations (\ref{Lag-beta0-IS}), (\ref{Lag-Wald-IS}) and (\ref{Lag-anomaly-IS}), we find that
total holographic entanglement entropy from ${\cal O}(\beta^0)$ and ${\cal O}(\beta)$ terms for the action (\ref{D=11_O(l_p^6)}) corresponding to island surface is given by:
\begin{equation}
\label{Lag-total-IS}
S_{\rm EE}^{\rm total}=\int dr {\cal L}_{\rm {Total}}^{\rm {IS}}=\int dr \left({\cal L}_{0}^{\rm IS}+ {\cal L}_{\cal W}^{\rm IS}+ {\cal L}_{\cal A}^{\rm IS}\right).
\end{equation}
Using appendix \ref{ISM}, $x(r)$ EOM $\frac{\delta {\cal L}_{\rm {Total}}^{\rm {IS}}}{\delta x(r)} - \frac{d}{dr}\left(\frac{\delta{\cal L}_{\rm {Total}}^{\rm {IS}}}{dx^\prime(r)}\right) + \frac{d^2}{dr^2}\left(\frac{\delta {\cal L}_{\rm {Total}}^{\rm {IS}}}{\delta x^{\prime\prime}(r)}\right) = 0$, turns out to be,
{\footnotesize
\begin{eqnarray}
\label{x(r)-EOM}
& & -\frac{N^{3/10} x(r) \left(F_1\left(r_h\right) \left(4 \left(r-r_h\right) x'(r)^2+x(r) \left(2 \left(r-r_h\right) x''(r)+x'(r)\right)\right)-2 N f_1\left(r_h\right)\right)}{2 \sqrt{r-r_h}} \nonumber\\
& & + \frac{\beta}{4 N^{7/10}
   (r-r_h)^{3/2} x'(r)^4}\Biggl[2 N^{7/10} F^\beta_4(r_h) x(r) \left(4 (r_h-r) x'(r)^2+x(r) \left(6 (r-r_h) x''(r)+x'(r)\right)\right) \nonumber\\
  & &  +(r-r_h) x'(r)^4 \left(3 {\cal F}^{\beta}_1(r_h)+3
   {\cal F}^{\beta}_2(r_h)+4 N^2 x(r) Y_1(r_h)\right)\Biggr]= 0.
\end{eqnarray}
}
Writing $x(r) = x_0(r) + \beta x_1(r)$, and making an ansatz:
\begin{equation}
\label{ansatz-x0(r)}
x_0(r)=\sqrt{\frac{{a_3}}{{a_2}}+r} \sqrt{\frac{F_1(r_h)}{N f_1(r_h)}}.
\end{equation}
Defining $X(r) \equiv \frac{x_0(r) \sqrt{\frac{F_1(r)}{N f_1(r_h)}}}{\sqrt{2}}$, one notes that $X(r)$ must satisfy:
{\footnotesize
\begin{eqnarray}
\label{X0(r)}
& & \hskip -0.3in X(r) \left(2 (r-r_h) X''(r)+X'(r)\right)+4 (r-r_h) X'(r)^2-1=\frac{a_3 \left(a_1^2 a_2-2\right)+{a_2} \left(2 r \left(a_1^2 a_2-1\right)-a_1^2 a_2r_h\right)}{2 ({a_2} r+{a_3})} = 0.\nonumber\\
\end{eqnarray}
}
If $a_1^2a_2=2$, then the LHS of (\ref{X0(r)}) becomes proportional to $\frac{r-r_h}{\frac{{a_3}}{{a_2}}+r}$, which in the deep IR, i.e., $r\sim r_h$ becomes negligible if $\frac{a_3}{a_2 r}\gg1$. Therefore,
\begin{equation}
\label{x0(r)-solution}
x_0(r)=\sqrt{\frac{{a_3}}{{a_2}}+r} \sqrt{\frac{F_1(r_h)}{N f_1(r_h)}}.
\end{equation}
One therefore sees,
{\footnotesize
\begin{eqnarray}
\label{x1(r)EOM-i}
& & \hskip -0.3in x_1(r) = \frac{N^{3/10} }{2 \sqrt{r-r_h}}\Biggl[2 N f_1(r_h) {x_1}(r)-{F_1}(r_h) \Biggl({x_0}(r) \left(2  (r-r_h) \left(2 {x_1}(r) {x_0}''(r)+{x_0}(r) {x_1}''(r)\right) 
   +{x_0}(r)
  {x_1}'(r)\right)
  \nonumber\\
   & & +2 {x_0}(r) {x_0}'(r) \left(4 (r-r_h) {x_1}'(r)+{x_1}(r)\right)+4(r-r_h) {x_1}(r) {x_0}'(r)^2\Biggr)\Biggr] \nonumber\\
   & & =\frac{2 {a_2} N^2{f_1}(r_h)^2 {x_1}(r)-{F_1}(r_h)^2 \left({x_1}'(r) (5 a_2 r-4 {a_2}r_h+{a_3})+2 (r-r_h) ({a_2} r+{a_3}) {x_1}''(r)+{a_2} {x_1}(r)\right)}{2{a_2} N^{7/10} {f_1}(r_h) \sqrt{r-r_h}}, \nonumber\\
\end{eqnarray}
}
and,
{\footnotesize
\begin{eqnarray}
\label{x1(r)EOM-ii}
& & \frac{\beta}{4 N^{7/10}
   (r-r_h)^{3/2} x_0'(r)^4}\Biggl(2 N^{7/10} F^\beta_4(r_h) x_0(r) \left(4 (r_h-r) x_0'(r)^2+x_0(r) \left(6 (r-r_h) x_0''(r)+x_0'(r)\right)\right) \nonumber\\
  & &  +(r-r_h) x_0'(r)^4 \left(3 {\cal F}^{\beta}_1(r_h)+3
   {\cal F}^{\beta}_2(r_h)+4 N^2 x_0(r) Y_1(r_h)\right)\Biggr) \nonumber\\
 & &  =\frac{4 \left(a_2 r+a_3\right) \sqrt{\frac{a_3}{a_2}+r} \left(a_2 \left(5 r_h-4 r\right)+a_3\right) F^\beta_4\left(r_h\right)}{a_2^2 \left(r-r_h\right){}^{3/2} \sqrt{\frac{F_1\left(r_h\right)}{N
   f_1\left(r_h\right)}}}+\frac{4 N^2 \sqrt{\frac{a_3}{a_2}+r} Y_1\left(r_h\right) \sqrt{\frac{F_1\left(r_h\right)}{N f_1\left(r_h\right)}}+3 {\cal F}^{\beta}_1(r_h)+3 {\cal F}^{\beta}_2(r_h)}{4 N^{7/10}
   \sqrt{r-r_h}}.\nonumber\\
\end{eqnarray}
}
Therefore equation of motion corresponding to embedding $x_1(r)$ for the island surface is given by:
{\footnotesize
\begin{eqnarray}
\label{x1(r)-EOM}
& & \frac{2 {a_2} N^2{f_1}(r_h)^2 {x_1}(r)-{F_1}(r_h)^2 \left({x_1}'(r) (5 a_2 r-4 {a_2}r_h+{a_3})+2 (r-r_h) ({a_2} r+{a_3}) {x_1}''(r)+{a_2} {x_1}(r)\right)}{2{a_2} N^{7/10} {f_1}(r_h) \sqrt{r-r_h}}\nonumber\\
& & +\frac{4 \left(a_2 r+a_3\right) \sqrt{\frac{a_3}{a_2}+r} \left(a_2 \left(5 r_h-4 r\right)+a_3\right) F^\beta_4\left(r_h\right)}{a_2^2 \left(r-r_h\right){}^{3/2} \sqrt{\frac{F_1\left(r_h\right)}{N
   f_1\left(r_h\right)}}}+\frac{4 N^2 \sqrt{\frac{a_3}{a_2}+r} Y_1\left(r_h\right) \sqrt{\frac{F_1\left(r_h\right)}{N f_1\left(r_h\right)}}+3 {\cal F}^{\beta}_1(r_h)+3 {\cal F}^{\beta}_2(r_h)}{4 N^{7/10}
   \sqrt{r-r_h}}=0,\nonumber\\
\end{eqnarray}
}
which near $r=r_h$ simplifies to:
\begin{eqnarray}
\label{x1(r)-EOM-final}
& & \frac{4 \sqrt{\frac{a_3}{a_2}+r_h} \left(a_2 r_h+a_3\right){}^2 F^\beta_4\left(r_h\right)}{a_2^2 \left(r-r_h\right){}^{3/2} \sqrt{\frac{F_1\left(r_h\right)}{N f_1\left(r_h\right)}}}+\frac{N^{13/10} x_1(r) f_1\left(r_h\right)-\frac{F_1\left(r_h\right){}^2 \left(\left(a_2 r_h+a_3\right) x_1'(r)+a_2 x_1(r)\right)}{2 a_2 N^{7/10} f_1\left(r_h\right)}}{\sqrt{r-r_h}}=0.\nonumber\\
\end{eqnarray}

The solution to (\ref{x1(r)-EOM-final}) is given by,
\begin{eqnarray}
\label{x1(r)-solution}
& & x_1(r) = \frac{8 \sqrt{\frac{a_3}{a_2}+r_h} \left(a_2 r_h+a_3\right) {F^\beta_4}\left(r_h\right) \log \left(r-r_h\right)}{a_2 N^{3/10} F_1\left(r_h\right) \left(\frac{F_1\left(r_h\right)}{N f_1\left(r_h\right)}\right){}^{3/2}},
   \end{eqnarray}
    implying,
   \begin{eqnarray}
   \label{x(r)-solution}
   & & x(r)\approx \sqrt{\frac{a_3}{a_2}+r} \sqrt{\frac{ F_1(r_h)}{N  f_1(r_h)}}+\beta N^{\frac{11}{5}}\left( \frac{8  \left(r_h+\frac{a_3}{a_2}\right)^{\frac{3}{2}} {F^\beta_4}\left(r_h\right) \log \left(r-r_h\right)}{F_1\left(r_h\right) \left(\frac{F_1\left(r_h\right)}{ f_1\left(r_h\right)}\right)^{3/2}}\right).
\end{eqnarray}
To determine the turning point via $r_T: 1/x'(r_T) = 0$, one therefore obtains:
\begin{eqnarray}
\label{rT-2}
& & \frac{32 \beta N^{11/5} \left(a_2 r_T+a_3\right) \sqrt{\frac{a_3}{a_2}+r_h} \left(a_2 r_h+a_3\right) F^\beta_4\left(r_h\right)}{a_2^2  \left(r_T-r_h\right) F_1\left(r_h\right)
   \left(\frac{F_1\left(r_h\right)}{ f_1\left(r_h\right)}\right)^{5/2}}-\frac{2 \sqrt{\frac{a_3}{a_2}+r_T}}{\sqrt{\frac{F_1(r_h)}{N
    f_1(r_h)}}}=0, \nonumber\\
\end{eqnarray}
whose solution is given by
\begin{eqnarray}
\label{rT-solution}
& & r_T = r_h + \frac{128 \beta^2 }{a_2^4 F_1\left(r_h\right){}^6}\Biggl({a_2  N^{17/5} \left(a_2 r_h+a_3\right)^3 f_1\left(r_h\right)^4 F^\beta_4\left(r_h\right)^2}+a_2 a_3^3  N^{17/5} f_1\left(r_h\right)^4
   F^\beta_4\left(r_h\right){}^2 \nonumber\\
  & & +a_2^4  N^{17/5} r_h^3 f_1\left(r_h\right){}^4 F^\beta_4\left(r_h\right){}^2+3 a_2^3 a_3  N^{17/5} r_h^2 f_1\left(r_h\right){}^4
   F^\beta_4\left(r_h\right){}^2+3 a_2^2 a_3^2  N^{17/5} r_h f_1\left(r_h\right){}^4 F^\beta_4\left(r_h\right){}^2\Biggr)\nonumber\\
   & & \hskip -0.2in \sim r_h +\frac{289 \left(\frac{3}{2}\right)^{2/3} \pi ^{17/3} \beta ^2 M^2 N^{17/5} (107-540 \log (2))^2 g_s^7 \kappa _{\sigma }^6 \kappa _{U_1}^2 \log ^2(N)\left(3 a_2 a_3^2 r_h+3 a_2^2 a_3 r_h^2+2 a_2^3
   r_h^3+a_3^3\right) }{50 a_2^3 \log ^2(2) (\log (64)-1)^2 N_f^{2/3} r_h^{20} \kappa _{\lambda _5}^2 \left(-\log \left(r_h\right)\right){}^{2/3}}\nonumber\\
& &  \equiv r_h + \delta \in{\rm IR},
\end{eqnarray}
in ${\cal R}_{D5/\overline{D5}}=1$-units. One therefore obtains the ${\cal O}(\beta^0)$ on-shell entanglement entropy:
{\footnotesize
\begin{eqnarray}
\label{on-shell-L-beta0}
& & S_{\rm EE}^{\beta^0, {\rm IS}} = \int_{r_h}^{r_h+\delta {\cal R}_{D5/\overline{D5}}} dr\Biggl(\frac{\kappa_{\lambda_5} {\log N} M
   N^{17/10} {N_f}^{4/3} {r_h}^2 x(r)^2 \log (r) \sqrt[3]{{\log N}-3 \log(r)} \sqrt{\alpha (r) \left(\sigma (r)+x'(r)^2\right)}}{
   r^4 \alpha _{\theta _1}^3 \alpha _{\theta _2}^2}\Biggr)\nonumber\\
& &\sim \int_{r_h}^{r_h+\delta}dr\Biggl(\frac{2\ 2^{2/3} M N^{3/10} \log (2) (\log (64)-1) N_f^{5/3} g_s^{10/3} \kappa _{\lambda _5} \log (N) \left(a_2 r_h+a_3\right) F_1\left(r_h\right) \log \left(r_h\right) \sqrt{\kappa _{\alpha } \kappa
   _{\sigma }} \left(\log (N)-3 \log \left(r_h\right)\right){}^{2/3}}{81\ 3^{5/6} \pi ^{11/12} a_2 \sqrt{r-r_h} r_h^{5/2} f_1\left(r_h\right)}\Biggr)\nonumber\\
& &  \sim \Biggl(\frac{ \sqrt{\delta } M N^{3/10} N_f^{5/3} g_s^{7/3} \sqrt{r_h} \kappa _{\alpha } \kappa _{\lambda _5} \log (N) \left(a_2 r_h+a_3\right) \log \left(r_h\right)
   \left| \log \left(r_h\right)\right| {}^{2/3}}{ a_2 \sqrt{\kappa _{\alpha } \kappa _{\sigma }}}\Biggr)\nonumber\\
& &   \sim \frac{\beta  M^2 N^2 N_f^{4/3} g_s^{35/6}  \log ^2(N) \left(a_2 r_h+a_3\right)   \left| \log \left(r_h\right)\right| {}^{4/3}}{a_2^{5/2} r_h^{19/2}}  \sqrt{3 a_2 a_3^2 r_h+3 a_2^2 a_3 r_h^2+2
   a_2^3 r_h^3+a_3^3}.
\end{eqnarray}
}
Now, this appears to present a contradiction - $ S_{\rm EE}^{\beta^0, {\rm IS}}\sim \beta$. The resolution, using (\ref{lp-rh-relation}), is that $\beta\propto r_h^{\frac{3}{2}}$. 

The Hawking BH entropy is given by:
\begin{eqnarray}
\label{SBH}
& & S_{\rm BH} \sim 
%\frac{g_s^{5/4} M N^{9/20} N_f^3 r_h^3 \log ^4(r_h) (2-\beta
   % ({\cal C}^{\rm BH}_{zz}-2 {\cal C}^{\rm BH}_{\theta_1 z}))}{\alpha _{\theta _1}^3 \alpha _{\theta_2}^2}\nonumber\\
 \frac{g_s^{7/4} M N_f^3 r_h^3 \log (N) \log ^4(r_h) (2-\beta
    ({\cal C}^{\rm BH}_{zz}-2 {\cal C}^{\rm BH}_{\theta_1 z}))}{N^{3/4}}, 
\end{eqnarray}
From the above equation we see that ${\cal O}(\beta^0)$ contribution to the black hole thermal entropy scale as $S_{\rm BH}^{\beta^0} \sim r_h^3 \log ^4(r_h)$. The ${\cal O}(\beta)$-corrections to the MQGP background of \cite{MQGP} as worked out in \cite{HD-MQGP} and as quoted in appendix ${\bf A}$, were worked out in the $\psi=2n\pi, n=0, 1, 2$-patches, by setting the ${\cal O}(\beta)$-corrections to the ${\cal M}$-theory three-form potential, to zero. This required ${\cal C}^{\rm BH}_{zz}=2 {\cal C}^{\rm BH}_{\theta_1 z}$. One therefore sees that the BH entropy receives no higher-derivative corrections at ${\cal O}(R^4)$.

From (\ref{on-shell-L-beta0}) and (\ref{SBH}), as well as $\beta = \kappa_\beta(g_s, N) \left(\frac{r_h}{{\cal R}_{D5/\overline{D5}}}\right)^{\frac{3}{2}}\alpha'\ ^3$ from the discussion beneath (\ref{on-shell-L-beta0}), one sees,
\begin{eqnarray}
\label{SEEISoverSBH}
 & &\frac{S^{\beta^0,\ IS}_{\rm EE}}{S_{\rm BH}} \sim \frac{\kappa_\beta(g_s, N)  {g_s}^{49/12} {\log N} M N^{11/4} ({a_2} {r_h}+{a_3}) \sqrt{3 {a_2}^2
   {a_3} {r_h}^2+2 {a_2}^3 {r_h}^3+3 {a_2} {a_3}^2 {r_h}+{a_3}^3}}{{a_2}^{5/2}
   {N_f}^{5/3} {r_h}^{11} | \log ({r_h})| ^{8/3}}.\nonumber\\
   & & 
\end{eqnarray}
If $a_2 r_h \gg a_3$ then,
\begin{eqnarray}
\label{SEEISoverSBH-arrh>>a3}
& &  \frac{S^{\beta^0,\ IS}_{\rm EE}}{S_{\rm BH}} \sim \frac{\kappa_\beta(g_s, N){g_s}^{49/12} {\log N} M N^{11/4}}{ {N_f}^{5/3} {r_h}^{17/2} | \log ({r_h})| ^{8/3}}\left[1 + \sum_{n=1}^\infty {\cal A}_n\left(\frac{a_3}{a_2 r_h}\right)^n\right], 
\end{eqnarray}
where, e.g., ${\cal A}_1 = \frac{7}{4}, {\cal A}_2 = \frac{39}{32}$, etc. Now, one expects:
\begin{equation}
\label{SEEISoverSBH-D-dims}
\frac{S^{\beta^0,\ IS}_{\rm EE}}{S_{\rm BH}} = 2 + \sum_{n=1}a_n\left(\frac{G_N^D}{r_h^{D-2}}\right)^n,
\end{equation}
 for a $D$-dimensional black-hole (the central charge for conformal backgrounds is absorbed into the $a_n$'s) \cite{SEEISoverSBHapprox2}; $a_3$ in (\ref{SEEISoverSBH-arrh>>a3}) is the non-conformal analog of the conformal charge "$c$'' figuring in \cite{SEEISoverSBHapprox2}. It is rather non-trivial to obtain a somewhat similar expression for the non-conformal backgrounds considered in this paper in (\ref{SEEISoverSBH-arrh>>a3}). To ensure that (\ref{SEEISoverSBH-arrh>>a3}) $\sim$ (\ref{SEEISoverSBH-D-dims}), one needs to cure the large-$N$ and IR(via small $r_h$) enhancements in the (\ref{SEEISoverSBH-arrh>>a3}). Utilizing the estimate in \cite{Bulk-Viscosity-McGill-IIT-Roorkee} of the $r=r_0\sim r_h: N_{\rm eff}(r_0=0)(N_{\rm eff}$ being the effective number of color $D3$-branes in \cite{metrics} - the type IIB dual of thermal QCD-like theories), and in particular the exponential $N$-suppression therein \footnote{To find an appropriate $r_0/r_h$ it would be easier to work with the type IIB side instead of its type IIA mirror as the mirror {\it a la} SYZ
keeps the radial coordinate unchanged. To proceed then, let us define an {\it effective} number of three-brane charge
as:
\begin{equation*}
\label{Neff}
N_{\rm eff}(r) = \int_{\mathbb{M}_5} F_5^{\rm IIB} + \int_{\mathbb{M}_5} B_2^{\rm IIB} \wedge F_3^{\rm IIB}, 
\end{equation*}
where $B_2^{\rm IIB}, F_3^{\rm IIB}$ and $F_5^{\rm IIB}$ are given in \cite{metrics}. The five-dimensional internal space $\mathbb{M}_5$,
with coordinates ($\theta_i, \phi_i, \psi$), is basically the base of the resolved warped-defomed conifold. As shown in \cite{Bulk-Viscosity-McGill-IIT-Roorkee}, 
\begin {eqnarray*}
& & N_{\rm eff}(r_0) = N + {3g_sM^2 \log~r\over 10 r^4} \bigg\{18\pi r (g_sN_f)^2 \log~N \sum_{k = 0}^1\left(18a^2 (-1)^k\log~r + r^2\right)\left({108a^2 \log~r\over 2k+1} + r\right) \\
 &+& 5 \left(3 a^2 ({g_s}-1)+r^2\right) (3 {g_s} {N_f} \log ~r+2\pi ) (9 {g_s} {N_f} \log ~r+4 \pi )
 \left[9 a^2 {g_s} {N_f} \log\left({e^2\over r^3}\right) +4 \pi  r^2\right]\biggr\}, \nonumber\\
 & = & N\left[1 + 6\pi\log~r\left(3 g_s N_f \log~r + 2\pi\right)\left(9 g_s N_f \log~r + 4\pi\right){g_sM^2\over N}  \right] +
 {\cal O}\left[{g_sM^2\over N}\left(g_sN_f\right)^2 \log~N\right]. \end {eqnarray*}
Note that the assumption of small $r_0$ is crucial here as the same implies the dominance of $g_sN_f|\log~r_0|$ over other constant pieces. Solving for $r_0: N_{\rm eff}(r_0)=0$ yields $r_0\sim r_h\sim e^{-\kappa_{r_h}(M, N_f, g_s)N^{\frac{1}{3}}}$.} ,  we therefore propose $\kappa_\beta(g_s, N) \sim e^{-\kappa_{l_p}N^{1/3}}$, i.e.,
\begin{eqnarray}
\label{scalings}
& & \beta \sim \left(g_s^{\frac{4}{3}}\alpha'^2e^{-\kappa_{l_p}N^{1/3}}\frac{r_h}{{\cal R}_{D5/\overline{D5}}}\right)^{3/2}.\nonumber\\
\end{eqnarray}
In particular then, 
\begin{eqnarray}
\label{SEE-IS-simp}
S_{\rm EE}^{\beta^0, {\rm IS}} \sim \frac{M^2 e^{-\frac{3\kappa_{l_p} N^{\frac{1}{3}}}{2}}
 N_f^{4/3} g_s^{35/6} \log ^2(N)  \left| \log \left(r_h\right)\right|^{4/3} }{r_h^{11/2}}.
\end{eqnarray}

\begin{figure}
\begin{center}
\includegraphics[width=0.45\textwidth]{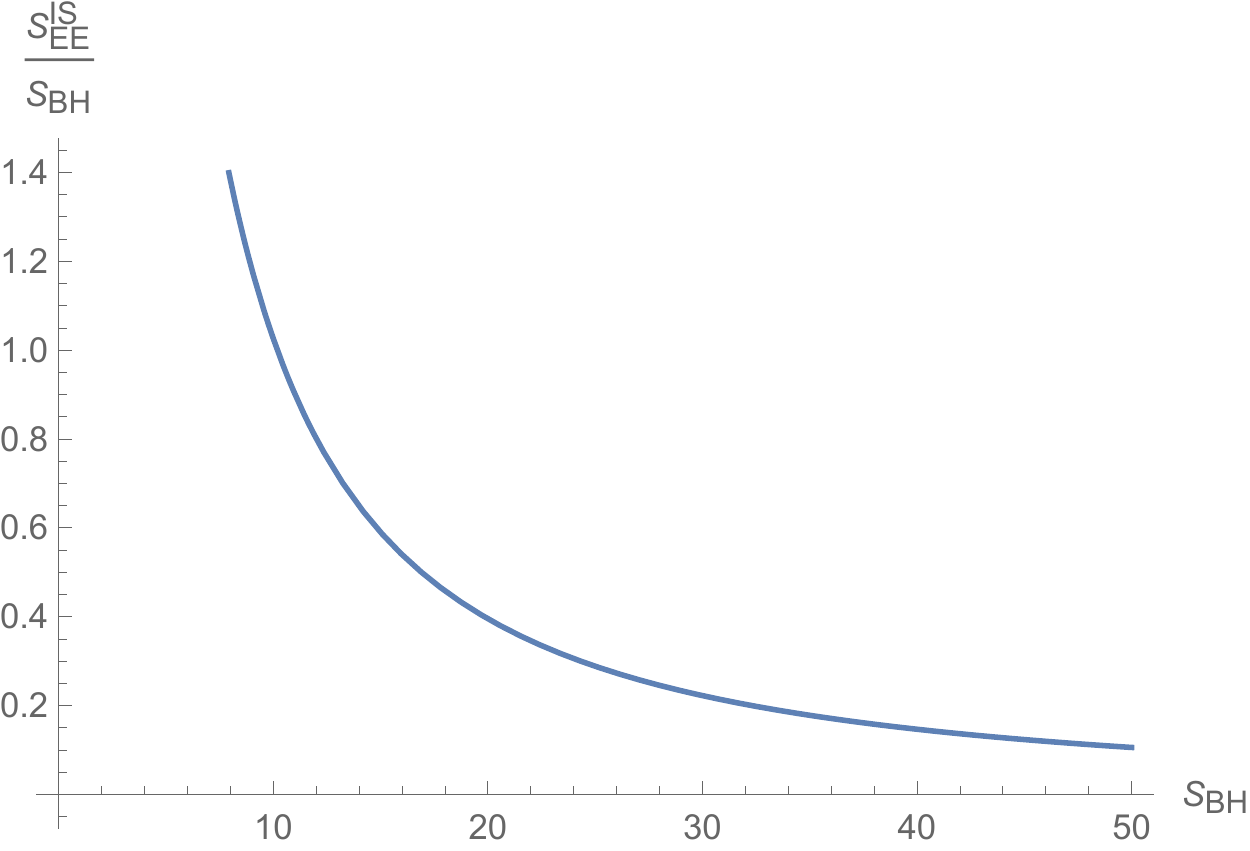}
\end{center}
\caption{$\frac{S^{\beta^0,\ IS}_{\rm EE}}{S_{\rm BH}}$-versus-$S_{\rm BH}$ for $N=10^{3.3}, M=N_f=3, g_s=0.1, \kappa_{l_p}=0.47$}
\label{SEEISoverSBH-vs-SBH}
\end{figure}

Setting $N=10^{3.3}, M=N_f=3, g_s=0.1$, one sees  (without worrying about numerical multiplicative constants in (\ref{SEE-IS-simp}) and (\ref{SBH})) from (\ref{SBH}), $11.4 r_h^3|\log r_h|^4 = S_{\rm BH}$. This is solved to yield,
\begin{equation}
\label{rh-SBH}
r_h = e^{\frac{4}{3}W(0.4 S_{\rm BH}^{\frac{1}{4}})}.
\end{equation}
 From Fig. \ref{SEEISoverSBH-vs-SBH} and using (\ref{rh-SBH}), it is evident that  (\ref{SEEISoverSBH-arrh>>a3}) $\sim$ (\ref{SEEISoverSBH-D-dims})  implies there will be a lower bound on  $r_h$, the non-extremality parameter in the ${\cal M}$-theory dual of large-$N$ thermal QCD.

From equation (\ref{SEE-IS-simp}) we find that entanglement entropy contribution from the island surface reaches a constant value. Now considering both the contribution together i.e. Hartman-Maldacena-like surface and island surface, we obtain the Page curve as follows. Initially entanglement entropy of the Hawking radiation is increasing linearly with time and will diverge at late times. But due to presence of the island surface entanglement entropy contribution from this surface dominates after the Page time and stops the linear time growth of the entanglement entropy of the Hawking radiation and reaches a constant value. In this way we obtain the Page curve of an eternal black hole from ${\cal M}$-theory dual where bath is a non-conformal theory.

\subsection{Page Curve and An Exponential(-in-$N$) Hierarchy of Entanglement Entropies up to ${\cal O}(R^4)$ Before/After the Page Time}
\label{Page-curve-plot-HD}
In this subsection we will obtain the Page curve of an eternal neutral black hole appearing on the gravity dual of thermal QCD-like theories at intermediate coupling, and show that a hierarchy in powers of $e^{-\kappa_{l_p}N^{1/3}}$ in the entanglement entropies arising from the EH+GHY terms and the ${\cal O}(R^4)$ corrections to the same, before and after the Page time, naturally arises. For this purpose we will use the results of subsections \ref{EE-HM-HD} and \ref{EE-IS-HD}.\par
Since we are equipped with all the results therefore we are going to obtain the Page curve in this subsection. For this purpose we are plotting the entanglement entropy contribution from the Hartman-Maldacena-like surface obtained in subsection \ref{EE-HM-HD} and entanglement entropy of the Hawking radiation coming from the island surface obtained in subsection \ref{EE-IS-HD}.\par
Entanglement entropy of the Hawking radiation corresponding to the Hartman-Maldacena-like surface is given in equation (\ref{SEE-HM-tb0}) and for the island surface it is given in equation (\ref{SEE-IS-simp}). For the numerical values of the parameters of the model we obtain the following plot.
\par
\begin{figure}
\begin{center}
\includegraphics[width=0.60\textwidth]{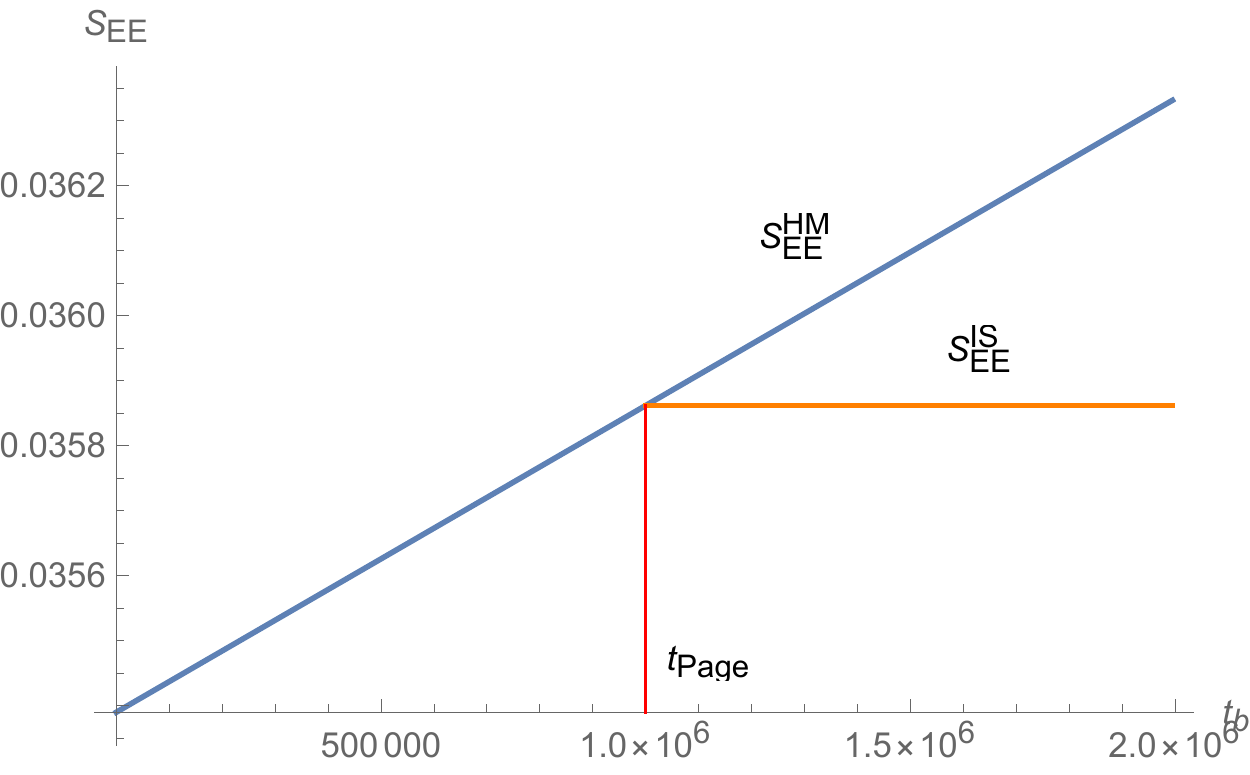}
\end{center}
\caption{Page curve of an eternal black hole from doubly holographic setup in ${\cal M}$-theory dual. Blue line in the graph corresponds to the entanglement entropy contribution from the Hartman-Maldacena-like surface and orange line corresponds to the entanglement entropy contribution from the island surface.}
\label{Page-curve-plot}
\end{figure}

We have plotted $S_{EE}^{\beta^0, {\rm HM}}$ and $S_{EE}^{\rm IS}$ in Fig. \ref{Page-curve-plot} for $N=10^{3.3},M=N_f=3,g_s=0.1, \kappa_{l_p}=0.47, c_2=-10^{8}, c_1=-10^3$.  
%\begin{eqnarray}
%\label{S_EE^HM-approx-linear-time}
%S_{EE}^{\beta^0,\ {\rm HM}} \sim 
%\left(1-\frac{4 t_{b_0}}{3 c_2}\right)\left(\frac{2}{3}\log \left(\frac{c_2}{c_1}\right) -n_{t_b} \log(N)\right)^{17/6}.
%\end{eqnarray}

From figure \ref{Page-curve-plot} it is clear that initially entanglement entropy of the Hawking radiation is increasing linearly with time. After the Page time entanglement entropy of the Hawking radiation coming from the island surface dominates therefore entanglement entropy stops increasing and we obtain the Page curve. The values of $S_{\rm EE}$ in Fig. \ref{Page-Curve-Areas} and Fig. \ref{Page-curve-plot} differ because, e.g., the latter did not include the factor of $\left(2\pi\right)^4$ arising from integrations w.r.t. $\phi_{1,2}, \psi, x^{10}$, etc.

\begin{figure}
\begin{center}
\includegraphics[width=0.60\textwidth]{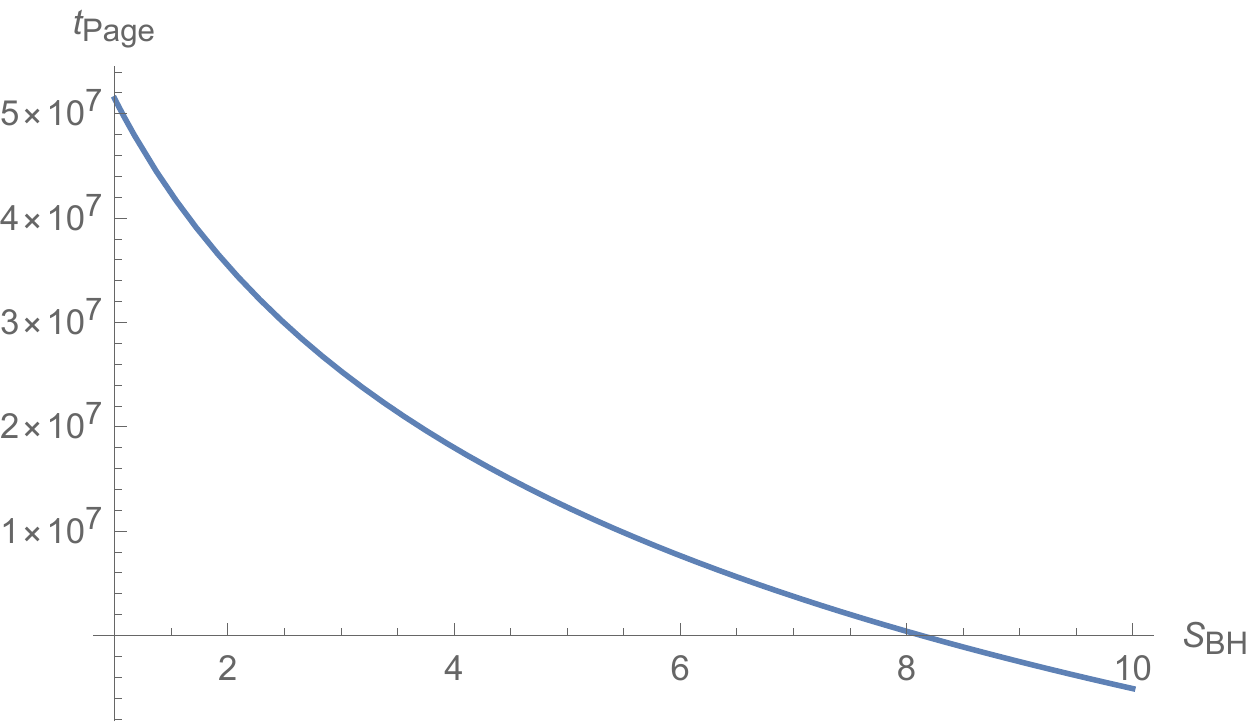}
\end{center}
\caption{Page time for $N=10^{3.3}, M=N_f=3, g_s=0.1,  c_2=-10^8, c_1=-10^{3}$}
\label{Page-time-vs-SBH-plot}
\end{figure}

The Page time is obtained by equality of the entanglement entropies for the Hartman-Maldacena-like surface and Island Surface (at the Page time). This yields:
\begin{eqnarray}
\label{Page-time}
& & t_{\rm Page} = \frac{3}{4} c_2 \left(1-\frac{9\ 3^{5/6} {g_s}^{35/6} M N^{7/10} \log ^2(N) | \log ({r_h})|
   ^{4/3}}{\sqrt[3]{{N_f}} {r_h}^{11/2} \left(-3 {n_{t_b}} \log (N)+2 \log
   \left(\frac{c_2}{c_1}\right)\right){}^{17/6}}\right).
\end{eqnarray}
The Page time, implicitly assumes choosing the time evolution of 
$S_{\rm EE}^{\rm HM}$ which is (approximately) linear in time up to the Page time beyond which, the constant contribution to the entanglement entropy from the island, takes over. The behavior of the Page time as a function of the Black-hole entropy (using (\ref{rh-SBH})) is given in Fig. \ref{Page-time-vs-SBH-plot}. This also shows that positivity of the Page time requires an upper bound on the black-hole entropy and therefore the IR cut-off $r_h$. Given that $r_h$, the non-extremality parameter, is essentially a constant of integration \cite{Klebanov+Buchel-et-al_rh-const-of-integration} and requiring the same (in units of ${\cal R}_{D5/\overline{D5}}$) to be less than unity, can be effected, e.g., by noting from (\ref{rh-SBH}) that $r_h$ is an increasing function of $S_{\rm BH}$, and therefore: $r_h(S_{\rm BH})\rightarrow \frac{r_h(S_{\rm BH})}{r_h(S^0_{\rm BH}:t_{\rm Page}(S^0_{\rm BH})>0)}$. Therefore, for the values of $N, M, g_s, c_1, c_2, n_{t_b}, \kappa_{l_p}$ chosen, this would imply $r_h\rightarrow\frac{r_h(S_{\rm BH})}{r_h(S_{\rm BH}\approx8)}$.

It appears we have disregarded $S_{\rm EE, HM}^\beta$ altogether. Let us now discuss why the same is justified and how a exponential-large-$N$-suppression hierarchy is generated in the process.

One can show that:
\begin{eqnarray}
\label{SEEHMbeta-i}
& & S_{\rm EE, HM}^\beta \sim \frac{\beta^{4/3}x_R^2g_s^{\frac{23}{6}}M^3N_f^{\frac{2}{3}}\log ^2({r_h}) (\log (N)-12 \log ({r_h})) (\log (N)-9 \log ({r_h}))}{r_h^{\frac{5}{2}}\left(\log N - 3 \log r_h\right)^{\frac{7}{6}}} \nonumber\\
& & \times  \left(-216
   \left(16+\sqrt{2}\right) {g_s}^4 \kappa_{\lambda_1} M^4 {N_f}^2 \log (N) \log ^3({r_h})+9
   \left(16+\sqrt{2}\right) {g_s}^4 \kappa_{\lambda_1} M^4 {N_f}^2 \log ^2(N) \log ^2({r_h})\right.\nonumber\\
   & & \left.+1296
   \left(16+\sqrt{2}\right) {g_s}^4 \kappa_{\lambda_1} M^4 {N_f}^2 \log ^4({r_h})+4096
   \left(4+\sqrt{2}\right) \pi ^4 {\kappa_{\lambda_2}}\right),
\end{eqnarray}
which in the $|\log r_h|\gg \log N$-limit, approximates to:
\begin{eqnarray}
\label{SEEHMbeta-ii}
& & \frac{\beta ^{4/3} {g_s}^{47/6} M^7 N^{3/10} {N_f}^{8/3} x_R^2 \log (2)
   (-\log ({r_h}))^{41/6}}{{r_h}^{5/2}}.
\end{eqnarray}
Using (\ref{scalings}), (\ref{SEEHMbeta-ii}) yields,
\begin{eqnarray}
\label{SEEHMbeta-iii}
& & \frac{{g_s}^{47/6} M^7 N^{3/10} {N_f}^{8/3} x_R^2 e^{-2 \kappa_{l_p}
   N^{\frac{1}{3}}} (-\log ({r_h}))^{41/6}}{\sqrt{{r_h}}}.
\end{eqnarray}
Similarly,
\begin{eqnarray}
\label{SEEISbeta-i}
& & S_{\rm EE, IS}^\beta \sim \frac{ \beta ^2 {g_s}^{15/2} \sqrt{{\kappa_\alpha}} {\kappa_\sigma}^{5/2} {\kappa_{U_1}} M^2 N \log ^2(N) ({a_2} {r_h}+{a_3}) \sqrt{3 {a_2}^2 {a_3} {r_h}^2+2 {a_2}^3
   {r_h}^3+3 {a_2} {a_3}^2 {r_h}+{a_3}^3}}{{a_2}^{5/2} {\kappa_{\lambda_5}} {r_h}^{35/2} (-\log ({r_h}))^{2/3}}\nonumber\\
& & \times  \left(6561 \sqrt[6]{2} \left(1+8 \sqrt{2}\right)
   {g_s}^{17/6} \kappa_{\lambda_3} M^6 {N_f}^{7/3} {r_h}^8 \log ^6({r_h})+1048576 \sqrt[3]{6} \pi
   ^{47/6} {\kappa_\alpha}^2 {\kappa_{Z_1}} N^2 (-\log ({r_h}))^{2/3}\right)\nonumber\\
& & \approx \frac{17 \beta ^2 {g_s}^{15/2} {\log N}^2 M^2 N^3 ({a_2} {r_h}+{a_3}) \sqrt{3 {a_2}^2 {a_3}
   {r_h}^2+2 {a_2}^3 {r_h}^3+3 {a_2} {a_3}^2 {r_h}+{a_3}^3}}{{a_2}^{5/2}
   {r_h}^{35/2}},
\end{eqnarray}
which in the $a_2r_h\gg a_3$-limit yields,
\begin{eqnarray}
\label{SEEISbeta-ii}
& & 
   \frac{\beta ^2 {g_s}^{15/2} {\log N}^2 M^2 N^3}{{r_h}^{15}}.
\end{eqnarray}
Using (\ref{scalings}), (\ref{SEEISbeta-ii}) yields:
\begin{eqnarray}
\label{SEEISbeta-iii}
& & \frac{{g_s}^{15/2} {\log N}^2 M^2 N^3 e^{-3 {\kappa_{l_p}} N^{\frac{1}{3}}}}{{r_h}^{12}}.
\end{eqnarray}
From (\ref{SEE-HM-tb0}), (\ref{SEE-IS-simp}), (\ref{SEEHMbeta-iii}) and (\ref{SEEISbeta-iii}), one sees the following hierarchy:
\begin{eqnarray}
\label{S_EE_hierarchy}
& & S_{\rm EE}^{\rm HM, \beta^0}:S_{\rm EE}^{\rm IS,\ \beta^0}:S_{\rm EE}^{\rm HM,\ \beta}:S_{\rm EE}^{\rm IS,\ \beta} \sim \gamma^{\frac{3}{2}}:\gamma^{\frac{3}{2}}:\gamma^2:\gamma^3,
\end{eqnarray}
where $\gamma\equiv e^{-\kappa_{l_p}N^{\frac{1}{3}}}$. From the above equation we can see that $O(\beta)$ corrections to entanglement entropies for the Hartman-Maldacena-like and Island surfaces are more exponentially large-$N$ suppressed. Therefore we have disregarded those contributions for the computation of Page curve.

\section{Massless Graviton - the Physical reason for Exponentially Suppressed Entanglement Entropies}
\label{massless_graviton}

We now aim at presenting a physical reason behind the exponential suppression of the entanglement entropies of the HM-like and Island surfaces in \ref{Page-curve-plot-HD}. The essence of the discussion is the following. We will show that despite the coupling of a non-gravitational bath to the ETW-"brane'', (imposition of Dirichlet boundary condition at the horizon on the radial profile of the graviton wave function) results in the quantization of the graviton mass, which (for an appropriate choice of the quantum number) can therefore be taken to be massless. Usually in the context of AdS$_{d+1}$ gravity duals of CFT$_d$ on $\partial$AdS$_{d+1}$ at zero temperature, massless graviton implies a vanishing angle between the ETW/KR-brane and $\partial$AdS$_{d+1}$, which further implies the islands cease to contribute \cite{G-bath,Island-IIB-2}. However, as shown/implied in \ref{ETW-sub}, the ETW-``brane" in our setup, $x^1=$constant, is orthogonal (in the $x^1-r$-plane) to the thermal bath/QCD-like theory (after having integrated out the angular directions of ${\cal M}_6(\theta_{1,2},\phi_{1,2},\psi,x^{10})$). Despite the same, we will show that one obtains a massless graviton in our setup \footnote{See \cite{Luest_et_al-massless-graviton} in the context of islands in black holes in $AdS_4$ coupled to an external bath embedded in type IIB string theory supporting light gravitons.}.  What is more interesting is that there is a comparable exponential-in-$N$ suppression in the HM-like entanglement entropy which is why the two can be compared at the Page time and one obtains a Page curve.
 
As in \cite{C. Bachas and J. Estes [2011]}, let us write the $D=11$ metric as: 
\begin{equation}
\label{Kounas+Estes-metric}
ds^2 = e^{2A(y)}\overline{g}_{\mu\nu}(x)dx^\mu dx^\nu + 
\hat{g}_{mn}dy^m dy^n.
\end{equation}
As the resolution parameter $a$, up to LO in $\frac{g_s M^2}{N}$, is proportional to $r_h$ \cite{Mia_Vaidya}, \cite{EPJC-2}, \cite{Bulk-Viscosity-McGill-IIT-Roorkee}, strictly speaking (\ref{Kounas+Estes-metric}) is applicable if one disregards ${\cal O}\left(\frac{a^4}{r^4}\right)$ terms (working in the IR-UV interpolating region would ensure the same). The perturbed metric that will be considered is: 
 $\widetilde{ds^2} = e^{2A(y)}\left(\overline{g}_{\mu\nu} + h_{\mu\nu}\right) dx^\mu dx^\nu + \hat{g}_{mn}dy^m dy^n$, $h_{\mu\nu}(x,y) = h^{[tt]}_{\mu\nu}(x)\psi(y)$ 
 (as ansatz under linear perturbations): $\overline{D}^\mu h_{\mu\nu}^{[tt]} = \overline{g}^{\mu\nu}h_{\mu\nu}^{[tt]}=0$, and \cite{C. Bachas and J. Estes [2011]} 
 \begin{equation}
 \label{EOM-psi}
 -\frac{e^{-2A(y)}}{\sqrt{|\hat{g}(y)|}}\partial_m\left(
 \sqrt{|\hat{g}(y)|}\hat{g}^{mn}e^{4A(y)}\partial_n\psi(y)\right) = m^2\psi(y).
 \end{equation}

Under the simplifying assumption of localization around, say, $(\theta_1,\theta_2)\sim\left(\frac{\alpha_{\theta_1}}{N^{\frac{1}{5}}},\frac{\alpha_{\theta_2}}{N^{\frac{3}{10}}}\right)$ near which one can show that $\psi(r,\theta_{1,2},\phi_{1,2},\psi,x^{10})\rightarrow\psi(r,\theta_1)$, 
the eigenvalue equation (\ref{EOM-psi}) up to LO in $N$ in the IR-UV interpolating region reduces to:
\begin {eqnarray}
\label{EOM-LO-N}
& & \hskip -0.4in -\frac{\partial^2\psi(r,\theta_1)}{\partial r^2} +\frac{16(r^4+r_h^4)}{(r^4-r_h^4)}
\frac{\partial\psi(r,\theta_1)}{\partial r}\nonumber\\
& & \hskip -0.4in + \kappa_{r\theta_1}\frac{N^{4/5}r^6}{g_s^3M^2N_f^2\left(r^2-3a^2\right)^2\left(r^4-r_h^4\right)\left(\log N - 9 \log r\right)^2\left(\log r\right)^2}\left(\frac{\partial^2\psi(r,\theta_1)}{\partial\theta_1^2} - 2\frac{\partial\psi(r,\theta_1)}{\partial\theta_1}\right)\nonumber\\
& & \hskip -0.4in - m^2\psi(r,\theta_1) = 0,
\end {eqnarray}
or, equivalently, using separation of variables $\psi(r,\theta_1) = \mathbb{R}(r)\Theta(\theta_1)$:
\begin {eqnarray}
\label{EOM-separation_of_variables}
& & \hskip -0.4in \left(-\frac{1}{{\cal R}(r)}\mathbb{R}''(r) + \frac{1}{\mathbb{R}(r)}\frac{16(r^4+r_h^4)}{(r^4-r_h^4)}\mathbb{R}'(r) - m^2\right)\nonumber\\
& & \hskip -0.4in \left(\kappa_{r\theta_1}\frac{N^{4/5}r^6}{g_s^3M^2N_f^2\left(r^2-3a^2\right)^2\left(r^4-r_h^4\right)\left(\log N - 9 \log r\right)^2\left(\log r\right)^2}\right)^{-1}\nonumber\\
& & \hskip -0.4in = \frac{1}{\Theta(\theta_1)}\left(\Theta''(\theta_1) - 2 \Theta'(\theta_1)\right) \equiv \lambda
\end {eqnarray}

 The solution to $\Theta(\theta_1)$ equation in (\ref{EOM-separation_of_variables}):
\begin {eqnarray}
\label{solution-psi}
e^{\theta_1\left(1\pm\sqrt{1+\lambda}\right)},
\end {eqnarray}
which can be meaningful if $\lambda=0$; we hence obtain: $\Theta(\theta_1)=$Constant.

The "EOM'' for the radial profile $\mathbb{R}$ from (\ref{EOM-separation_of_variables}), is given as under:
\begin{equation}
\label{EOM-R}
-\mathbb{R}''(r) + \left(\frac{8}{r-r_h} - \frac{4}{r_h} + {\cal O}(r-r_h)\right)\mathbb{R}'(r) - m^2 \mathbb{R}(r) = 0.
\end{equation}
 
As EOM (\ref{EOM-psi}) is homogeneous, $\psi\rightarrow m^2\psi(r)$ is also a valid solution which yields:
\begin {eqnarray}
\label{solution-psi-TricomiU}
& & m^2c_1 U\left(-\frac{8-5 \sqrt{4-m^2 {r_h}^2}}{\sqrt{4-m^2 {r_h}^2}},10,\frac{2
   r \sqrt{4-m^2 {r_h}^2}}{{r_h}}-2 \sqrt{4-m^2 {r_h}^2}\right)\nonumber\\
& & \times \exp
   \left(\frac{r \left(-\sqrt{4-m^2 {r_h}^2}-2\right)+9 {r_h} \log
   ({r_h}-r)}{{r_h}}\right)\nonumber\\
& & +m^2c_2 L_{\frac{8-5 \sqrt{4-m^2
   {r_h}^2}}{\sqrt{4-m^2 {r_h}^2}}}^9\left(\frac{2 r \sqrt{4-m^2
   {r_h}^2}}{{r_h}}-2 \sqrt{4-m^2 {r_h}^2}\right)\nonumber\\
& & \times \exp \left(\frac{r
   \left(-\sqrt{4-m^2 {r_h}^2}-2\right)+9 {r_h} \log
   ({r_h}-r)}{{r_h}}\right).
\end {eqnarray}

 If $c_2=0$ then
\begin {eqnarray}
\label{R-exp-rh}
& & m^2 U\left(-\frac{8-5 \sqrt{4-m^2 {r_h}^2}}{\sqrt{4-m^2 {r_h}^2}},10,\frac{2
   r \sqrt{4-m^2 {r_h}^2}}{{r_h}}-2 \sqrt{4-m^2 {r_h}^2}\right)\nonumber\\
& & \times \exp
   \left(\frac{r \left(-\sqrt{4-m^2 {r_h}^2}-2\right)+9 {r_h} \log
   ({r_h}-r)}{{r_h}}\right) \nonumber\\
& &  = \frac{e^{-2-\sqrt{4-m^2r_h^2}}}{\sum_{l=0}^3c_l (m r_h)^{2l}\Gamma\left(-4 - \frac{8}{\sqrt{4-m^2r_h^2}}\right)}\left[1 + {\cal O}\left((r-r_h)^2\right) \right].
\end {eqnarray}

If one imposes Dirichlet b.c.\footnote{Neumann b.c., $\psi'(r=r_h)=0$, is identically satisfied $\forall m$.} at $r=r_h: \psi(r=r_h)=0$ then one sees that one would require
$ -4 - \frac{8}{\sqrt{4-m^2r_h^2}} = - n, n\in\mathbb{Z}^+$, i.e., 
\begin{equation}
\label{solution-m}
m = \frac{2\sqrt{n(n-8)}}{4r_h(1-n)}\stackrel{n=8}{\rightarrow}0.
\end{equation}
One notes that,
\begin{eqnarray}
\label{m_0}
& & U\left(-\frac{8-5 \sqrt{4-m^2 {r_h}^2}}{\sqrt{4-m^2 {r_h}^2}},10,\frac{2
   r \sqrt{4-m^2 {r_h}^2}}{{r_h}}-2 \sqrt{4-m^2 {r_h}^2}\right)\nonumber\\
& & \times \exp
   \left(\frac{r \left(-\sqrt{4-m^2 {r_h}^2}-2\right)+9 {r_h} \log
   ({r_h}-r)}{{r_h}}\right) \stackrel{m=0}{\rightarrow}-\frac{c_1 {r_h}^9 \Gamma \left(9,\frac{4 r}{{r_h}}-4\right)}{262144 e^4}\nonumber\\
   & & = -\frac{315 \left(c_1 {r_h}^9\right)}{2048 e^4}+\frac{c_1 (r-r_h )^9}{9 e^4} + {\cal O}\left((r-{r_h})^{10}\right).
\end{eqnarray}

If $c_1=0$ then \\ $c_2 L_{\frac{8-5 \sqrt{4-m^2
   {r_h}^2}}{\sqrt{4-m^2 {r_h}^2}}}^9\left(\frac{2 r \sqrt{4-m^2
   {r_h}^2}}{{r_h}}-2 \sqrt{4-m^2 {r_h}^2}\right) \exp \left(\frac{r
   \left(-\sqrt{4-m^2 {r_h}^2}-2\right)+9 {r_h} \log
   ({r_h}-r)}{{r_h}}\right)$\\
    satisifies Dirichlet/Neumann b.c. $\forall m$. In particular for $m=0$, the above yields $c_2 e^{-\frac{4 r}{{r_h}}} ({r_h}-r)^9 L_{-1}^9\left(\frac{4 r}{{r_h}}-4\right)$. Interestingly, $\frac{d^n}{dr^n}L_{-1}^9\left(\frac{4 r}{{r_h}}-4\right)=(-4)^n\frac{1}{r_h^n}L_{-n-1}^{n+9}\left(\frac{4 r}{r_h }-4\right)$, implying \\ $\lim_{r\rightarrow r_h}L_{-n-1}^{n+9}\left(\frac{4 r}{r_h }-4\right) = \binom{8}{n+9}=0. $ 
 
In the UV, the EOM up to LO in $N, N_f^{\rm UV}, M^{\rm UV}$,  is:
\begin {eqnarray}
\label{EOM-UV}
& & -\frac{\partial^2\psi_{\rm UV}(r,\theta_1)}{\partial r^2} + \frac{16}{r}\frac{\partial\psi_{\rm UV}(r,\theta_1)}{\partial\theta_1}\nonumber\\
& & -2\kappa_{r\theta_1}\frac{N^{4/5}}{g_s^3M_{\rm UV}^2N_{f\ rm UV}^2\left(\log N - 9 \log r\right)^2\left(\log r\right)^2r^2}\frac{\partial\psi_{\rm UV}}{\partial\theta_1}-m^2\psi_{\rm UV} ,
\end {eqnarray}
whose solution is given as:
\begin {eqnarray}
\label{EOM-UV-sol-i}
& & \psi_{\rm UV}\sim\frac{1}{m^{17/2}}\Biggl[c_1\sum_{l_1=0}^3a_{l_1} (m r)^{2l_1+1}
\cos(m r) \nonumber\\
& & + \Biggl(c_1\sum_{l_2=0}^4a_{l_2}(m r)^{2l_2} + c_2 \sum_{l_1=0}^3
(m r)^{2l_1+1}\Biggr)\sin (m r)\Biggr].
\end {eqnarray}
 
As the above, for $m\neq0$, is ill-behaved in the UV, one is required to set $m=0$ for which
\begin {eqnarray}
\label{EOM-UV-sol-ii}
\mathbb{R}_{\rm UV}(r)=c_1^{\rm UV}r^{17} + c_2^{\rm UV}.
\end {eqnarray}
For $\psi(r)$ to be normalizable, one has to set $c_1^{\rm UV}=0$ implying, as expected a constant graviton wave-function in the UV.

One thus sees the physical/intuitive reason for the exponentially suppressed entanglement entropy for Island Surface (\ref{SEE-IS-simp}) is that the Laplace-Beltrami equation for the internal coordinates (\ref{EOM-psi}) permits a vanishing graviton mass - this is also related to the fact that our computations are in the "near-horizon" limit ($r<\left(4\pi g_s N\right)^{1/4}$) wherein even the UV cut-off $r_{\rm UV}\stackrel{\sim}{<}(4\pi g_s N)^{1/4}$ and hence the internal manifold is not non-compact \cite{C. Bachas and J. Estes [2011]}. This is what is responsible for the very small (involving exponential-in-$N$ suppression) entanglement entropy of the Island surface. What is non-trivial and therefore extremely interesting is a comparable entanglement entropy of the Hartman-Maldacena-like surface in (\ref{SEE-HM-tb0}) near the Page time. The vanishing graviton mass further manifests itself at ${\cal O}(\beta)$ in the entanglement entropies of the IS-vs-HM-like surfaces in (\ref{S_EE_hierarchy}) - the former is suppressed relative to the latter. 

\begin{figure}
\begin{center}
\includegraphics[width=0.60\textwidth]{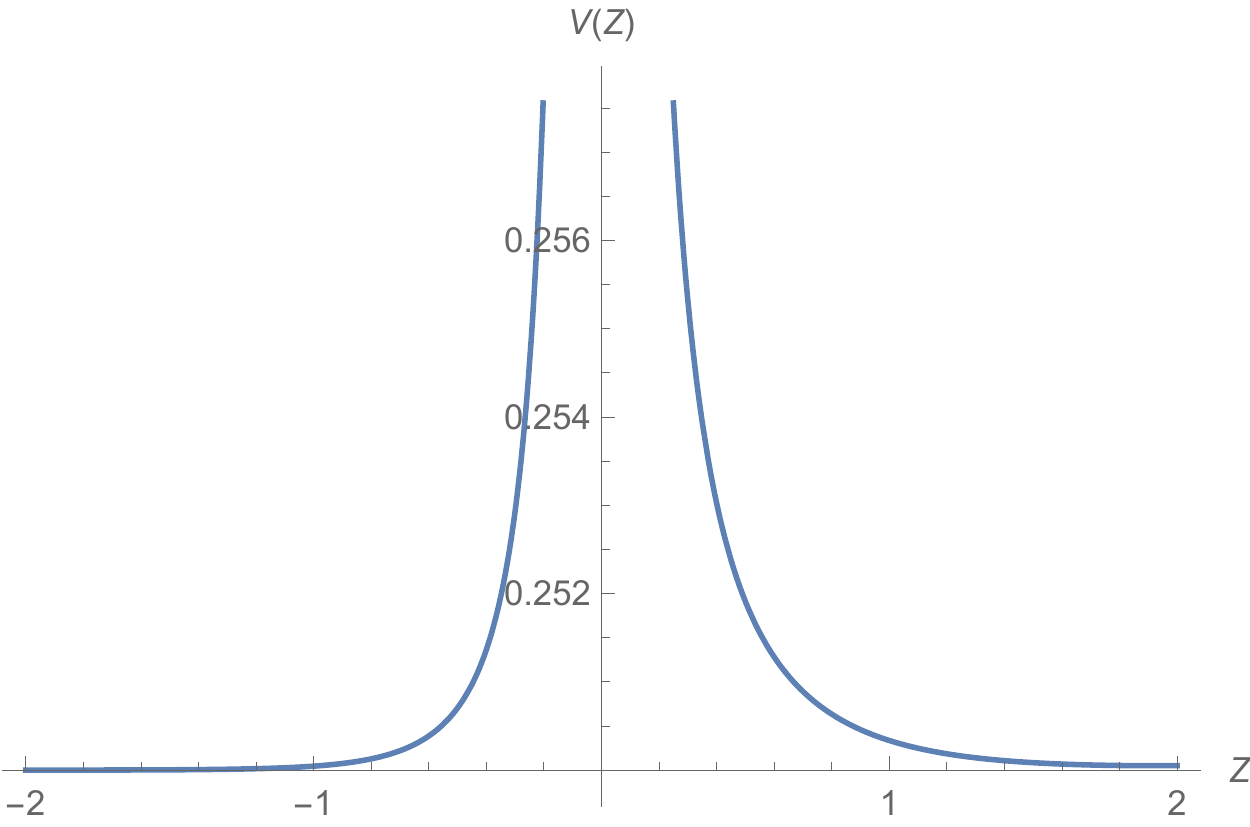}
\end{center}
\caption{Volcano potential for massless graviton from (\ref{V[Z]})}
\label{V-Potential}
\end{figure}

Now using $r=r_h e^Z$\footnote{Advantage of using redefined radial coordinate is that $r \in (0,\infty)$ maps to $Z \in (-\infty,\infty)$ so that we can see a nice volcano-like potential on both sides of $Z=0$.}, we can convert equation (\ref{EOM-separation_of_variables}) into the following Schr\"odinger-like equation
\begin{eqnarray}
\label{S-like-EOM-Z}
-R''(Z)+ V(Z)R(Z) = 0,
\end{eqnarray}
where 
\begin{equation}
\label{R-calR}
R(Z) = \mathbb{R}(Z) e^{r_h\left(-8 e^Z + \frac{Z}{2r_h} + 8 \tan^{-1}(e^Z) + 8 \tanh^{-1}(e^Z)\right)}\approx \mathbb{R}(Z) e^{\frac{Z}{2}},
\end{equation}
and,
\begin{eqnarray}
\label{V[Z]}
V(Z)=-\frac{e^{2 Z} r_h^2 \left(\left(m^2-64\right) e^{8 Z}-2 \left(m^2+64\right) e^{4 Z}+m^2-64\right)}{\left(e^{4 Z}-1\right)^2}+\frac{64 e^{5 Z} r_h}{\left(e^{4 Z}-1\right)^2}+\frac{1}{4}
\end{eqnarray}
The above potential for massless graviton $(m=0)$ is plotted in figure \ref{V-Potential};  this potential is ``volcano''-like  with the massless graviton localized near the horizon on the ETW ``brane''.  This is similar to \cite{KR1} wherein one can localize gravity on end-of-the-world(ETW) brane for non-zero brane tension due to appearance of ``crater'' in ``volcano'' potential appearing in Schr\"odinger-like equation of motion of graviton wave function. 

\begin{figure}
\begin{center}
\includegraphics[width=0.60\textwidth]{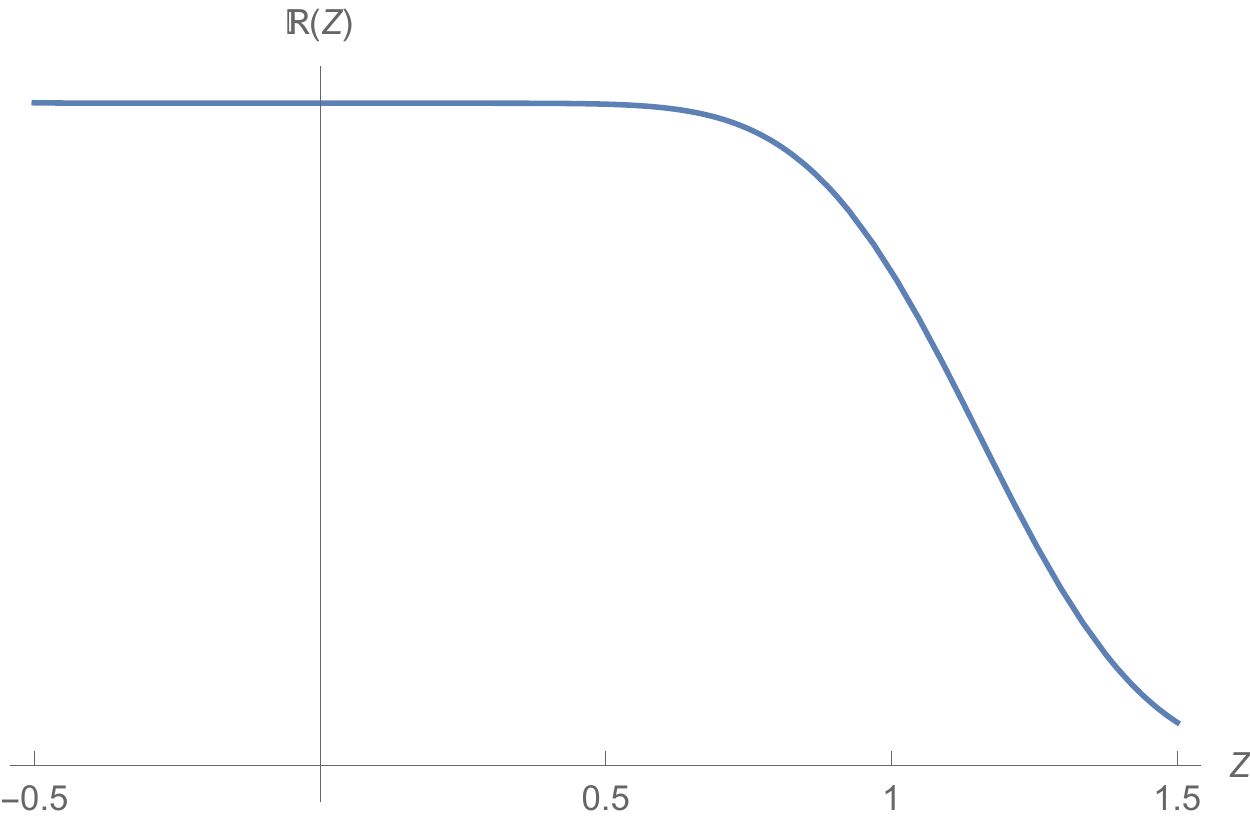}
\end{center}
\caption{Localization of the Graviton Wave Function near the horizon from the $m=0$-limit of the solution (\ref{m_0}) of (\ref{EOM-R})}
\label{grav-wf-localized}
\end{figure}

In our setup, the ETW-``brane'' has non-zero ``tension'' (\ref{T_ETW}) and therefore it is possible to localize gravity on ETW-``brane''. Using (\ref{m_0}), one sees that in the massless-limit of the graviton, the graviton wave-function is indeed localized near the horizon as shown in figure \ref{grav-wf-localized}.
Alternatively, expanding $V(Z)$ of (\ref{V[Z]}) around the horizon $Z=0$ and obtaining: $V(Z\sim0)\sim \frac{4r_h(1+4r_h)}{Z^2} + \frac{4r_h(1+8r_h)}{Z}
-\frac{1}{12}(-3+40r_h-896r_h^2) + {\cal O}(Z)$, one obtains:
{\footnotesize
\begin{eqnarray}
\label{wave-function-Schroedinger}
& & \hskip -0.4in R(Z) = c_1 M_{-\frac{4 \sqrt{3} r_h  (8 r_h +1)}{\sqrt{896 r_h ^2-40 r_h +3}},4
   r_h +\frac{1}{2}}\left(\frac{\sqrt{896 r_h ^2-40 r_h +3} Z}{\sqrt{3}}\right)+c_2 W_{-\frac{4 \sqrt{3}
   r_h  (8 r_h +1)}{\sqrt{896 r_h ^2-40 r_h +3}},4 r_h +\frac{1}{2}}\left(\frac{\sqrt{896
   r_h ^2-40 r_h +3} Z}{\sqrt{3}}\right).\nonumber\\
   & & 
\end{eqnarray}
}
Now, both Whittaker functions are complex for $Z<0$, i.e., $r<r_h$. Setting $c_1=0$, one obtains $\mathbb{R}(Z) = e^{-\frac{Z}{2}}W_{-\frac{4 \sqrt{3}
   r_h  (8 r_h +1)}{\sqrt{896 r_h ^2-40 r_h +3}},4 r_h +\frac{1}{2}}\left(\frac{\sqrt{896
   r_h ^2-40 r_h +3} Z}{\sqrt{3}}\right)$, and hence the following plot which indiates a decay of the graviton wave-function away from the horizon.
  \begin{figure}
\begin{center}
\includegraphics[width=0.60\textwidth]{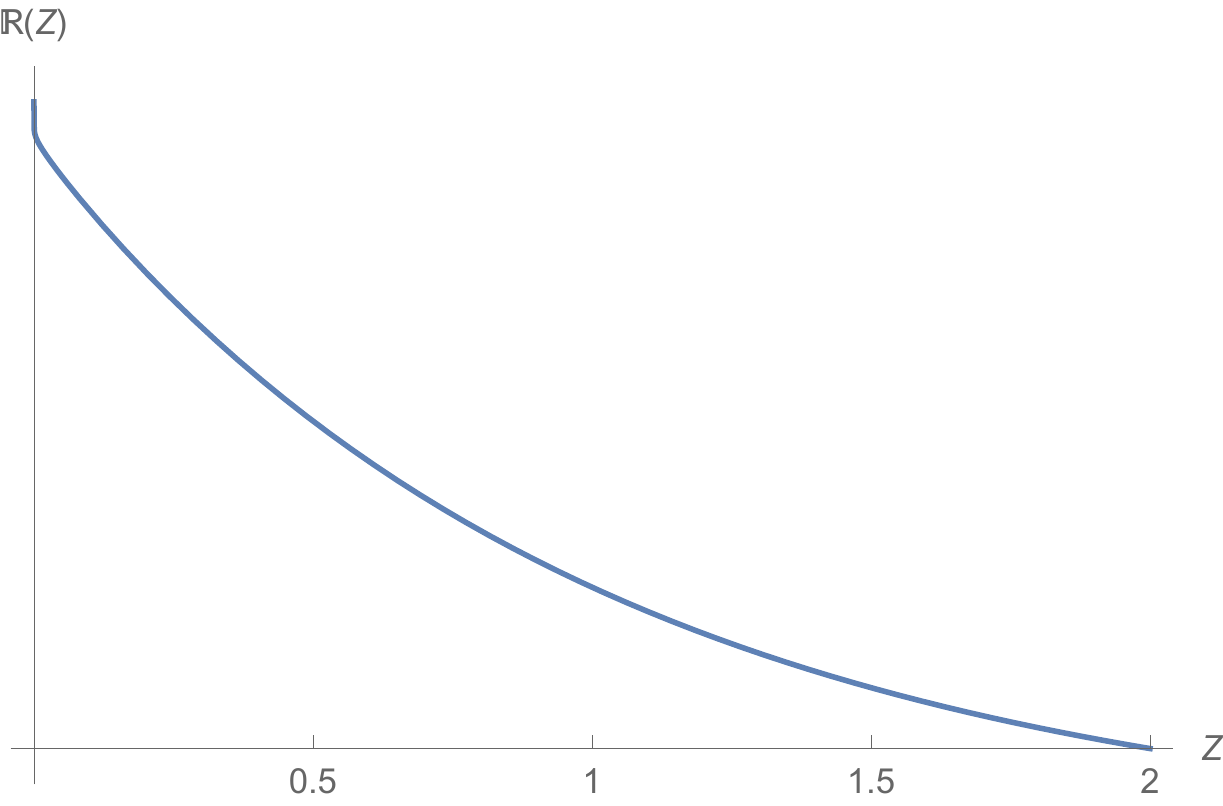}
\end{center}
\caption{Localization of the Graviton Wave Function near the horizon from the solution of the Schr\"odinger-like radial wave equation (\ref{S-like-EOM-Z})}
\label{grav-wf-localized-Schroedinger}
\end{figure}  

\section{Summary}
\label{summary}
Following is the summary of the key steps and results that were obtained in this paper. \par
\begin{itemize}
\item We discussed the doubly holographic setup in the context of ${\cal M}$ theory background in \ref{DHS} summarized in Fig. \ref{Flowchart}. The main idea is that we consider an ETW-``brane" at $x=0$ in figure \ref{Doubly Holographic Setup}, which contains the black hole and then we couple the ETW-``brane" to a non-conformal bath (QCD in 3 dimensions after integrating out the angular coordinates and wick rotation along $x^3$). We then discussed how we can calculate the entanglement entropy of Hawking radiation using the prescription of \cite{Dong} for higher derivative gravity theories. We derived the ETW-``brane" embedding in ${\cal M}$-theory background in section \ref{ETW-sub}. First we considered the Einstein Hilbert term and the GHY term along with the tension of the ETW-``brane" and obtained the embedding  (\ref{ETW_iii}). By using the induced metric (\ref{ETW_ii}) on the ETW, we computed various terms appearing in equation (\ref{ETW_iii}) and found that at ${\cal O}(\beta^0)$ the embedding of the ETW-``brane" will be given by constant $x$ (equation (\ref{ETW_vii})). Given that in the MQGP limit of (\ref{MQGP_limit}), the $J_0$ dominates over the $E_8$ and $t_8^2G^2R^3$ terms in the ${\cal O}(R^4)$ terms in (\ref{D=11}), we work with only the $J_0$ term. We found that the ETW-``brane" embedding obtained with just EH-GHY terms, continues to hold even with the inclusion of the $J_0$ term as restricted to the same (embedding) the boundary term involving covariant derivative of the metric variation (that one would have had to cancel by construction of an appropriate boundary term relevant to the $J_0$ term), vanishes.

\item Given the robustness of the Islands picture, the ${\cal M}$-theory dual of large-$N$ thermal QCD at high temperature constructed in \cite{MQGP}, \cite{HD-MQGP} (based on \cite{metrics}) too is thus expected to generate a Page curve. The aforementioned top-down ${\cal M}$-theory dual however offers the following additional conceptual insights:
\begin{itemize}
\item There are very few papers (e.g., \cite{NGB}) wherein the authors discuss doubly holographic setup in higher derivative theory of gravity because of absence of the knowledge of explicit forms of boundary terms on the ETW-brane with the inclusion of higher derivative terms. As we explicitly show in \ref{ETW-sub-sub-ii}, no boundary terms are generated from the inclusion of the ${\cal O}(R^4)$ terms. 

\item	To the best of our knowledge, our model/(aforementioned) ${\cal M}$-theory dual is either the only one (from ${\cal M}$ theory) or one of the very few top-down models that obtain(s) the Page curve for massless graviton\footnote{See \cite{critical-islands} in which author has discussed that one can get Page curve for massless graviton in critical Randall-Sundrum II model from bottom-up approach.} in  \ref{massless_graviton}. 

\item In our top-down ${\cal M}$-theory dual, we find that ETW-``brane'' to be a fluxed hypersurface ${\cal W}$ that is a warped product of an asymptotic $AdS_4$ and a six-fold $M_6$ where $M_6$ is a warped product of the ${\cal M}$-theory circle and a non-Einstenian generalization of $T^{1,1}$; the hypersurface ${\cal W}$, can also be thought of as an effective ETW-``brane'' corresponding to fluxed intersecting $M5$-brane wrapping a homologous sum of $S^3\times[0,1]$ and
$S^2\times S^2$ in a warped product of $\mathbb{R}^2$ and an $SU(4)/Spin(7)$-structure eight-fold. The ETW-``brane", ${\cal W}$, then has non-zero ``tension'' and a massless graviton localized near the horizon by a ``volcano''-like potential.

\item	Further, unlike almost all the papers in the literature wherein the Page curve computation is done for a CFT bath, in our model, the external bath is a non-CFT bath (thermal QCD). 

\item	Entanglement entropy contribution from Hartman-Maldacena (HM)-like surface which is responsible for increase of Einstein-Rosen bridge in time, also exhibits a Swiss-Cheese structure in Large-$N$ scenario (\ref{Swiss-Cheese-i}).

\item With the inclusion of ${\cal O}(R^4)$ terms in the action, our aforementioned ${\cal M}$-theory dual yields a hierarchy in the entanglement entropies of the HM-like and Island surface (IS) with respect to a large-$N$ exponential suppression factor, physically arising from the existence of massless graviton mode on the ETW-brane. This suppression further implies that the Page curve is unaffected by the inclusion of higher derivative - ${\cal O}(R^4)$ in particular - terms.

\item To tame the IR- and large-$N$ enhancement in the IS entanglement entropy per unit BH entropy to ensure the same is around 2 in \ref{EE-HM-WHD}, along with dimensional considerations of \ref{EE-HM-WHD}, a connection between the Planckian length and the non-extremality parameter (the horizon radius) is shown to arise in \ref{EE-HM-WHD}.

\end{itemize}

\item In subsections \ref{EE-HM-WHD} and \ref{EE-IS-WHD}, using the Ryu-Takayanagi formula (since we are not considering higher derivative terms in the gravitational action in this part), we calculated the entanglement entropies of HM-like and Islands Surfaces (IS) by computing the areas of the co-dimension two surfaces for the HM-like and island surfaces. The entanglement entropy for HM-like surface was calculated in (\ref{Page-curve-beta0}) using (\ref{SHM}), which shows linear time dependence. If one does not consider the island surface then (\ref{Page-curve-beta0}) implies that at late times (i.e., beyond the Page time) we have infinite amount of Hawking radiation. The entanglement entropy of the island surface (\ref{EE-IS-simp}) was calculated using (\ref{EE-IS-WHD-TERMS})). If we are below the Page time then entanglement entropy of HM-like surface dominates which shows linear time dependence and after the Page time the entanglement entropy of island surface, which is constant, dominates. Therefore combination of the two produces the Page curve of an eternal black hole in figure \ref{Page-Curve-Areas}. 

In ${\cal R}_{D5/\overline{D5}}=1$-units, using (\ref{Page-curve-beta0}), (\ref{EE-IS-simp}), (\ref{r_T-iii}) and $ r_h\sim e^{-\kappa_{r_h}N^{\frac{1}{3}}}$ \cite{Bulk-Viscosity-McGill-IIT-Roorkee}, the aforementioned is succinctly captured by:
\begin{eqnarray}
\label{SEE-beta0-Area-Summary-i}
& & S^{\rm HM} = \frac{A_{\rm HM}}{4G_N^{(11)}}\sim \frac{{\cal O}(1)\times10^{-4}M^2  N_f^6 g_s^{15/4} N^{34/15} {\bf e^{-6\kappa_{r_h}N^{1/3}}}}{ G_N^{(11)}}t_b,\ t_b\leq t_{\rm Page};\nonumber\\
& & {\cal S}^{IS} = {\cal S}^{IS}\left(\tilde{r}_T \equiv \frac{r_T}{r_h} = (1+\delta_2); r_h\right),\  t_b\geq t_{\rm Page},
\end{eqnarray}
where $\delta_2$ in the turning point $r_T$ is as given in (\ref{r_T-iii}).

\item In section \ref{Page-curve-HD}, we obtained the Page curve of an eternal black hole in the presence of ${\cal O}(R^4)$ terms in eleven dimensional supergravity action. For this purpose  use the prescription of \cite{Dong} as given in (\ref{HD-Entropy}) to calculate the entanglement entropies of HM-like and island surfaces with the inclusion of ${\cal O}(R^4)$ terms. Since in the working action (\ref{D=11_O(l_p^6)}) only $J_0$ term is considered (as it is the most dominant in the MQGP limit) therefore we need to calculate two type of terms appearing in (\ref{HD-Entropy}). These two type of terms are given in (\ref{Wald-J0-c-r}) and (\ref{second-der-J0}) where we have used $z=x e^{it}$. In subsection \ref{EE-HM-HD}, the entanglement entropy for HM-like surface is calculated and it turns out that for Einstein-Hilbert term Dong's prescription \cite{Dong} produces (\ref{Lag-wald-beta0}). For the $J_0$ term there are two types of terms mentioned earlier and these terms are given in (\ref{Lag-wald-HM}) and (\ref{Lag-anomaly-HM}). Now if we combine the Einstein-Hilbert and $J_0$ terms's contribution then we obtain the total entanglement entropy (\ref{Lag-total-HM}) in terms of the embedding function of HM-like surface $t(r)$ and its derivatives. We obtained the on-shell entanglement entropy (\ref{SEEHM-tb0}) for HM-like surface at ${\cal O}(\beta^0)$ by computing the solution to the EOM for the HM-like surface embedding $t(r)$ (\ref{t(r)}) and substituting back into (\ref{Lag-total-HM}). Interestingly (\ref{SEEHM-tb0}) also has linear time dependence similar to (\ref{Page-curve-beta0}).

In \ref{Swiss-Cheese-i}, with $c_1$ and $c_2$ being two parameters in the family of HM-like surface embeddings, curiously, $\frac{\delta S_{\rm EE}^{\rm HM, \beta^0}}{\delta |c_2|}>0,\ \frac{\delta S_{\rm EE}^{\rm HM, \beta^0}}{\delta |c_1|}<0$, which along with $|c_2|\sim e^{\kappa_{c_2}|c_1|^{\frac{1}{3}}},\ |c_1|\sim N$, provides a perfect "Swiss-Cheese'' (of the "single-big-divisor-single-small-divisor'' variety) structure to $S_{\rm EE}^{\rm HM, \beta^0}(|c_1|, |c_2|)$ (which essentially is a co-dimension-two volume) wherein $\log|c_2|$ plays the role of the "big divisor'' volume and $\log|c_1|$ plays the role of a solitary "small divisor'' volume, realizing what could be dubbed as a "Large N Scenario''(LNS). Alternatively, the entanglement entropy for the Hartman-Maldacena-like surface  can be viewed as a Swiss-Cheese-like open surface in the two-dimensional (in the IR) space of family of HM-like embeddings $\mathbb{R}_+^2\left(|c_1|, |c_2|\right)$ augmented by the entanglement entropy that coordinatizes  $\mathbb{R}_{\geq0}(S_{\rm EE}^{\rm HM, \beta^0})$. 

In subsection \ref{EE-IS-HD},  we found that entanglement entropy contributions from Einstein-Hilbert and $J_0$ terms as given in (\ref{Lag-beta0-IS}),(\ref{Lag-Wald-IS}) and (\ref{Lag-anomaly-IS}). Combining all these contributions we obtained the total entanglement entropy in (\ref{Lag-total-IS}) in terms of embedding function of island surface and its derivatives. We solved the embedding EOM for the island surface similar to HM-like surface and the solution is given in (\ref{x(r)-solution}). Now, using the solution we  obtained the entanglement entropy for island surface at the level of EH-GHY terms by substituting back in (\ref{Lag-total-IS}). This yields that the entanglement entropy will be of the form given in (\ref{on-shell-L-beta0}). Given that the turning point lies deep in the IR, the entanglement entropy for HM-like and island surfaces at ${\cal O}(\beta^0)$ came out to be proportional to $\beta$, an obvious but apparent contradiction. This was resolved by proposing a $\beta/l_p-r_h$ relation as discussed above (\ref{SEEISoverSBH}) and its explicit form as given in (\ref{scalings}). The on-shell entanglement entropy for island surface turns out to be (\ref{SEE-IS-simp}). Upon plotting the entanglement entropy contributions from HM-like (\ref{SEEHM-tb0}) and island surfaces (\ref{SEE-IS-simp}) we obtained the Page curve shown in figure \ref{Page-curve-plot}. 

The aforementioned is neatly summarized below. 
Using the HM-like embedding (\ref{t(r)}), and (\ref{SEE-HM-simp-beta0}) and (\ref{SEE-HM-tb0}), the on-shell Entanglement Entropy (EE) for HM-like surface is given by:
\begin{eqnarray}
\label{SEE-beta0-Wald-Summary-i}
& & \hskip -0.3in S_{\rm EE}^{\beta^0, {\rm HM}} \sim   {\bf e^{-\frac{3\kappa_{l_p}N^{\frac{1}{3}}}{2}}}M N^{13/10} N_f^{5/3}\left(1-\frac{4 t_{b_0}}{3 c_2}\right)\left(\frac{2}{3}\log \left(\frac{c_2}{c_1}\right) -n_{t_b} \log(N)\right)^{\frac{17}{6}},\ \kappa_{l_p}\equiv\frac{1}{{\cal O}(1)},\nonumber\\
& & \hskip -0.3in c_{1,2}<0, |c_2|\gg |c_1|; |c_2|\sim e^{\kappa_{c_2}|c_1|^{1/3}},\ \kappa_{c_1}\equiv{\cal O}(1)\nonumber\\
& & \hskip -0.3in t\leq t_{\rm Page},
\end{eqnarray}
and using the embedding (\ref{x(r)-solution}), and
(\ref{SEE-IS-simp}), (\ref{Page-time}), (\ref{SEEHMbeta-ii}) and (\ref{SEEISbeta-iii}) , the on-shell EE for Island Surface (IS) is given by: 
\begin {eqnarray}
& & \hskip -0.3in S_{\rm EE}^{\beta^0, {\rm IS}} \sim \frac{M^2 {\bf e^{-\frac{3\kappa_{l_p} N^{\frac{1}{3}}}{2}}}
 N_f^{4/3} g_s^{35/6} \log ^2(N)  \left| \log \left(r_h\right)\right|^{4/3} }{r_h^{11/2}}, \nonumber\\
 & & \hskip -0.3in t\geq t_{\rm Page}.
\end {eqnarray}
The above can be shown to be consistent with the RT computation summarized earlier \footnote{Compatibility of (\ref{SEE-beta0-Area-Summary-i}) and (\ref{SEE-beta0-Wald-Summary-i}) (using footnote 8) requires $\kappa_{r_h},\ \kappa_{l_p},\ \kappa_{c_2}$:\\ $M^2N_f^6g_s^{15/4}N^{34/15}e^{-6\kappa_{r_h}N^{1/3}}\sim M N^{13/10}N_f^{5/3}\left(\frac{2}{3}\biggl[\kappa_{c_2}N^{1/3} - \log |c_1|\biggr] -n_{t_b} \log(N)\right)^{\frac{17}{6}}e^{-\frac{3\kappa_{l_p}N^{1/3}}{2}}$\\ (for compatibility of $S^{\beta^0, {\rm HM}}_{\rm EE}$ obtained from computation of the area of the HM-like surface and from Wald entanglement entropy) and $11\kappa_{r_h} < 3\kappa_{l_p}$ (for $S^{\beta^0, {\rm IS}}_{\rm EE}$ to be well-defined in the large-$N$ limit).
}.

\item
Positivity of the Page time - worked out in (\ref{Page-time}) - is shown to provide an upper bound on the non-extremality parameter - the black-hole horizon radius $r_h$. 

\item We computed the ${\cal O}(\beta)$ contributions to the HM-like and island surfaces as given in (\ref{SEEHMbeta-ii}) and (\ref{SEEISbeta-ii}) and we  show in (\ref{S_EE_hierarchy}) that there is hierarchy between the ${\cal O}(\beta^0)$ and ${\cal O}(\beta)$ contributions. To be a bit more specific, as the turning point for the Hartman-Maldacena-like surface $r_*: r_*-r_h\sim \beta^{\frac{2}{3}}$, and that for Island surface $r_T: r_T-r_h\sim \beta^2$, one obtains the following exponential hierarchy:
$S_{\rm EE}^{\rm HM, \beta^0}:S_{\rm EE}^{\rm IS,\ \beta^0}:S_{\rm EE}^{\rm HM,\ \beta}:S_{\rm EE}^{\rm IS,\ \beta} \sim \gamma^{\frac{3}{2}}:\gamma^{\frac{3}{2}}:\gamma^2:\gamma^3,$ where $\gamma\equiv e^{-\kappa_{l_p}N^{\frac{1}{3}}}$. Therefore we can neglect the ${\cal O}(\beta)$ contributions to the entanglement entropy for both extremal surfaces. The exponentially suppressed $S_{\rm EE}^{\rm IS, \beta^0}$ and the suppression of $S_{\rm EE}^{\rm IS, \beta}$ relative to $S_{\rm EE}^{\rm HM, \beta}$ is argued in Section \ref{massless_graviton} to be due to the existence of a massless graviton.  It is also rather non-trivial to have a similar exponential suppression in $S_{\rm EE}^{\rm HM}$ so that the same could be matched with $S_{\rm EE}^{\rm IS}$ at the Page time, and a Page curve obtained. 

\item
Motivated by the requirement $\frac{S^{\beta^0,\ IS}_{\rm EE}}{S_{\rm BH}} \sim e^{-\kappa_{l_p}N^{1/3}}{\cal F}\left(M, N, N_f; r_h\right)\left[1 + \sum_{n=1}^\infty {\cal A}_n\left(\frac{a_3}{a_2 r_h}\right)^n\right]$ (as obtained in (\ref{SEEISoverSBH-D-dims})) to be less than or equal to 2 up to leading order in the dimensionless "$G_N^{(11)}/r_h^9$'' \cite{SEEISoverSBHapprox2}, which for us is $\frac{a_3}{a_2 r_h}$ ($a_{2,3}$ parametrizing the IS-surface embedding (\ref{x(r)-solution})), one needs to cure the large-$N$ and IR(via small $r_h$) enhancements in $\frac{S^{\beta^0,\ IS}_{\rm EE}}{S_{\rm BH}}$. Utilizing the estimate in \cite{Bulk-Viscosity-McGill-IIT-Roorkee} of the $r=r_0\sim r_h: N_{\rm eff}(r_0=0)$, and in particular the exponential $N$-suppression therein we therefore propose 
$\beta \sim \left(g_s^{\frac{4}{3}}\alpha'^2e^{-\kappa_{l_p}N^{1/3}}\frac{r_h}{{\cal R}_{D5/\overline{D5}}}\right)^{3/2}$. This further requires (see Fig. \ref{SEEISoverSBH-vs-SBH}) a lower bound on the non-extremality parameter - the black-hole horizon. With the inclusion of the ${\cal O}(R^4)$ terms in ${\cal M}$ theory, the fact the turning point associated with the Hartman-Maldacena-like surface is in the deep IR, also require the aforementioned $\beta$- or $l_p-r_h$ relation.

\end{itemize}
 Some open questions related to our work are:
 \begin{itemize}
 \item How the holographic complexity will be modified in the presence of ${\cal O}(R^4)$ terms in the gravitational action \cite{HC}?
 
 \item What will be the effect of higher derivative (${\cal O}(R^4)$) terms on the reflected entropy \cite{RE}?
 \end{itemize}
We will return to these issues in our future work.

\section*{Acknowledgements}

GY is supported by a Senior Research Fellowship (SRF) from the Council of Scientific and Industrial Research, Govt. of India. AM is partly supported by a Core Research Grant number SER-1829-PHY from the Science and Engineering Research Board, Govt. of India. One of us (GY) would like to thank Arpan Bhattacharyya, Xi Dong and Juan Maldacena for helpful clarifications. Some of the results were presented by one of us (GY) at the workshop, {\it "Gravity: Current challenges in black hole physics and cosmology"}, organised by Yukawa Institute for Theoretical Physics (YITP), Kyoto University, Kyoto, Japan.

\appendix

\section{HM-Like/IS Analytics/Numerics}
\label{HM+IS-analyt+num}
\setcounter{equation}{0}\seceqaa

In this appendix, we discuss (i) the angular integrations which have been used in the paper and have written the expression for the area of Hartman-Maldacena-like surface, and (ii) how to obtain an estimate on the turning point in the context of Island Surface  entanglement entropy.

\subsection{HM-Like Surface Area}
\label{a-1}
The angular integrations, in promoting the delocalized results around $(\theta_1,\theta_2)=\left(\frac{\alpha_{\theta_1}}{N^{\frac{1}{5}}},\frac{\alpha_{\theta_2}}{N^{\frac{3}{10}}}\right)$, to global results, disregarding all contributions of 
${\cal O}\left(\frac{{\cal O}(1)}{N^\alpha}\right), \alpha>1$, have been performed as under. Promoting $(x, y, z)\rightarrow (\phi_1, \phi_2, \psi)$ via \cite{MQGP}:
\begin{eqnarray}
\label{local-to-global}
& &
dx = \sqrt{h_2}\left(g_sN\right)^{\frac{1}{4}}\Biggl[1 + {\cal O}\left(\frac{g_sM^2}{N}\right) + {\cal O}\left(\frac{(g_sM^2)(g_sN_f)}{N}\right)\Biggr]\sin\theta_1d\phi_1, \nonumber\\
& & dy = \sqrt{h_4}\left(g_sN\right)^{\frac{1}{4}}\Biggl[1 + {\cal O}\left(\frac{g_sM^2}{N}\right) + {\cal O}\left( \frac{(g_sM^2)(g_sN_f)}{N}\right)\Biggr]\sin\theta_2d\phi_2, \nonumber\\
& & dz = \sqrt{h_1}\left(g_sN\right)^{\frac{1}{4}}\Biggl[1 + {\cal O}\left(\frac{g_sM^2}{N}\right) + {\cal O}\left( \frac{(g_sM^2)(g_sN_f)}{N}\right)\Biggr]d\psi,
\end{eqnarray}
where $h_1 = \frac{1}{9} + {\cal O}\left(\frac{g_sM^2}{N}\right),\ h_2 = \frac{1}{6} + {\cal O}\left(\frac{g_sM^2}{N}\right),\ h_4 = h_2 + \frac{4a^2}{r^2}$ \cite{metrics}, \cite{MQGP}, 
\begin{eqnarray}
\label{angular-integrations-2}
& & \frac{1}{\alpha_{\theta_1}^3\alpha_{\theta_2}^2}\rightarrow\lim_{\epsilon_{1,2}\rightarrow0}
\int_{\epsilon_2}^{\pi-\epsilon_2}\sqrt{g_sN}d\theta_2\sin\theta_2\int_{\epsilon_1}^{\pi-\epsilon_1}d\theta_1\sin\theta_1\frac{1}{\left(N^{\frac{1}{5}}\sin{\theta_1}\right)^3\left(N^{\frac{3}{10}}\sin{\theta_2}\right)^2}  \sim \lim_{\epsilon_{1,2}\rightarrow0}\frac{|\log \epsilon_2|}{N^{\frac{7}{10}}\epsilon_1}.
\nonumber\\
\end{eqnarray}
The principal(${\cal P}$) value of (\ref{angular-integrations-2}) is obtained by approximating $|\log \epsilon_2|$ by $\log \mathbb{P} + |\log \epsilon_2|, \mathbb{P}\in\mathbb{Z}^+$ and requiring $\log \mathbb{P} + \log \epsilon_2 = - \epsilon_1$, or
$\epsilon_2=\frac{e^{-\epsilon_1}}{\sqrt{\mathbb{P}}}, \mathbb{P}>1$.
Similarly,
\begin{eqnarray}
\label{angular-integrations-1}
& &  \frac{1}{\alpha_{\theta_1}\alpha_{\theta_2}^6}\rightarrow\lim_{\epsilon_{1,2}\rightarrow0}
\int_{\epsilon_2}^{\pi-\epsilon_2}\sqrt{g_sN}d\theta_2\sin\theta_2\int_{\epsilon_1}^{\pi-\epsilon_1}d\theta_1\sin\theta_1\frac{1}{\left(N^{\frac{1}{5}}\sin{\theta_1}\right)\left(N^{\frac{3}{10}}\sin{\theta_2}\right)^6} \nonumber\\
& & \stackrel{{\cal P}}{\rightarrow} \frac{\pi  (540 \log (2)-107) \sqrt{g_s}}{720 N^{3/2}}.
\nonumber\\
\end{eqnarray}
Assuming $r_*\in$IR, approximating $\log r\approx\log r_h$ for $r = \upsilon r_h\forall r\in$IR, $\upsilon={\cal O}(1)$, 
one obtains the following expression after radial integration of equation (\ref{AHM_i}):
{
\begin{eqnarray}
\label{AHM_ii}
& &  A_{\rm HM} \sim \Biggl[E M^2 \sqrt[10]{N} N_f^6 g_s^{17/4} \log ^2\left(r_h\right) \left(\log (N)-3 \log \left(r_h\right)\right){}^4 \nonumber\\
& &\hskip 0.5in \times \Biggl(3 r_*^5 \left(\log (N)-9 \log \left(r_h\right)\right){}^2 -10 r_*^3 r_h^2 \log (N) \left(\log (N)-9 \log \left(r_h\right)\right)\nonumber\\
& & \hskip 0.5in +15 r_* r_h^4 \log ^2(N)-r_h^5 \left(36 \log (N) \log \left(r_h\right)+243 \log ^2\left(r_h\right)+8 \log ^2(N)\right)\Biggr) \Biggr] .\nonumber\\
   & & 
\end{eqnarray}
}

\subsection{IS Turning Point}
\label{a-2}
In the context of IS entanglement entropy (\ref{EE-IS-simp}), to get an estimate of the turning point $\tilde{r}_T$, one notes that: $\frac{H(r_T)}{\sigma(r_T)}={\cal C}^2$, up to LO in $N$, that obtains:
{\footnotesize
\begin{eqnarray}
\label{r_T-i}
& & 1458{g_s}^{5/4} M^2 N^{2/5} {N_f}^6 r_T^4 \left(r_T^4-{r_h}^4\right) \log ^7(r_T)
   \left(7\log N  r_T^2-3 a^2 \log N \right)-6561{g_s}^{5/4} M^2 N^{2/5} {N_f}^6 r_T^6 \left(r_T^4-{r_h}^4\right)
   \log ^8(r_T)  \nonumber\\
   & &  + {\cal O}\left(\frac{1}{(\log r_h)^6}\right) = \tilde{\cal C}^2.
\end{eqnarray}
}
Writing $\tilde{r}_T = 1 + \delta_2$, upon substitution into (\ref{r_T-i}), and assuming $0<\delta_2\ll1$ and thereby approximating $\log r\approx \log r_h$, yields the following:
{\footnotesize
\begin{eqnarray} 
\label{r_T-ii}
& & 1458 \delta_2  {r_h}^{10} \log ^7({r_h}) (4 (47 \delta_2 +6) \log N -9 (15
   \delta_2 +2) \log ({r_h}))-5832 \sqrt{3} \delta_2  (11 \delta_2 +2)
   \epsilon  \log N  {r_h}^{10} \log ^7({r_h})=\frac{M^2 N^{2/5} {N_f}^6
   \tilde{\cal C}^2}{{g_s}^{5/4}}.\nonumber\\
   & & 
\end{eqnarray}
}
Assuming again $|\log r_h|\gg \log N$, the relevant solution to (\ref{r_T-ii}) is:
\begin{eqnarray}
\label{r_T-iii}
& & \delta_2\approx \frac{M^2 N^{2/5} {N_f}^6 \tilde{\cal C}^2}{26244 {g_s}^{5/4} {r_h}^{10} \log
   ^8({r_h})}-\frac{2}{15}.
\end{eqnarray}
Guided by the estimate of $r=r_0\sim r_h: N_{\rm eff}(r_0)=0$ obtained in \cite{Bulk-Viscosity-McGill-IIT-Roorkee}, writing $r_h\sim e^{-\kappa_{r_h}N^{\frac{1}{3}}}$, and assuming $\tilde{\cal C}^2=e^{-\kappa_{\cal C}N^{\frac{1}{3}}}$. Numerically, for $M=N_f=3, g_s=0.1, N=10^{3.3}$ and assuming $\kappa_{\cal C}=1.37, \kappa_{r_h}=0.1, \tilde{\cal C}=3\times10^{-8}$ one obtains $\delta_2=5.6\times10^{-4}$.

\section{Hartman-Maldacena-like Surface Miscellania}
\label{HM}
\setcounter{equation}{0}\seceqbb
In this appendix we have listed various $r$ dependent functions appearing in the entanglement entropy expression for the Hartman-Maldacena-like surface at ${\cal O}(\beta^0)$ and ${\cal O}(\beta)$. Additionally we have also worked out the equation of motion for the embedding function of the Hartman-Maldacena-like surface.
\begin{itemize}
\item
The $r$ dependent functions appearing in equation (\ref{Lag-wald-beta0}) are,
\begin{eqnarray}
\label{alpha-sigma-lambda}
& & \alpha(r)= \kappa_{\alpha}\frac{ r^2 \left(N_f g_s (2 \log (N)-6 \log (r))\right){}^{2/3}}{ \sqrt{N} g_s^{3/2}},\nonumber\\
& & \sigma(r)=\kappa_{\sigma}\frac{ N g_s-3 M^2 N_f g_s^3 \log (r) (\log (N)-6 \log (r))}{ \left(r^4-{r_h}^4\right)}, \nonumber\\
& & \lambda (r)=\kappa_{\lambda}\Biggl(-\frac{ M N^{17/10} N_f^{4/3} g_s^{5/2} \left(r^2-{r_h}^2\right) \log (r) (\log (N)-9 \log (r)) \sqrt[3]{\log (N)-3 \log (r)}}{ r^4 \alpha _{\theta _1}^3 \alpha_{\theta _2}^2}\Biggr) \nonumber\\
& & \sim -\frac{M N^{21/20} N_f^{4/3} g_s^{13/4} \left(r^2-r_h^2\right) \kappa _{\lambda } \log (r) (\log (N)-9 \log (r)) \sqrt[3]{\log (N)-3 \log (r)}}{3 r^4},\nonumber\\
\end{eqnarray}
where $\kappa_{\alpha},\kappa_{\sigma}$ and $\kappa_{\lambda}$ are numerical pre-factors.

\item
In Wald entanglement entropy term (\ref{Lag-wald-HM}) for HM-like surface, we have the following $r$ dependent functions:
\begin{eqnarray}
\label{lambda1-2}
& & 
\hskip -.5in \lambda_1(r)= \kappa_{\lambda_1}\left(\frac{ M^7 N^{7/10} N_f^{7/3} g_s^7 \log ^4(r) (\log (N)-12 \log (r))^3 (\log (N)-9 \log (r))}{ r^2 \alpha _{\theta _1}^3 \alpha
   _{\theta _2}^2 (\log (N)-3 \log (r))^{5/3}}\right) \nonumber\\
   & & \sim \frac{M^7 \sqrt[20]{N} N_f^{7/3} g_s^{31/4} \kappa _{\lambda _1} \log ^4(r) (\log (N)-12 \log (r))^3 (\log (N)-9 \log (r))}{3 r^2 (\log (N)-3 \log (r))^{5/3}}
   ,\nonumber\\
   & & 
  \hskip -.5in \lambda_2(r)=\kappa_{\lambda_2}\left(\frac{ M^3 N^{7/10} \sqrt[3]{N_f} g_s^3 \log ^2(r) (\log (N)-12 \log (r)) (\log (N)-9 \log (r))}{ r^2 \alpha _{\theta _1}^3 \alpha _{\theta _2}^2 (\log(N)-3 \log (r))^{5/3}}\right) \nonumber\\
  & & \sim \frac{M^3 \sqrt[20]{N} \sqrt[3]{N_f} g_s^{15/4} \kappa _{\lambda _2} \log ^2(r) (\log (N)-12 \log (r)) (\log (N)-9 \log (r))}{3 r^2 (\log (N)-3 \log (r))^{5/3}}.
\end{eqnarray}
$\kappa_{\lambda_1}$ and $\kappa_{\lambda_2}$ in the above equation are numerical pre-factors.

\item
In the anomaly term (\ref{Lag-anomaly-HM}) following are the $r$ dependent functions: 
\begin{eqnarray}
\label{L-terms-HM}
& &   {\cal L}_1= {\cal L}_2=\frac{\sqrt{\alpha (r) \left(\sigma (r)-\left(1-\frac{r_h^4}{r^4}\right) t'(r)^2\right)} }{t'(r)^2 \left(\sigma (r)-\left(1-\frac{r_h^4}{r^4}\right) t'(r)^2\right)^4} \nonumber\\
& & \times \left(\alpha '(r) \left(\sigma (r)-\left(1-\frac{r_h^4}{r^4}\right) t'(r)^2\right)+\alpha (r) \left(\sigma '(r)-t'(r) \left(\frac{4 r_h^4}{r^5} t'(r)+2
   \left(1-\frac{r_h^4}{r^4}\right) t''(r)\right)\right)\right)^2; \nonumber\\
   & &   {\cal L}_3= {\cal L}_4=\frac{\sqrt{\alpha (r) \left(\sigma (r)-\left(1-\frac{r_h^4}{r^4}\right) t'(r)^2\right)}}{t'(r)^2}.
\end{eqnarray}

\item
Use has been made of the following for the computation of first and second term in entanglement entropy computation in the presence of higher derivative terms:
{\footnotesize
\begin{eqnarray}
\label{dJ0overdR+d2J0overdR2}
& &
\frac{\delta J_0}{\delta R_{M_1 N_1 P_1 Q_1}}=G^{M_1 M_1}G^{N_1 N_1}G^{P_1 P_1}G^{Q_1 Q_1}\left( R_{P N_1 P_1 Q}  R^Q_{\ \ RSQ_1}  +\frac{1}{2}R_{PQP_1 Q_1}  R^Q_{\ \ RSN_1}\right)R_{M_1}^{\ \ RSP} \nonumber\\
& &
+\left(\delta^{Q_1}_Q R^{HN_1 P_1 K}+\frac{1}{2}\delta^{N_1}_Q R^{HKP_1 Q_1}\right)R_{H}^{\ \ RSM_1} R^Q_{\ \ RSK} \nonumber\\
& & + G^{N_1 N_1}G^{P_1 P_1}G^{Q_1 Q_1} \left( R_{Q_1 MNQ}R^{M_1 MNK}+\frac{1}{2}R_{Q_1 QMN} R^{M_1 KMN}\right)R^Q_{\ \ N_1 P_1 K}
 \nonumber\\
& & +G^{M_1 M_1}\left( R_{PMNM_1} R^{HMN Q_1}+\frac{1}{2}  R_{PM_1 MN} R^{H Q_1 MN}\right)R_H^{\ \ N_1 P_1 P}
\end{eqnarray}
}
and 
\begin{eqnarray}
& &
\frac{\delta^2 J_0}{\delta R_{M_2 N_2 P_2 Q_2} \delta R_{M_1 N_1 P_1 Q_1}}=A_1+A_2+A_3+A_4,
\end{eqnarray}
where
{\footnotesize
\begin{eqnarray}
& &
A_1=G^{M_1 M_1}G^{N_1 N_1}G^{P_1 P_1}G^{Q_1 Q_1}\Biggl[\delta^{M_2}_{M_1} G^{N_2 N_2}G^{P_2 P_2}G^{Q_2 Q_2}\left(R_{Q_2 N_1 P_1 Q}R^Q_{\ \ N_2 P_2 Q_1}+\frac{1}{2}R_{Q_2 Q P_1 Q_1}R^Q_{\ \ N_2 P_2 N_1}\right)\nonumber\\
& & + \left(\delta^{M_2}_P\delta^{N_2}_{N_1}\delta^{P_2}_{P_1}\delta^{Q_2}_{Q} R^Q_{\ \ RSQ_1}R_{M_1}^{\ \ RSP}+\frac{1}{2}\delta^{M_2}_P\delta^{N_2}_{Q}\delta^{P_2}_{P_1}\delta^{Q_2}_{Q_1} R^Q_{\ \ RSN_1}R_{M_1}^{\ \ RSP} \right)\nonumber\\
& & + \left(G^{M_2M_2}\delta^{N_2}_{R}\delta^{P_2}_{S}\delta^{Q_2}_{Q_1} R_{M_1}^{\ \ RSP}R_{PN_1P_1M_2}+\frac{1}{2}G^{M_2M_2}\delta^{N_2}_{R}\delta^{P_2}_{S}\delta^{Q_2}_{N_1} R_{M_1}^{\ \ RSP}R_{PM_2P_1Q_1}\right)\Biggr] \nonumber\\
& & \equiv G^{M_1 M_1}G^{N_1 N_1}G^{P_1 P_1}G^{Q_1 Q_1}\Biggl[\delta^{M_2}_{M_1} G^{N_2 N_2}G^{P_2 P_2}G^{Q_2 Q_2}\left(R_{Q_2 N_1 P_1 Q}R^Q_{\ \ N_2 P_2 Q_1}+\frac{1}{2}R_{Q_2 Q P_1 Q_1}R^Q_{\ \ N_2 P_2 N_1}\right)\nonumber\\
& & + \left(\delta^{N_2}_{N_1}\delta^{P_2}_{P_1} R^{Q_2}_{\ \ RSQ_1}R_{M_1}^{\ \ RSM_2}+\frac{1}{2}\delta^{P_2}_{P_1}\delta^{Q_2}_{Q_1} R^{N_2}_{\ \ RSN_1}R_{M_1}^{\ \ RSM_2} \right)\nonumber\\
& & + \left(G^{M_2M_2}\delta^{Q_2}_{Q_1} R_{M_1}^{\ \ N_2 P_2 P}R_{PN_1P_1M_2}+\frac{1}{2}G^{M_2M_2}\delta^{Q_2}_{N_1} R_{M_1}^{\ \ N_2 P_2 P}R_{PM_2P_1Q_1}\right)\Biggr],
\nonumber\\
& & A_2= \Biggl[\left(G^{M_2M_2}G^{Q_2Q_2}G^{N_2N_1}G^{P_2P_1}R^{Q_1}_{\ \ RS Q_2}+\frac{1}{2}G^{M_2M_2}G^{N_2N_2}G^{P_2P_1}G^{Q_2Q_1}R^{N_1}_{\ \ RS N_2} \right)R_{M_2}^{\ \ RS M_1} 
\nonumber\\
& & +\left(G^{Q_2M_1}G^{N_2N_2}G^{P_2P_2}R^{M_2 N_1 P_1 K}+\frac{1}{2}G^{Q_2M_1}G^{N_2N_2}G^{P_2P_2}R^{M_2 K P_1 Q_1} \right)R^{N_1}_{\ \ N_2 P_2 K}\nonumber\\
& & +\left(G^{M_2Q_1}R^{H N_1 P_1 Q_2}+\frac{1}{2}G^{M_2N_1}R^{H Q_2 P_1 Q_1} \right)R_{H}^{\ \ N_2 P_2 M_1}\Biggr],\nonumber\\
& & 
A_3= \Biggl[G^{N_1 N_1}G^{P_1 P_1}G^{Q_1 Q_1}\Biggl[\Biggl(\delta^{M_2}_{Q_1}\delta^{Q_2}_Q R^{M_1 N_2 P_2 K}+G^{M_2 M_1}G^{N_2 N_2}G^{P_2 P_2}G^{Q_2 K} R_{Q_1 N_2 P_2 Q} \nonumber\\
& & + \frac{1}{2}\delta^{M_2}_{Q_1}\delta^{N_2}_Q R^{M_1 K P_2 Q_2}+\frac{1}{2}G^{M_2 M_1}G^{N_2 K}G^{P_2 P_2}G^{Q_2 Q_2} R_{Q_1 Q P_2 Q_2}\Biggr)R^Q_{\ \ N_1 P_1 K} \nonumber\\
& & + G^{M_2 M_2} \delta^{N_2}_{N_1}\delta^{P_2}_{P_1}\left(R_{Q_1 MN M_2}R^{M_1 MN Q_2}+\frac{1}{2}R_{Q_1 M_2 MN}R^{M_1 Q_2 MN}\right)
\Biggr] \Biggr], \nonumber\\
& & A_4=G^{M_1 M_1}\Biggl[\Biggl(\delta^{M_2}_P \delta^{Q_2}_{M_1}R^{H N_2 P_2 Q_1}+G^{M_2 H}G^{N_2 N_2}G^{P_2 P_2}G^{Q_2 Q_1}R_{P N_2 P_2 M_1}\nonumber\\
& & + \frac{1}{2} \delta^{M_2}_P \delta^{N_2}_{M_1}R^{HQ_1 P_2 Q_2}+\frac{1}{2}G^{M_2 H}G^{N_2 Q_1}G^{P_2 P_2}G^{Q_2 Q_2}R_{P M_1 P_2 Q_2}\Biggr)R_H^{\ \ N_1 P_1 P} \nonumber\\
& & + G^{N_2 N_1}G^{P_2 P_1}G^{Q_2 Q_2}\left(R_{Q_2 MN M_1}R^{M_2 MN Q_1}+\frac{1}{2}R_{Q_2  M_1 MN}R^{M_2 Q_1 MN} \right)\Biggr].
\end{eqnarray}
}
 
 \item
The $r$-dependent functions appearing in equation (\ref{Lag-anomaly-HM}) are given below:
\begin{eqnarray}
\label{Z-W-U-V}
& & Z(r)=-\kappa_Z \frac{  M N^{37/10} g_s^{35/6} \log (r) (\log (N)-9 \log (r))}{ r^{14} \alpha _{\theta _1}^3 \alpha _{\theta _2}^2 N_f^{4/3} (\log (N)-3 \log (r))^{7/3}} \nonumber\\
& & \sim -\frac{\pi ^{23/6} M N^{61/20} g_s^{79/12} \kappa _Z \log (r) (\log (N)-9 \log (r))}{r^{14} N_f^{4/3} (\log (N)-3 \log (r))^{7/3}}
, \nonumber\\
& &  W(r)=-\kappa_W \frac{ M N^{37/10} g_s^{35/6} \log (r) (\log (N)-9 \log (r))}{ r^{14} \alpha _{\theta _1}^3 \alpha _{\theta _2}^2 N_f^{4/3} (\log (N)-3 \log (r))^{7/3}} \nonumber\\
& & \sim -\frac{M N^{61/20} g_s^{79/12} \kappa _W \log (r) (\log (N)-9 \log (r))}{r^{14} N_f^{4/3} (\log (N)-3 \log (r))^{7/3}}
, \nonumber\\
& & U(r)=-\kappa_U \frac{ M^2 N N_f g_s^3 \log (r) (\log (N)-45 \log (r))}{r^4 \alpha _{\theta _1} \alpha _{\theta _2}^6 (\log (N)-3 \log (r))^3} \nonumber\\
& & \sim -\frac{M^2 N_f g_s^{15/4} \kappa _U \log (r) (\log (N)-45 \log (r))}{\sqrt[4]{N} r^4 (\log (N)-3 \log (r))^3}
, \nonumber\\
& & V(r)=\kappa_V\frac{ M^2 N N_f g_s^3}{ r^4 \alpha _{\theta _1} \alpha _{\theta _2}^6 \log ^2(r)}  \sim \frac{M^2 N_f g_s^{15/4} \kappa _V}{\sqrt[4]{N} r^4 \log ^2(r)},
\end{eqnarray}
where $\kappa_Z,\kappa_W,\kappa_U$ and $\kappa_V$ are the numerical factors including $\left(\frac{8}{q_\alpha +1}\right)$.

\item
To obtain the EOM for $t(r)$ we need to calculate the following derivatives from equations (\ref{Lag-wald-beta0}), (\ref{Lag-wald-HM}), (\ref{Lag-anomaly-HM}) and (\ref{Lag-total-HM}):
{\footnotesize
\begin{eqnarray}
\label{DLag-tprime}
& &\frac{\delta {\cal L}^{\rm HM}_{\rm Total}}{\delta t'(r)} =\beta  \Biggl(\frac{\left(r-r_h\right){}^{3/2} p_4^{\beta }\left(r_h\right) \left({{\cal A}_1} t'(r)+{{\cal A}_2} t''(r)\right)}{N^{7/10} t'(r)^2}+\frac{p_5^{\beta }\left(r_h\right) \left({{\cal A}_3}
   t'(r)+{{\cal A}_4} t''(r)\right)}{N^{7/10} t'(r)^2}+\frac{\left(r-r_h\right){}^{13/2} p_7^{\beta }\left(r_h\right)}{N^{57/10} t'(r)^4}\nonumber\\& &
 \hskip 0.6in  +\frac{ p_6^{\beta
   }\left(r_h\right)}{\left(r-r_h\right)N^{7/10} t'(r)} +\frac{\left(r-r_h\right){}^{3/2} p_1^{\beta }\left(r_h\right)}{N^{5/4} t'(r)}+\frac{N^{3/10} p_8^{\beta }\left(r_h\right)}{\sqrt{r-r_h} t'(r)^3}+\frac{N^{3/10}
   p_9^{\beta }\left(r_h\right)}{\sqrt{r-r_h} t'(r)^3}+\frac{p_2^{\beta }\left(r_h\right)}{\sqrt{r-r_h} t'(r)^3}\nonumber\\& & \hskip 0.6in
   +\frac{p_3^{\beta }\left(r_h\right)}{\sqrt{r-r_h} t'(r)^3}
   +N^{3/10}
   \left(r-r_h\right){}^{3/2} {Y_5}\left(r_h\right) t'(r)\Biggr)+N^{3/10} \left(r-r_h\right){}^{5/2} p_1\left(r_h\right) t'(r),
\end{eqnarray}
}
and 
\begin{eqnarray}
\label{DLag-tprimeprime}
& &\frac{\delta {\cal L}^{\rm HM}_{\rm Total}}{\delta t''(r)} =\frac{\beta  {\cal F}^{\beta}({r_h}) (r-{r_h})^{5/2}}{N^{7/10} t'(r)},
\end{eqnarray}
where 
{\footnotesize
\begin{eqnarray}
\label{fi[rh]-HM}
& &
p_1(r_h)=\kappa_{p_1}\left(\frac{M N_f^{5/3} g_s^{7/3} \sqrt{\kappa _{\alpha }} \kappa _{\lambda } \log \left(r_h\right) \left(\log (N)-9 \log \left(r_h\right)\right) \left(\log (N)-3 \log\left(r_h\right)\right){}^{5/6}}{r_h^{3/2} \sqrt{\kappa _{\sigma }}}\right),\nonumber\\
   & & p_1^\beta(r_h)=-\kappa_{p_1^\beta}\left(\frac{M^2 \sqrt[4]{N} x_R^2 N_f^{4/3} g_s^{17/6} \sqrt{\kappa _{\alpha }} \kappa _U \log \left(r_h\right) \left(\log (N)-45 \log \left(r_h\right)\right)}{r_h^{5/2} \sqrt{\kappa _{\sigma }} \left(\log
   (N)-3 \log \left(r_h\right)\right){}^{5/2}}\right),\nonumber\\
   & & p_2^\beta(r_h)= \kappa_{p_2^\beta} \left(\frac{M^2 x_R^2 N_f^{4/3} g_s^{23/6} \kappa _U \log \left(r_h\right) \sqrt{\kappa _{\alpha } \kappa _{\sigma }} \left(\log (N)-45 \log \left(r_h\right)\right)}{r_h^{9/2} \left(\log (N)-3 \log
   \left(r_h\right)\right){}^{5/2}}\right),\nonumber\\
   & & p_3^\beta(r_h)=-\kappa_{p_3^\beta}\Biggl(\frac{M^2 x_R^2 (540 \log (2)-107) N_f^{4/3} g_s^{23/6} \sqrt{\kappa _{\alpha } \kappa _{\sigma }}}{r_h^{9/2} \log ^2\left(r_h\right) \left(\log (N)-3 \log\left(r_h\right)\right){}^{5/2}} \nonumber\\
   & & \times \left(28 \kappa _V \left(\log (N)-3 \log \left(r_h\right)\right){}^3-27 \kappa _U \log ^3\left(r_h\right) \left(\log (N)-45 \log \left(r_h\right)\right)\right)\Biggr),\nonumber\\
   & & p_4^\beta(r_h)=\kappa_{p_4^\beta}F(g_s,r_h,N_f,M),\nonumber\\
   & & p_5^\beta(r_h)=\kappa_{p_5^\beta}F(g_s,r_h,N_f,M),\nonumber\\
   & & p_6^\beta(r_h)=-\kappa_{p_6^\beta}\left(\frac{M x_R^4 \sqrt[3]{N_f} g_s^2 \kappa _W \log \left(r_h\right) \left(\kappa _{\alpha } \kappa _{\sigma }\right){}^{5/2} \left(\log (N)-9 \log \left(r_h\right)\right) \sqrt[6]{\log (N)-3 \log
   \left(r_h\right)}}{ r_h^{5/2} \kappa _{\sigma }^5}\right),\nonumber\\
   & & p_7^\beta(r_h)=-\kappa_{p_7^\beta}\left(\frac{M \sqrt[3]{N_f} r_h^{25/2} \kappa _Z \log \left(r_h\right) \left(\kappa _{\alpha } \kappa _{\sigma }\right){}^{5/2} \left(\log (N)-9 \log \left(r_h\right)\right) \sqrt[6]{\log (N)-3 \log
   \left(r_h\right)}}{g_s^3 \kappa _{\sigma }^{10}}\right),\nonumber\\
   & & p_8^\beta(r_h)=\kappa_{p_8^\beta}\left(\frac{M x_R^4 \sqrt[3]{N_f} g_s^3 \kappa _W \log \left(r_h\right) \left(\kappa _{\alpha } \kappa _{\sigma }\right){}^{5/2} \left(\log (N)-9 \log \left(r_h\right)\right) \sqrt[6]{\log (N)-3 \log
   \left(r_h\right)}}{r_h^{9/2} \kappa _{\sigma }^4}\right),\nonumber\\
   & & p_9^\beta(r_h)=\kappa_{p_9^\beta}\left(\frac{M \sqrt[3]{N_f} g_s^3 \kappa _Z \log \left(r_h\right) \left(\kappa _{\alpha } \kappa _{\sigma }\right){}^{5/2} \left(\log (N)-9 \log \left(r_h\right)\right) \sqrt[6]{\log (N)-3 \log
   \left(r_h\right)}}{r_h^{9/2} \kappa _{\sigma }^4}\right),\nonumber\\
   & & {\cal F}^\beta=-\kappa_{{\cal F}^\beta}\left(\frac{M \sqrt[3]{N_f} g_s^2 \kappa _{\alpha }^{5/2} \log \left(r_h\right) \left(\log (N)-9 \log \left(r_h\right)\right) \sqrt[6]{\log (N)-3 \log \left(r_h\right)} \left(x_R^4 \kappa _W+2 \kappa
   _Z\right)}{r_h^{5/2} \kappa _{\sigma }^{5/2}}\right),
\end{eqnarray}
}
where $\kappa_{p_1},\kappa_{p_{i=1,..,9}^\beta}$ and $\kappa_{{\cal F}^\beta}$ are the numerical factors. From equation (\ref{Lag-total-HM}) and using equations (\ref{DLag-tprime}) and (\ref{DLag-tprimeprime}), the EOM corresponding to the embedding $t(r)$, given by $\frac{d \left(\frac{\delta {\cal L}}{\delta t'(r)}\right)}{dr} = \frac{d^2\left(\frac{\delta{\cal L}}{\delta t''(r)}\right)}{dr^2}$ turns out to be:
{\scriptsize
\begin{eqnarray}
\label{EOM-HM}
& &\hskip -0.3in \beta  \left(-\frac{N^{3/10} p_8^{\beta }\left(r_h\right)}{2 (r-r_h)^{3/2} t'(r)^3}-\frac{N^{3/10} p_9^{\beta }\left(r_h\right)}{2 (r-r_h)^{3/2} t'(r)^3}-\frac{3 N^{3/10} t''(r) p_8^{\beta
   }\left(r_h\right)}{\sqrt{r-r_h} t'(r)^4}-\frac{3 N^{3/10} t''(r) p_9^{\beta }\left(r_h\right)}{\sqrt{r-r_h} t'(r)^4}\right)+N^{3/10} (r-r_h)^{5/2} p_1\left(r_h\right)
   t''(r) \nonumber\\
 & & \hskip -0.3in  +\frac{5}{2} N^{3/10} (r-r_h)^{3/2} p_1\left(r_h\right) t'(r)=-\frac{\beta  M \sqrt[3]{N_f} g_s^2 \sqrt{r-r_h} \kappa _{\alpha }^{5/2} \log \left(r_h\right) \left(\log (N)-9 \log
   \left(r_h\right)\right) \sqrt[6]{\log (N)-3 \log \left(r_h\right)} \left(x_R^4 \kappa _W+2 \kappa _Z\right)}{4 N^{7/10} r_h^{5/2} \kappa _{\sigma }^{5/2} t'(r)^3}\nonumber\\
   & & \times \left(8 \left(r-r_h\right){}^2 t''(r)^2-4 \left(r-r_h\right) t'(r) \left(\left(r-r_h\right)
   t^{(3)}(r)+5 t''(r)\right)+15 t'(r)^2\right).
\end{eqnarray}
} 
\end{itemize}

\section{Island Surface Miscellania}
\label{ISM}
\setcounter{equation}{0}\seceqcc
In this appendix we are listing the various functions appearing in the entanglement entropy of the island surface at ${\cal O}(\beta)$. We have also computed the derivatives of the Lagrangian of the island surface here which have been used in obtaining the equation of motion of the embedding function of the island surface.

\begin{itemize}
\item
$\lambda_{3,4}(r)$ appearing in the Wald entanglement entropy term (\ref{Lag-Wald-IS}) of the island surface are given below: 
\begin{eqnarray}
\label{lambda3-4}
& & 
\hskip -.5in \lambda_3(r)= \kappa_{\lambda_3}\frac{ M^7 N^{7/10} N_f^{7/3} g_s^7 r_h^{14} \log (N) \log ^7(r) (5 \log (N)-12 \log (r))^3}{ r^{16} \alpha _{\theta _1}^3 \alpha
   _{\theta _2}^2 (\log (N)-3 \log (r))^{14/3}} \nonumber\\
   & & \sim \frac{M^7 \sqrt[20]{N} \log (2) (\log (64)-1) N_f^{7/3} g_s^{31/4} r_h^{14} \kappa _{\lambda _3} \log (N) \log ^7(r) (5 \log (N)-12 \log (r))^3}{r^{16} (\log (N)-3 \log (r))^{14/3}},\nonumber\\
   & & 
  \hskip -.5in \lambda_4(r)=\kappa_{\lambda_4}\frac{ M^3 N^{7/10} \sqrt[3]{N_f} g_s^3 r_h^{14} \log (N) \log ^3(r) (5 \log (N)-12 \log (r))}{ r^{16} \alpha _{\theta _1}^3 \alpha _{\theta _2}^2
   (\log (N)-3 \log (r))^{8/3}} \nonumber\\
 & & \sim  \frac{M^3 \sqrt[20]{N} \log (2) (\log (64)-1) \sqrt[3]{N_f} g_s^{15/4} r_h^{14} \kappa _{\lambda _4} \log (N) \log ^3(r) (5 \log (N)-12 \log (r))}{r^{16} (\log (N)-3 \log (r))^{8/3}},
\end{eqnarray}
where $\kappa_{\lambda_3}$ and $\kappa_{\lambda_4}$ are the numerical factors. 

\item
In the anomaly term (\ref{Lag-anomaly-IS}) following are the $r$ dependent functions:
\begin{eqnarray}
\label{Z-W-U-V-1}
& & 
Z_1(r)=\kappa_{Z_1}\frac{ M N^{37/10} g_s^{35/6} r_h^2 \log (N) \log (r)}{ r^{16} \alpha _{\theta _1}^3 \alpha _{\theta _2}^2 N_f^{4/3} (\log (N)-3 \log (r))^{7/3}} \nonumber\\
& & \sim \frac{M N^{61/20} \log (2) (\log (64)-1) g_s^{79/12} r_h^2 \log (N) \log (r)  \kappa _{Z_1}}{r^{16} N_f^{4/3} (\log (N)-3 \log
   (r))^{7/3}},\nonumber\\
   & & W_1(r)=\kappa_{W_1}\frac{ M N^{37/10} g_s^{35/6} \log (N) \log (r)}{ r^8 \alpha _{\theta _1}^3 \alpha _{\theta _2}^2 N_f^{4/3} r_h^6 (\log (N)-3 \log (r))^{7/3}} \nonumber\\
  & & \sim \frac{ M N^{61/20} \log (2) (\log (64)-1) g_s^{79/12} \kappa _{W_1} \log (N) \log (r)}{ r^8 N_f^{4/3} r_h^6 (\log (N)-3 \log (r))^{7/3}},
   \nonumber\\
& & U_1(r)=\kappa_{U_1}\frac{M^2 N N_f g_s^3 r_h^{12} \log ^2(N) \log ^3(r) (\log (N)-17 \log (r)) (\log (N)-9 \log (r))}{ r^{15} \alpha _{\theta _1} \alpha _{\theta _2}^6 (\log (N)-3 \log (r))^4},\nonumber\\
& & \sim \frac{M^2 N_f g_s^{15/4} r_h^{12} \kappa _{U_1} \log ^2(N) \log ^3(r) (\log (N)-17 \log (r)) (\log (N)-9 \log (r))}{  \sqrt[4]{N} r^{15} (\log (N)-3 \log (r))^4},\nonumber\\
& & V_1(r)=-\kappa_{V_1}\frac{M^2 N N_f g_s^3 r_h^8 \log ^5(N) \log ^3(r) (\log (N)-6 \log (r))}{ r^{11} \alpha _{\theta _1} \alpha _{\theta _2}^6 (\log (N)-9 \log (r))^3 (\log (N)-3 \log (r))^3} \nonumber\\
& & \sim -\frac{M^2  N_f g_s^{15/4} r_h^8 \kappa _{V_1} \log ^5(N) \log ^3(r) (\log (N)-6 \log (r))}{\sqrt[4]{N} r^{11} (\log (N)-9 \log (r))^3 (\log (N)-3 \log (r))^3},
\end{eqnarray}
where $\kappa_{Z_1},\kappa_{W_1},\kappa_{U_1}$ and $\kappa_{V_1}$ are the numerical pre-factors which also includes $\left(\frac{8}{q_\alpha +1}\right)$.

\item
Using equations (\ref{Lag-beta0-IS}), (\ref{Lag-Wald-IS}),(\ref{Lag-anomaly-IS}), (\ref{L1234}) and (\ref{Lag-total-IS}) we obtain:
{\footnotesize
\begin{eqnarray}
\label{dLagoverdx}
& & \frac{\delta {\cal L}_{\rm {Total}}^{\rm {IS}}}{\delta x(r)} = \beta  \left(\frac{N^{3/10} f^\beta_1(r_h) x(r)}{\sqrt{r-r_h}}+\left(\frac{ f^\beta_2(r_h)}{\sqrt{r-r_h}}+\frac{
   f^\beta_3(r_h)}{N^{1/4}\sqrt{r-r_h}}\right)\frac{x(r)}{x'(r)^2}+\frac{N^{3/10} f^\beta_4(r_h)
   x(r)^3}{\sqrt{r-r_h} x'(r)^2}+\frac{N^{13/10} x(r) Y_1(r_h)}{\sqrt{r-r_h}}\right) \nonumber\\
  & &  \hskip 0.6in +\frac{N^{13/10}
   f_1(r_h) x(r)}{\sqrt{r-r_h}},\nonumber\\
   & & 
\end{eqnarray}
}
{\footnotesize
\begin{eqnarray}
\label{dLagoverdxprime}
& & \frac{\delta{\cal L}_{\rm {Total}}^{\rm {IS}}}{\delta x'(r)} = \beta  \Biggl[\frac{{F^\beta_1}(r_h) \sqrt{r-r_h} x(r)^2 x'(r)}{N^{7/10}} +\frac{F^\beta_{10}(r_h) \sqrt{r-r_h} x(r)^4}{N^{7/10} x'(r)}+\frac{F^\beta_{11}(r_h)
   \sqrt{r-r_h}}{N^{7/10} x'(r)}+\frac{N^{3/10} F^\beta_{12}(r_h) x(r)^4}{\sqrt{r-r_h}
   x'(r)^3}\nonumber\\
& &+\frac{N^{3/10} F^\beta_{13}(r_h)}{\sqrt{r-r_h} x'(r)^3}   +\frac{F^\beta_2(r_h)
   \sqrt{r-r_h} x(r)^2}{N x'(r)} +\frac{ {F^\beta_3}(r_h) \sqrt{r-r_h}
   x(r)^2}{N^{5/4} x'(r)}+\frac{{F^\beta_4}(r_h) x(r)^2}{\sqrt{r-r_h} x'(r)^3}+\frac{
   {F^\beta_5}(r_h) x(r)^2}{N^{1/4}\sqrt{r-r_h} x'(r)^3}\nonumber\\
  & & +\frac{{F^\beta_6}(r_h) (r-r_h)^{3/2}
   x(r)^4 \left(2 x'(r)+r x''(r)\right)}{N^{7/10} x'(r)^2}
   +\frac{{F^\beta_7}(r_h) (r-r_h)^{3/2} \left(2
   x'(r)+r x''(r)\right)}{N^{7/10} x'(r)^2}\nonumber\\
   & &+\frac{{F^\beta_8}(r_h) \sqrt{r-r_h} x(r)^4}{N^{7/10}
   x'(r)}+\frac{{F^\beta_9}(r_h) \sqrt{r-r_h}}{N^{7/10} x'(r)} +N^{3/10} x(r)^2{Y_3}(r_h) x'(r)\Biggr]+N^{3/10} {F_1}(r_h) \sqrt{r-r_h} x(r)^2 x'(r),
\end{eqnarray}
}
and,
\begin{eqnarray}
\label{dLagoverdxprimexprime}
& & \frac{\delta{\cal L}_{\rm {Total}}^{\rm {IS}}}{\delta x^{\prime\prime}(r)} = \beta   \left(\frac{{\cal F}^\beta_1(r_h) (r-r_h)^{3/2}}{N^{7/10}}+\frac{{\cal F}^\beta_2(r_h)
   (r-r_h)^{3/2}}{N^{7/10}}\right),
   \end{eqnarray}
 where we have defined $F_1(r_h),F^\beta_{i=1,..,13}(r_h),f_1(r_h),f^\beta_{i=1,..,4}(r_h)$ and $Y_{i=1,3}(r_h)$ for the island surface as: 
 {\footnotesize
 \begin{eqnarray}
 \label{Fbetai[rh]s}
 & & F_1(r_h)=\frac{\kappa_{F_1} M \log (2) (\log (64)-1) N_f^{5/3} g_s^{7/3} \sqrt{r_h} \kappa _{\alpha } \kappa _{\lambda _5} \log (N) \log \left(r_h\right) \left(\log (N)-3 \log \left(r_h\right)\right){}^{2/3}}{\sqrt{\kappa _{\alpha } \kappa _{\sigma }}},\nonumber\\
 & & F_1^\beta(r_h) =\frac{\kappa_{F^\beta_1}M^3 \log (2) (\log (64)-1) N_f^{2/3} g_s^{17/6} \sqrt{r_h} \kappa _{\alpha } \log (N) \log ^3\left(r_h\right) \left(5 \log (N)-12 \log \left(r_h\right)\right)}{\sqrt{\kappa _{\alpha } \kappa _{\sigma }} \left(\log (N)-3 \log \left(r_h\right)\right){}^{13/3}} \nonumber\\
   & & \times \left( 81 \left(16+\sqrt{2}\right) M^4 N_f^2 g_s^4 \kappa _{\lambda _3} \log ^4\left(r_h\right) \left(5 \log (N)-12 \log \left(r_h\right)\right){}^2+4096 \left(4+\sqrt{2}\right) \pi ^4 \kappa _{\lambda _4}
   \left(\log (N)-3 \log \left(r_h\right)\right){}^2\right),\nonumber\\
 & & F_2^\beta(r_h) = -\frac{\kappa_{F^\beta_2}M^2 g_s^{3/2} \kappa _{\alpha } \kappa _{U_1} \log ^2(N) \log ^3(r) \left(\log (N)-17 \log \left(r_h\right)\right) \left(\log (N)-9 \log \left(r_h\right)\right) \left(N_f g_s
   \left(\log (N)-3 \log \left(r_h\right)\right)\right){}^{4/3}}{\sqrt{r_h} \sqrt{\kappa _{\alpha } \kappa _{\sigma }} \left(\log (N)-3 \log
   \left(r_h\right)\right){}^5},\nonumber\\
 & & F_3^\beta(r_h) =\frac{\kappa_{F^\beta_3}M^4 N_f^{7/3} g_s^{79/12} \kappa _{\alpha } \log ^4(N) \log ^6\left(r_h\right)}{ r_h^{7/2} \sqrt{\kappa _{\alpha } \kappa _{\sigma }}
   \left(\log (N)-9 \log \left(r_h\right)\right){}^6 \left(\log (N)-3 \log \left(r_h\right)\right){}^{23/3}} \nonumber\\
   & & \times \left(\left(\kappa _{V_1} \log ^3(N) \left(-9 \log (N) \log \left(r_h\right)+18 \log ^2\left(r_h\right)+\log ^2(N)\right)-6 \kappa _{U_1} \left(\log (N)-17 \log \left(r_h\right)\right) \left(\log (N)-9 \log
   \left(r_h\right)\right){}^4\right){}^2\right)
   ,\nonumber\\
 & & F_4^\beta(r_h) =\frac{\kappa_{F^\beta_4}M^2 N_f^{4/3} g_s^{23/6} \kappa _{U_1} \log ^2(N) \log ^3\left(r_h\right) \sqrt{\kappa _{\alpha } \kappa _{\sigma }} \left(\log (N)-17 \log \left(r_h\right)\right) \left(\log
   (N)-9 \log \left(r_h\right)\right)}{r_h^{7/2} \left(\log (N)-3 \log \left(r_h\right)\right){}^{11/3}},\nonumber\\
   & &  F_5^\beta(r_h) = -\frac{\kappa_{F^\beta_5}M^4  g_s^{21/4} \log ^4(N) \sqrt{\kappa _{\alpha } \kappa _{\sigma }} \left(N_f g_s \left(\log (N)-3 \log \left(r_h\right)\right)\right){}^{7/3}}{r_h^{13/2} \left(\log (N)-9 \log \left(r_h\right)\right){}^6 \left(\log (N)-3 \log \left(r_h\right)\right){}^{10}}  \nonumber\\
   & &\hskip -0.2in \left(\kappa _{V_1} \log ^3(N) \log ^3\left(r_h\right) \left(-9 \log (N) \log \left(r_h\right)+18 \log ^2\left(r_h\right)+\log ^2(N)\right)
   -6 \kappa _{U_1} \log ^3(r) \left(\log (N)-17 \log
   \left(r_h\right)\right) \left(\log (N)-9 \log \left(r_h\right)\right){}^4\right){}^2;\nonumber
   \end{eqnarray}
   \begin{eqnarray}
   & & F_6^\beta(r_h) =-\frac{\kappa_{F^\beta_6} M  \sqrt[3]{N_f} g_s^2 \kappa _{W_1} \log (N) \log \left(r_h\right) \left(\kappa _{\alpha } \kappa _{\sigma }\right){}^{5/2}}{r_h^{5/2} \kappa _{\sigma }^5 \left(\log (N)-3 \log \left(r_h\right)\right){}^{2/3}},\nonumber\\
    & &   F_7^\beta(r_h) =-\frac{\kappa_{F^\beta_7} M \log (2) (\log (64)-1) \sqrt[3]{N_f} g_s^2 \kappa _{Z_1} \log (N) \log \left(r_h\right) \left(\kappa _{\alpha } \kappa _{\sigma }\right){}^{5/2}}{r_h^{5/2} \kappa _{\sigma }^5 \left(\log (N)-3 \log \left(r_h\right)\right){}^{2/3}}, \nonumber\\   
 & & F_8^\beta(r_h) =\frac{\kappa_{F^\beta_8} M \log (2) (\log (64)-1) \sqrt[3]{N_f} g_s^2 \kappa _{W_1} \log (N) \log \left(r_h\right) \left(\kappa _{\alpha } \kappa _{\sigma }\right){}^{5/2}}{r_h^{3/2} \kappa _{\sigma }^5 \left(\log (N)-3 \log \left(r_h\right)\right){}^{2/3}},\nonumber\\
 & & F_9^\beta(r_h) = \frac{\kappa_{F^\beta_9} M \log (2) (\log (64)-1) \sqrt[3]{N_f} g_s^2 \kappa _{Z_1} \log (N) \log \left(r_h\right) \left(\kappa _{\alpha } \kappa _{\sigma }\right){}^{5/2}}{ r_h^{3/2} \kappa _{\sigma }^5 \left(\log (N)-3 \log \left(r_h\right)\right){}^{2/3}},\nonumber\\
 & & F_{10}^\beta(r_h) = -\frac{\kappa_{F^\beta_{10}} M \log (2) (\log (64)-1) \sqrt[3]{N_f} g_s^2 \kappa _{W_1} \log (N) \log \left(r_h\right) \left(\kappa _{\alpha } \kappa _{\sigma }\right){}^{5/2}}{r_h^{3/2} \kappa _{\sigma }^5 \left(\log (N)-3 \log \left(r_h\right)\right){}^{2/3}};\nonumber
 \end{eqnarray}
 }
 {\footnotesize
 \begin{eqnarray} 
 & & F_{11}^\beta(r_h) = -\frac{\kappa_{F^\beta_{11}}M \log (2) (\log (64)-1) \sqrt[3]{N_f} g_s^2 \kappa _{Z_1} \log (N) \log \left(r_h\right) \left(\kappa _{\alpha } \kappa _{\sigma }\right){}^{5/2}}{r_h^{3/2} \kappa _{\sigma }^5 \left(\log (N)-3 \log \left(r_h\right)\right){}^{2/3}}, \nonumber\\
   & &F_{12}^\beta(r_h) =-\frac{\kappa_{F^\beta_{12}} M \log (2) (\log (64)-1) \sqrt[3]{N_f} g_s^3 \kappa _{W_1} \log (N) \log \left(r_h\right) \left(\kappa _{\alpha } \kappa _{\sigma }\right){}^{5/2}}{r_h^{9/2} \kappa _{\sigma }^4 \left(\log (N)-3 \log \left(r_h\right)\right){}^{2/3}},
   \nonumber\\
   & & F_{13}^\beta(r_h) =-\frac{\kappa_{F^\beta_{13}} M \log (2) (\log (64)-1) \sqrt[3]{N_f} g_s^3 \kappa _{Z_1} \log (N) \log \left(r_h\right) \left(\kappa _{\alpha } \kappa _{\sigma }\right){}^{5/2}}{r_h^{9/2} \kappa _{\sigma }^4 \left(\log (N)-3 \log \left(r_h\right)\right){}^{2/3}};
 \end{eqnarray}
 }
{\footnotesize
\begin{eqnarray}
\label{fi[rh]s}
& & f_1(r_h)=\frac{\kappa_{f_1} M \log (2) (\log (64)-1) N_f^{5/3} g_s^{10/3} \kappa _{\lambda _5} \log (N) \log \left(r_h\right) \sqrt{\kappa _{\alpha } \kappa _{\sigma }} \left(\log (N)-3 \log
   \left(r_h\right)\right){}^{2/3}}{r_h^{5/2}},\nonumber\\
   & & f^\beta_1(r_h)=\frac{\kappa_{f^\beta_1}M^3  N_f^{2/3} g_s^{23/6} \log (N) \log ^3\left(r_h\right) \sqrt{\kappa _{\alpha } \kappa _{\sigma }} \left(5 \log (N)-12 \log \left(r_h\right)\right)}{ r_h^{5/2} \left(\log (N)-3 \log \left(r_h\right)\right){}^{13/3}} \nonumber\\
   & & \times \left(\frac{M^3 \log (2) (\log (64)-1) N_f^{2/3} g_s^{23/6} \log (N) \log ^3\left(r_h\right) \sqrt{\kappa _{\alpha } \kappa _{\sigma }} \left(5 \log (N)-12 \log \left(r_h\right)\right)}{1572864 \sqrt[3]{2}
   3^{5/6} \pi ^{77/12} r_h^{5/2} \left(\log (N)-3 \log \left(r_h\right)\right){}^{13/3}}\right),\nonumber\\
   & & f^\beta_2(r_h)=-\frac{\kappa_{f^\beta_2}M^2  N_f^{4/3} g_s^{23/6} \kappa _{U_1} \log ^2(N) \log ^3\left(r_h\right) \sqrt{\kappa _{\alpha } \kappa _{\sigma }} \left(\log (N)-17 \log \left(r_h\right)\right) \left(\log
   (N)-9 \log \left(r_h\right)\right)}{r_h^{7/2} \left(\log (N)-3 \log \left(r_h\right)\right){}^{11/3}},\nonumber\\
   & & f^\beta_3(r_h)=\frac{\kappa_{f^\beta_3}M^4 g_s^{21/4} \log ^4(N) \sqrt{\kappa _{\alpha } \kappa _{\sigma }} \left(N_f g_s \left(\log (N)-3 \log \left(r_h\right)\right)\right){}^{7/3}}{r_h^{13/2} \left(\log (N)-9 \log \left(r_h\right)\right){}^6 \left(\log (N)-3 \log \left(r_h\right)\right){}^{10}} \nonumber\\
   & & \times \left(\kappa _{V_1} \log ^3(N) \log ^3\left(r_h\right) \left(-9 \log (N) \log \left(r_h\right)+18 \log ^2\left(r_h\right)+\log ^2(N)\right)-6 \kappa _{U_1} \log ^3(r) \left(\log (N)-17 \log
   \left(r_h\right)\right) \left(\log (N)-9 \log \left(r_h\right)\right){}^4\right){}^2,\nonumber\\
   & & f^\beta_4 (r_h)=\frac{\kappa_{f^\beta_4} M \log (2) (\log (64)-1) \sqrt[3]{N_f} g_s^3 \kappa _{W_1} \log (N) \log \left(r_h\right) \left(\kappa _{\alpha } \kappa _{\sigma }\right){}^{5/2}}{r_h^{9/2} \kappa _{\sigma }^4 \left(\log (N)-3 \log \left(r_h\right)\right){}^{2/3}}; \nonumber
\end{eqnarray} 
}
{\footnotesize
\begin{eqnarray} 
\label{FD-beta-i}
   & & {\cal F}^\beta_1(r_h)=-\frac{\kappa_{{\cal F}^\beta_1}M \log (2) (\log (64)-1) \sqrt[3]{N_f} g_s^2 \kappa _{\alpha }^{5/2} \kappa _{W_1} \log (N) \log \left(r_h\right)}{ r_h^{3/2} \kappa _{\sigma }^{5/2}
   \left(\log (N)-3 \log \left(r_h\right)\right){}^{2/3}},\nonumber\\
   & &{\cal F}^\beta_2(r_h) =-\frac{\kappa_{{\cal F}^\beta_1} M \log (2) (\log (64)-1) \sqrt[3]{N_f} g_s^2 \kappa _{\alpha }^{5/2} \kappa _{Z_1} \log (N) \log \left(r_h\right)}{r_h^{3/2} \kappa _{\sigma }^{5/2}\left(\log (N)-3 \log \left(r_h\right)\right){}^{2/3}},
\end{eqnarray}
}
and,  
{\footnotesize
 \begin{eqnarray}
 \label{Yi[rh]s}
 & & Y_1(r_h) = \frac{\kappa_{Y_1}{\cal C}_{\theta_1 x}M N_f^{5/3} g_s^{10/3} \log ^2\left(r_h\right) \sqrt{\kappa _{\alpha } \kappa _{\sigma }} \left(\log (N)-3 \log \left(r_h\right)\right){}^{2/3}}{ r_h^{3/2}},\nonumber\\
   & & Y_3(r_h) =\frac{\kappa_{Y_3} {\cal C}_{\theta_1 x} M  N_f^{5/3} g_s^{7/3} r_h^{3/2} \kappa _{\alpha } \log ^2\left(r_h\right) \left(\log (N)-3 \log \left(r_h\right)\right){}^{2/3}}{ \sqrt{\kappa _{\alpha } \kappa _{\sigma }}} ,
 \end{eqnarray}
 }
 where $\kappa_{F_1},\kappa_{F^\beta_{i=1,..,13}},\kappa_{f_1},\kappa_{f^\beta_{i=1,..,4}},\kappa_{{\cal F}^\beta_{1,2}},\kappa_{Y_{1,3}}$ are numerical factors and ${\cal C}_{\theta_1 x}$ is the constant of integration appearing in the solution to ${\cal O}(l_p^6)$ correction to the metric component along the toroidal analog $T^2(\theta_1,x)$ of the vanishing two-cycle $S^2(\theta_1,\phi_1)$.
 \end{itemize}

\section{Possible Terms Appearing in Holographic Entanglement Entropies}
\label{PT}
\setcounter{equation}{0}\seceqdd

In this appendix we have listed all the possible terms that we obtain from differentiating Lagrangian of the ${\cal M}$-theory dual inclusive of ${\cal O}(R^4)$ corrections. In appendix \ref{PT-HM}, we have listed all the terms for Hartman-Maldacena-like surface and in appendix \ref{PT-IS} we have listed all the possible terms for the island surface. 
\subsection{Hartman-Maldacena-like Surface}
\label{PT-HM}
Square root of the determinant of the induced metric (\ref{metric-HM-t(r)}) for the Hartman-Maldacena-like surface scale as $N^{7/10}$ and is given below:
{\scriptsize
\begin{eqnarray}
\label{sqrtminusg-HM}
& &
\sqrt{-g}\sim \Biggl(\frac{M N^{7/10} \sqrt{g_s} \left(N_f (\log (N)-3 \log (r))\right){}^{5/3} \sqrt{\alpha (r) \left(\sigma (r)-\left(1-\frac{r_h^4}{r^4}\right) t'(r)^2\right)} }{\alpha _{\theta _1}^3 \alpha _{\theta _2}^2}\nonumber\\
& & \times \left(N_f g_s \left(r^2 (\log (N) (2 \log
   (r)+1)+3 (1-6 \log (r)) \log (r))-r_h^2 \left(\log (N) (2 \log (r)+1)-18 (6 r+1) \log ^2(r)\right)\right)+8 \pi  \left(r_h^2-r^2\right) \log (r)\right)\Biggr).\nonumber\\
\end{eqnarray}
}
There are four terms that we obtain from $\frac{\partial J_0}{\partial R_{txtx}}$:
{\footnotesize
\begin{eqnarray}
\label{Wald-J0-i}
& & \hskip -0.3in (i)\int d{\cal V}_9 \sqrt{-g}\left( \left(G^{xx}\right)^2 \left(G^{tt}\right)^2\left(R_{PxtQ} + \frac{1}{2}R_{PQtx}\right)R_{t}^{\ \ RSP}R^Q_{\ \ RSx}\right)\nonumber\\
& & \sim \int d{\cal V}_9 \sqrt{-g} \Biggl(-\frac{r_h^8 \left(-9 \log ^2(N) (\log (N)-6 \log (r))^2-13 (\log (N)-3 \log (r))^4-21 (\log (N)-3 \log (r))^2\right)}{\sqrt{N} r^{12} N_f^{10/3} \sqrt{g_s} (\log (N)-3 \log (r))^{22/3}}\Biggr);\nonumber\\
& & \hskip -0.3in (ii)\int d{\cal V}_9 \sqrt{-g}\left( R^{HxtK}R_H^{\ \ RSt}R^x_{\ \ RSK} + \frac{1}{2}R^{HKtx}R_H^{\ \ RSt}R^Q_{\ \ RSK} \right)\nonumber\\
& & \sim \int d{\cal V}_9 \sqrt{-g} \Biggl(\frac{1}{\sqrt{N} r^{12} N_f^{22/3} \sqrt{g_s} r_h^8 (\log (N)-3 \log (r))^{22/3}}\Biggr) ;
\nonumber\\
& & \hskip -0.3in (iii)\int d{\cal V}_9 \sqrt{-g}  \left(\left(G^{xx}\right)^2 G^{tt}\left(R_{PxtQ} + \frac{1}{2}R_{PQtx}\right)R_{t}^{\ \ RSP}R^Q_{\ \ RSx}\right)\nonumber\\
& & \hskip -.5in \sim \int d{\cal V}_9 \Biggl(\lambda_1(r) \sqrt{\alpha (r) \left(\sigma (r)-\left(1-\frac{r_h^4}{r^4}\right) t'(r)^2\right)}\Biggr) ;\nonumber\\
& & \hskip -0.3in (iv)\int d{\cal V}_9 \sqrt{-g} \left( G^{tt} \left(R^{HMNx}R_{PMNt} + \frac{1}{2}R^{HxMN}R_{PtMN}\right)R_H^{\ \ xtP}\right)\nonumber\\
& & \sim \int d{\cal V}_9 \Biggl(\lambda_2(r) \sqrt{\alpha (r) \left(\sigma (r)-\left(1-\frac{r_h^4}{r^4}\right) t'(r)^2\right)}\Biggr).\nonumber\\
\end{eqnarray}
}
From equation (\ref{lambda1-2}) it is clear that (iii) and (iv) terms scale as $N^{7/10}$ whereas from equations (\ref{sqrtminusg-HM}) and (\ref{Wald-J0-i}) we can see that (i) and (ii) terms scale as $N^{1/5}$. Therefore (iii) and (iv) terms are the most dominant terms in the large-$N$ limit.\\

We obtain thirteen terms from {\bf $\frac{\partial^2 J_0}{\partial R_{xixj}\partial R_{xmxl}} K_{tij}K_{tml}$} which are listed below:
{\scriptsize
\begin{eqnarray}
& & \hskip -0.3in (i) \int d{\cal V}_9 \sqrt{-g} \left(G^{xx}\right)^2 \left(G^{mm}\right)\left(G^{ll}\right) \delta^i_m R_r^{~TSx} R^j_{~TSl}K_{tij}K_{tml}\nonumber\\ 
& & \sim \int d{\cal V}_9 \sqrt{-g}\Biggl(-\frac{M^2 N N_f g_s^3 \log ^3(r) (23 \log (N)-72 \log (r))^2 (29 \log (N)-72 \log (r)) (2 \log (N)-3 \log (r)) \sqrt{\alpha (r) \left(\sigma (r)-t'(r)^2\right)}}{r^4 \alpha _{\theta _1} \alpha _{\theta
   _2}^6 (\log (N)-9 \log (r))^3 (\log (N)-3 \log (r))^6 t'(r)^2} \Biggr),\nonumber\\
& & \hskip -0.3in (ii)\int d{\cal V}_9 \sqrt{-g} \left(G^{xx}\right)^3 \left(G^{ii}\right)\left(G^{jj}\right)\left(G^{mm}\right)\left(G^{ll}\right) R_{jmxQ} R^Q_{~~ixl}K_{tij}K_{tml} \sim 0, \nonumber\\
& & \hskip -0.3in (iii)\int d{\cal V}_9 \sqrt{-g}\frac{1}{2}\left(G^{xx}\right)^2 \left(G^{mm}\right)\left(G^{jl}\right)R_r^{~TSx} R^i_{~TSx}K_{tij}K_{tml} \sim 0,\nonumber\\
& & \hskip -0.3in (iv)\int d{\cal V}_9 \sqrt{-g}\frac{1}{2}\left(G^{xx}\right)^2\left(G^{ii}\right)\left(G^{jl}\right) R_r^{~TSx} R^m_{~~TSi} K_{tij}K_{tml}\nonumber\\
& &\sim \int d{\cal V}_9 \sqrt{-g}\Biggl( \frac{M^3 N^{13/10} \alpha _{\theta _1} N_f^2 g_s^{9/2} \log ^3(N) (\log (N)-23 \log (r))^3 \left(\log (N)+18 \log ^2(r)\right) \sqrt{\alpha (r) \left(\sigma (r)-t'(r)^2\right)}}{r^4 \alpha _{\theta
   _2}^{10} \log ^2(r) (\log (N)-9 \log (r))^3 (\log (N)-3 \log (r))^5 t'(r)^2}\Biggr),\nonumber\\
& & \hskip -0.3in (v)\int d{\cal V}_9 \sqrt{-g}\frac{1}{2} \left(G^{xx}\right)^3\left(G^{ii}\right)\left(G^{ll}\right)\left(G^{mm}\right)\left(G^{jj}\right) R_{jQxl} R^Q_{~~ixm}K_{tij}K_{tml} \sim \int d{\cal V}_9 \left(\frac{Z(r)}{2}{\cal L}_1 \right),\nonumber\\
& & \hskip -0.3in (vi)\int d{\cal V}_9 \sqrt{-g}\frac{1}{2}  \left(G^{xx}\right)^3\left(G^{jm}\right)\left(G^{ll}\right) R_{Pxxl} R_x^{~ixP}K_{tij}K_{tml} 
\nonumber\\
& & \sim \int d{\cal V}_9 \sqrt{-g}\Biggl(-\frac{M^2 N N_f g_s^3 \log ^6(N) (\log (N)-51 \log (r)) (\log (N)-23 \log (r)) \sqrt{\alpha (r) \left(1-\sigma (r) r'(t)^2\right)}}{r^4 \alpha _{\theta _1} \alpha _{\theta _2}^6 \log ^3(r) (\log (N)-9
   \log (r))^3 (\log (N)-3 \log (r))^5 t'(r)^2} \Biggr),\nonumber\\
   & & \hskip -0.3in (vii)\int d{\cal V}_9 \sqrt{-g} \left(G^{xx}\right)^2\left(G^{im}\right)\left(G^{ll}\right) R^{xMNj} R_{lMNx}K_{tij}K_{tml} \sim 0, \nonumber\\
   & & \hskip -0.3in (viii)\int d{\cal V}_9 \sqrt{-g}\left(G^{xx}\right)^2 \left(G^{jj}\right)\left(G^{im}\right) R^{xMNl} R_{jMNx}K_{tij}K_{tml} \sim 0,\nonumber
\end{eqnarray}
}
{\scriptsize
\begin{eqnarray}
& & \hskip -0.3in (ix)\int d{\cal V}_9 \sqrt{-g} \left(G^{xx}\right)^3 \left(G^{ii}\right)\left(G^{jj}\right)\left(G^{ll}\right)\left(G^{mm}\right) R^Q_{~~mxj} R_{lixQ}K_{tij}K_{tml}\sim 0, \nonumber\\
& & \hskip -0.3in (x)\int d{\cal V}_9 \sqrt{-g}\frac{1}{2} \left(G^{xx}\right)^2 \left(G^{im}\right)\left(G^{ll}\right) R^{xjMN} R_{lxMN}K_{tij}K_{tml}\nonumber\\
& & \sim \int d{\cal V}_9 \sqrt{-g}\Biggl[-\frac{M N^{7/10} r^2 N_f^{8/3} g_s^{3/2} \log (r) (\log (N)-9 \log (r)) (\log (N)-3 \log (r))^{5/3} \sqrt{\alpha (r) \left(\sigma (r)-t'(r)^2\right)}}{\alpha _{\theta _1}^3 \alpha _{\theta _2}^2
   t'(r)^2} \nonumber\\
   & & \times \biggl( 76800 {G_1}(r)-\frac{29 \sqrt[3]{6} M^4 g_s^4 \log ^2(N) \log ^2(r) (\log (N)-12 \log (r))^2 (\log (N)-6 \log (r))^2 }{\pi ^{4/3} r^4 \alpha (r)^2 N_f^{2/3} (\log (N)-3 \log (r))^{20/3} \left(\sigma (r)-t'(r)^2\right)^2} \nonumber\\
   & & \hskip 0.5in \times \left(\sigma (r) \alpha '(r)-\alpha '(r) t'(r)^2+\alpha (r) \left(\sigma '(r)-2
   t'(r) t''(r)\right)\right)^2 \Biggr) \Biggr],\nonumber\\
& & \hskip -0.3in (xi)\int d{\cal V}_9 \sqrt{-g}\frac{1}{2} \left(G^{xx}\right)^2 \left(G^{im}\right)\left(G^{jj}\right) R^{xlMN} R_{jxMN}K_{tij}K_{tml}\nonumber\\
& & \sim \int d{\cal V}_9 \sqrt{-g}\Biggl[-\frac{M N^{7/10} r^2 N_f^{8/3} g_s^{3/2} \log (r) (\log (N)-9 \log (r)) (2 \log (N)-\log (r))^{5/3} \sqrt{\alpha (r) \left(\sigma (r)-t'(r)^2\right)}}{129600 \sqrt{2} 3^{5/6} \pi ^{19/6} \alpha _{\theta
   _1}^3 \alpha _{\theta _2}^2 t'(r)^2}\nonumber\\
   & &  \times \Biggl( 450 {G_2}(r)-\frac{87 \sqrt[3]{3} M^4 g_s^4 \log ^2(N) \log ^2(r) (\log (N)-12 \log (r))^2 (\log (N)-6 \log (r))^2 (\log (N)-3 \log (r))^2 }\nonumber\\
   & & \times \left(\sigma (r) \alpha '(r)-\alpha '(r) t'(r)^2+\alpha (r)
   \left(\sigma '(r)-2 t'(r) t''(r)\right)\right)^2}{\pi ^{4/3} r^4 \alpha (r)^2 N_f^{2/3} (2 \log (N)-\log (r))^{26/3} \left(\sigma (r)-t'(r)^2\right)^2\Biggr) \Biggr],\nonumber\\
& & \hskip -0.3in (xii)\int d{\cal V}_9 \sqrt{-g}\frac{1}{2}\left(G^{xx}\right)^3 \left(G^{ii}\right)\left(G^{jj}\right)\left(G^{ll}\right)\left(G^{mm}\right)R^Q_{~~mxi} R_{lQxj}K_{tij}K_{tml} \sim \int d{\cal V}_9 \left(\frac{Z(r)}{2}{\cal L}_1 \right),\nonumber\\
& & \hskip -0.3in (xiii)\int d{\cal V}_9 \sqrt{-g}\frac{1}{2}\left(G^{xx}\right)^3 \left(G^{jj}\right)\left(G^{il}\right)R_x^{~mxP} R_{Pxxj}K_{tij}K_{tml} \nonumber\\
& & \sim \int d{\cal V}_9 \sqrt{-g}\Biggl(-\frac{M^2 N N_f g_s^3 \log ^5(N) (5 \log (N)-211 \log (r)) (\log (N)-23 \log (r)) \sqrt{\alpha (r) \left(\sigma (r)-t'(r)^2\right)}}{r^4 \alpha _{\theta _1} \alpha _{\theta _2}^6 \log (r) (\log (N)-9
   \log (r))^3 (\log (N)-3 \log (r))^5 t'(r)^2} \Biggr),
\end{eqnarray}
}
where $G_1(r)$ and $G_2(r)$ are $N$ independent functions. Out of all the terms listed above (v) and (xii) terms are the most dominant in the large $N$ limit.\\

Similarly find that there are thirteen terms arising from {\bf $\frac{\partial^2 J_0}{\partial R_{titj}\partial R_{tmtl}} K_{tij}K_{tml}$} which are listed below:
{\scriptsize
\begin{eqnarray}
& & \hskip -0.3in (i)\int d{\cal V}_9 \sqrt{-g}\left(G^{tt}\right)^2 \left(G^{mm}\right)\left(G^{ll}\right) \delta^i_m R_t^{~TSt} R^j_{~TSl}K_{tij}K_{tml}\nonumber\\
& & \sim \int d{\cal V}_9 \sqrt{-g}\Biggl(-\frac{M^2 N N_f g_s^3 \log ^5(N) (23 \log (N)-72 \log (r))^2 (\log (N)-27 \log (r)) \sqrt{\alpha (r) \left(\sigma (r)-t'(r)^2\right)}}{r^4 \alpha _{\theta _1} \alpha _{\theta _2}^6 (\log (N)-9 \log
   (r))^4 (\log (N)-3 \log (r))^6 t'(r)^2} \Biggr),\nonumber\\
& & \hskip -0.3in (ii)\int d{\cal V}_9 \sqrt{-g} \left(G^{tt}\right)^3 \left(G^{ii}\right)\left(G^{jj}\right)\left(G^{mm}\right)\left(G^{ll}\right) R_{jmtQ} R^Q_{~~itl}K_{tij}K_{tml}\sim 0,\nonumber\\
& & \hskip -0.3in (iii)\int d{\cal V}_9 \sqrt{-g}\frac{1}{2}\left(G^{tt}\right)^2 \left(G^{mm}\right)\left(G^{jl}\right)R_t^{~TSt} R^i_{~TSt}K_{tij}K_{tml} \sim 0,\nonumber\\
& & \hskip -0.3in (iv)\int d{\cal V}_9 \sqrt{-g}\frac{1}{2}\left(G^{tt}\right)^2\left(G^{ii}\right)\left(G^{jl}\right) R_t^{~TSt} R^m_{~~TSi} K_{tij}K_{tml}\nonumber\\
& & \sim \int d{\cal V}_9 \sqrt{-g}\Biggl(\frac{M^3 N^{13/10} \alpha _{\theta _1} N_f^2 g_s^{9/2} \log ^3(N) (\log (N)-23 \log (r))^3 \left(\log (N)+18 \log ^2(r)\right) \sqrt{\alpha (r) \left(\sigma (r)-t'(r)^2\right)}}{1728 \sqrt{6} \pi ^{7/2}
   r^4 \alpha _{\theta _2}^{10} \log ^2(r) (\log (N)-9 \log (r))^3 (\log (N)-3 \log (r))^5 t'(r)^2} \Biggr),\nonumber\\
& & \hskip -0.3in (v)\int d{\cal V}_9 \sqrt{-g}\frac{1}{2} \left(G^{tt}\right)^3\left(G^{ii}\right)\left(G^{ll}\right)\left(G^{mm}\right)\left(G^{jj}\right) R_{jQtl} R^Q_{~~itm}K_{tij}K_{tml} \sim \int d{\cal V}_9\left(\frac{W(r)}{2}{\cal L}_2 \right),\nonumber\\
\end{eqnarray}
}
{\scriptsize
\begin{eqnarray}
& & \hskip -0.3in (vi)\int d{\cal V}_9 \sqrt{-g}\frac{1}{2}  \left(G^{tt}\right)^3\left(G^{jm}\right)\left(G^{ll}\right) R_{Pttl} R_t^{~itP} K_{tij}K_{tml}
\nonumber\\
& & \sim \int d{\cal V}_9 \sqrt{-g}\Biggl(\frac{M^2 N N_f g_s^3 \log ^5(N) (5 \log (N)-211 \log (r)) (\log (N)-23 \log (r)) \sqrt{\alpha (r) \left(\sigma (r)-t'(r)^2\right)}}{ r^4 \alpha _{\theta _1} \alpha _{\theta _2}^6
   \log (r) (\log (N)-9 \log (r))^3 (\log (N)-3 \log (r))^5 t'(r)^2} \Biggr),\nonumber\\
& & \hskip -0.3in (vii)\int d{\cal V}_9 \sqrt{-g} \left(G^{tt}\right)^2\left(G^{im}\right)\left(G^{ll}\right) R^{tMNj} R_{lMNt} K_{tij}K_{tml} \sim 0,\nonumber\\
& & \hskip -0.3in (viii)\int d{\cal V}_9 \sqrt{-g}\left(G^{tt}\right)^2 \left(G^{jj}\right)\left(G^{im}\right) R^{tMNl} R_{jMNt}K_{tij}K_{tml} \sim 0,\nonumber\\
& & \hskip -0.3in (ix)\int d{\cal V}_9 \sqrt{-g} \left(G^{tt}\right)^3 \left(G^{ii}\right)\left(G^{jj}\right)\left(G^{ll}\right)\left(G^{mm}\right) R^Q_{~~mtj} R_{litQ}K_{tij}K_{tml}\sim 0,\nonumber\\
& & \hskip -0.3in (x)\int d{\cal V}_9 \sqrt{-g}\frac{1}{2} \left(G^{tt}\right)^2 \left(G^{im}\right)\left(G^{ll}\right) R^{tjMN} R_{ltMN}K_{tij}K_{tml}\nonumber\\
& & \sim \int d{\cal V}_9 \sqrt{-g}\Biggl[\frac{M^5 N^{7/10} N_f^2 g_s^{11/2} {G_3}(r) \log ^3(N) \log ^3(r) (\log (N)-24 \log (r)) (\log (N)-9 \log (r)) (\log (N)-6 \log (r))^2}{r^2 \alpha _{\theta _1}^3 \alpha _{\theta _2}^2 \alpha (r)^2
   (\log (N)-3 \log (r))^5}\nonumber\\
   & & \times \frac{\sqrt{\alpha (r) \left(\sigma (r)-t'(r)^2\right)} \left(\sigma (r) \alpha '(r)-\alpha '(r) t'(r)^2+\alpha (r) \left(\sigma '(r)-2 t'(r) t''(r)\right)\right)^2}{t'(r)^4 \left(\sigma
   (r)-t'(r)^2\right)^2} \Biggr],\nonumber\\
& & \hskip -0.3in (xi)\int d{\cal V}_9 \sqrt{-g}\frac{1}{2} \left(G^{tt}\right)^2 \left(G^{im}\right)\left(G^{jj}\right) R^{tlMN} R_{jtMN}K_{tij}K_{tml}\nonumber\\
& & \sim \int d{\cal V}_9 \sqrt{-g}\Biggl[-\frac{M N^{7/10} r^2 N_f^{8/3} g_s^{3/2} \log (r) (\log (N)-9 \log (r)) (2 \log (N)-\log (r))^{5/3} \sqrt{\alpha (r) \left(\sigma (r)-t'(r)^2\right)}}{\alpha _{\theta _1}^3 \alpha _{\theta _2}^2
   t'(r)^2} \nonumber\\
& & \times \left( {G_4}(r)-\frac{52 \sqrt[3]{6} M^4 g_s^4 \log ^2(r) (\log (N)-12 \log (r))^2 \left(\sigma (r) \alpha '(r)-\alpha '(r) t'(r)^2+\alpha (r) \left(\sigma '(r)-2 t'(r) t''(r)\right)\right)^2}{\pi
   ^{4/3} r^4 \alpha (r)^2 N_f^{2/3} (\log (N)-3 \log (r))^{8/3} \left(\sigma (r)-t'(r)^2\right)^2} \right)\Biggr],\nonumber\\
& & \hskip -0.3in (xii)\int d{\cal V}_9 \sqrt{-g}\frac{1}{2}\left(G^{tt}\right)^3 \left(G^{ii}\right)\left(G^{jj}\right)\left(G^{ll}\right)\left(G^{mm}\right)R^Q_{~~mti} R_{lQtj}K_{tij}K_{tml}\sim \int d{\cal V}_9 \left(\frac{W(r)}{2}{\cal L}_2 \right),\nonumber\\
& & \hskip -0.3in (xiii)\int d{\cal V}_9 \sqrt{-g}\frac{1}{2}\left(G^{tt}\right)^3 \left(G^{jj}\right)\left(G^{il}\right)R_t^{~mtP} R_{Pttj}K_{tij}K_{tml} \nonumber\\
& & \sim \int d{\cal V}_9 \sqrt{-g}\Biggl(-\frac{M^2 N N_f g_s^3 \log ^6(N) (\log (N)-51 \log (r)) (\log (N)-23 \log (r)) \sqrt{\alpha (r) \left(\sigma (r)-t'(r)^2\right)}}{ r^4 \alpha _{\theta _1} \alpha _{\theta _2}^6 \log
   ^3(r) (\log (N)-9 \log (r))^3 (\log (N)-3 \log (r))^5 t'(r)^2}\Biggr),
\end{eqnarray}
}
where $G_3(r)$ and $G_4(r)$ are $N$ independent functions. Out of all the thirteen terms listed above (v) and (xii) terms are the most dominant terms in the large-$N$ limit.\\
We find that at ${\cal O}(R^4)$, there are six terms coming from
$\frac{\partial^2 J_0}{\partial R_{titj}\partial R_{xmxl}}K_{tij}K_{tml}$ and listed below:
{\footnotesize
\begin{eqnarray}
& & \hskip -0.3in (i)\int d{\cal V}_9 \sqrt{-g}\left(G^{xx}\right)^2 \left(G^{mm}\right)\left(G^{ll}\right) \delta^i_m \delta^t_P \delta^t_R \delta^j_Q R_x^{~TSP} R^Q_{~TSl}K_{tij}K_{tml} \sim \int d{\cal V}_9 \left(U(r){\cal L}_3 \right),\nonumber\\
& & \hskip -0.3in (ii)\int d{\cal V}_9 \sqrt{-g} \left(G^{xx}\right)^2 \left(G^{ii}\right)\left(G^{jj}\right)\left(G^{mm}\right)\left(G^{ll}\right)\left(G^{tt}\right)\delta^t_x R_{jmxQ} R^Q_{~~itl}K_{tij}K_{tml} \sim 0,\nonumber\\
& & \hskip -0.3in (iii)\int d{\cal V}_9 \sqrt{-g}\left(G^{xx}\right)^2 \left(G^{mm}\right)\left(G^{ll}\right)\left(G^{tt}\right) \delta^i_T \delta^t_S \delta^j_l  R_{Pmxt} R_x^{~TSP}K_{tij}K_{tml} \sim 0,\nonumber\\
& & \hskip -0.3in (iv)\int d{\cal V}_9 \sqrt{-g}\frac{1}{2}\left(G^{xx}\right)^2\left(G^{mm}\right)\left(G^{ll}\right)\left(G^{tt}\right)\delta^i_T \delta^t_S \delta^j_m  R_{Ptxl} R_x^{~~TSP} K_{tij}K_{tml} \sim 0,\nonumber\\
& & \hskip -0.3in (v)\int d{\cal V}_9 \sqrt{-g} \left(G^{xx}\right)\left(G^{tt}\right)^2\left(G^{ii}\right)\left(G^{jl}\right) R_t^{mxP} R_{Pitx}K_{tij}K_{tml} \sim 0,\nonumber\\
& & \hskip -0.3in (vi)\int d{\cal V}_9 \sqrt{-g}\frac{1}{2}  \left(G^{tt}\right)^2\left(G^{xx}\right)\left(G^{jj}\right)\left(G^{iK}\right)\delta^l_K R_t^{mxP} R_{Pxtj} K_{tij}K_{tml} \sim 0.
\end{eqnarray}
}
From the above terms we can see that it is the only first term which is non-zero. Therefore only this term will contribute to the Lagrangian.\\
Similarly we found that at ${\cal O}(R^4)$, there are four terms coming from
$\frac{\partial^2 J_0}{\partial R_{tixj}\partial R_{xmtl}}K_{tij}K_{tml}$ which are listed below:
{\footnotesize
\begin{eqnarray}
\label{list-second-der-J0-mix}
& & \hskip -0.3in (i)\int d{\cal V}_9 \sqrt{-g}\left(G^{xx}\right)^2 \left(G^{mm}\right)\left(G^{ll}\right) \delta^i_m  R_x^{~TSx} R^j_{~TSl}K_{tij}K_{tml}\sim \int d{\cal V}_9 \left(U(r){\cal L}_4 \right),\nonumber\\
& & \hskip -0.3in (ii)\int d{\cal V}_9 \sqrt{-g} \left(G^{xx}\right)^3 \left(G^{ii}\right)\left(G^{jj}\right)\left(G^{ll}\right)\left(G^{mm}\right) R_{jmxQ} R^Q_{~~ixl}K_{tij}K_{tml}
 \sim 0,\nonumber\\
& & \hskip -0.3in (iii)\int d{\cal V}_9 \sqrt{-g}\frac{1}{2}\left(G^{xx}\right) \left(G^{tt}\right)^2\left(G^{jm}\right)\left(G^{ll}\right)  R_x^{~ixP} R_{Pttl}K_{tij}K_{tml} \sim \int d{\cal V}_9 \left( V(r){\cal L}_4 \right),\nonumber\\
& & \hskip -0.3in (iv)\int d{\cal V}_9 \sqrt{-g}\frac{1}{2}\left(G^{xx}\right)^2\left(G^{tt}\right)\left(G^{jj}\right)\left(G^{il}\right)  R_t^{~mtP} R_{Pxxj} K_{tij}K_{tml} \sim \int d{\cal V}_9 \left( V(r){\cal L}_4 \right)
\end{eqnarray}
}
From equation (\ref{Z-W-U-V}) it is clear that all the three non-zero terms scale as $N$. Therefore we will keep all non-zero terms in the Lagrangian.
\subsection{Island Surface}
\label{PT-IS}
Square root of minus of determinant of the induced metric (\ref{induced-metric-IS}) for the island surface scale as $N^{7/10}$ is given below:
{\footnotesize
\begin{eqnarray}
& &
\sqrt{-g}=\frac{M N^{7/10} \sqrt{g_s} \left(N_f (\log (N)-3 \log (r))\right){}^{5/3} \sqrt{\alpha (r) \left(\sigma (r)+x'(r)^2\right)}}{144\ 6^{5/6} \pi ^{19/6} \alpha _{\theta _1}^3 \alpha _{\theta
   _2}^2} \nonumber\\
   & & \times  \Biggl(18 N_f g_s \log ^2(r) \left(r^2-3 b^2 (6 r+1) r_h^2\right)+\log (r)
   \left(8 \pi  \left(r^2-3 b^2 r_h^2\right)-3 r^2 N_f g_s\right)-N_f g_s \log (N) (2 \log (r)+1) \left(r^2-3 b^2 r_h^2\right)\Biggr).\nonumber\\
\end{eqnarray}
}
Similar to Hartman-Maldacena-like surface, there are four terms that we obtain from $\frac{\partial J_0}{\partial R_{txtx}}$:\\
{\footnotesize
\begin{eqnarray}
\label{Wald-J0-i-IS}
& & \hskip -0.3in (i)\int d{\cal V}_9 \sqrt{-g}\left( \left(G^{xx}\right)^2 \left(G^{tt}\right)^2\left(R_{PxtQ} + \frac{1}{2}R_{PQtx}\right)R_{t}^{\ \ RSP}R^Q_{\ \ RSx}\right)\nonumber\\
& & \sim\int d{\cal V}_9 \sqrt{-g} \Biggl(\frac{r_h^8 \left(13 (\log (N)-3 \log (r))^4+21 (\log (N)-3 \log (r))^2+9 \log (r) (6 \log (N)-9 \log (r))^2\right)}{\sqrt{N} r^{12} N_f^{10/3} \sqrt{g_s} (\log (N)-3 \log (r))^{22/3}}\Biggr) ;\nonumber\\
& & \hskip -0.3in (ii)\int d{\cal V}_9 \sqrt{-g}\left( R^{HxtK}R_H^{\ \ RSt}R^x_{\ \ RSK} + \frac{1}{2}R^{HKtx}R_H^{\ \ RSt}R^Q_{\ \ RSK} \right)\nonumber\\
& & \sim \int d{\cal V}_9 \sqrt{-g}\Biggl(\frac{r_h^8 \left(-2 g_s \log ^2(N) (\log (N)-12 \log (r))^2-5 (\log (N)-9 \log (r))^2 (\log (N)-3 \log (r))^2\right)}{\sqrt{N} r^{12} N_f^{10/3} \sqrt{g_s} (\log (N)-9 \log (r))^2 (\log (N)-3 \log
   (r))^{16/3}}\Biggr) ;
\nonumber\\
& & \hskip -0.3in (iii)\int d{\cal V}_9 \sqrt{-g}  \left(\left(G^{xx}\right)^2 G^{tt}\left(R_{PxtQ} + \frac{1}{2}R_{PQtx}\right)R_{t}^{\ \ RSP}R^Q_{\ \ RSx}\right)\nonumber\\
& &  \sim \int d{\cal V}_9 \left( \lambda_3(r)  \sqrt{\alpha (r) \left(\sigma (r)+x'(r)^2\right)}\right);\nonumber\\
& & \hskip -0.3in (iv)\int d{\cal V}_9 \sqrt{-g} \left( G^{tt} \left(R^{HMNx}R_{PMNt} + \frac{1}{2}R^{HxMN}R_{PtMN}\right)R_H^{\ \ xtP}\right)\nonumber\\
& & \sim \int d{\cal V}_9\left( \lambda_4(r) \sqrt{\alpha (r) \left(\sigma (r)+x'(r)^2\right)}\right).\nonumber\\
\end{eqnarray}
}
We find that (iii) and (iv) terms are the most dominant terms in the large $N$ limit. There are thirteen terms that obtain from  $\frac{\partial^2 J_0}{\partial R_{xixj}\partial R_{xmxl}} K_{xij}K_{xml}$ which are listed below:
{\scriptsize
\begin{eqnarray}
& & \hskip -0.3in (i) \int d{\cal V}_9 \sqrt{-g} \left(G^{xx}\right)^2 \left(G^{mm}\right)\left(G^{ll}\right) \delta^i_m R_r^{~TSx} R^j_{~TSl}K_{xij}K_{xml}\nonumber\\
& & \sim  \int d{\cal V}_9 \sqrt{-g} \Biggl(-\frac{M^2 N N_f g_s^3 \log ^2(r) (23 \log (N)-72 \log (r))^3}{ r^4 \alpha _{\theta _1} \alpha _{\theta _2}^6 (\log (N)-9 \log (r))^2 (\log (N)-3 \log (r))^6 x'(r)^2} \Biggr), \nonumber\\
& & \hskip -0.3in (ii)\int d{\cal V}_9 \sqrt{-g} \left(G^{xx}\right)^3 \left(G^{ii}\right)\left(G^{jj}\right)\left(G^{mm}\right)\left(G^{ll}\right) R_{jmxQ} R^Q_{~~ixl}K_{xij}K_{xml} \sim 0, \nonumber\\
& & \hskip -0.3in (iii)\int d{\cal V}_9 \sqrt{-g}\frac{1}{2}\left(G^{xx}\right)^2 \left(G^{mm}\right)\left(G^{jl}\right)R_r^{~TSx} R^i_{~TSx}K_{xij}K_{xml}\sim 0, \nonumber\\
& & \hskip -0.3in (iv)\int d{\cal V}_9 \sqrt{-g})\frac{1}{2}\left(G^{xx}\right)^2\left(G^{ii}\right)\left(G^{jl}\right) R_r^{~TSx} R^m_{~~TSi} K_{xij}K_{xml}\nonumber\\
& & \sim  \int d{\cal V}_9 \sqrt{-g} \Biggl(\frac{M^3 N^{13/10} \alpha _{\theta _1} N_f^5 g_s^{15/2} \log (r) (23 \log (N)-72 \log (r))^3 \left(\log (N)+18 \log ^2(r)\right) \sqrt{\alpha (r) \left(\sigma (r)+x'(r)^2\right)}}{ r^4 \alpha _{\theta _2}^{10} (\log (N)-3 \log (r))^5 x'(r)^2 \left(4 \pi -N_f g_s (\log (N)-9 \log (r))\right){}^3} \Biggr),\nonumber\\
& & \hskip -0.3in (v)\int d{\cal V}_9 \sqrt{-g}\frac{1}{2} \left(G^{xx}\right)^3\left(G^{ii}\right)\left(G^{ll}\right)\left(G^{mm}\right)\left(G^{jj}\right) R_{jQxl} R^Q_{~~ixm}K_{xij}K_{xml} \sim  \int d{\cal V}_9 \left(\frac{Z_1(r)}{2}{\cal L}_1\right),\nonumber\\
& & \hskip -0.3in (vi)\int d{\cal V}_9 \sqrt{-g}\frac{1}{2}  \left(G^{xx}\right)^3\left(G^{jm}\right)\left(G^{ll}\right) R_{Pxxl} R_x^{~ixP} K_{xij}K_{xml}\nonumber\\
& & \sim  \int d{\cal V}_9 \sqrt{-g} \Biggl(\frac{M^2 N N_f g_s^3 \log ^3(N) (5 \log (N)-196 \log (r)) (23 \log (N)-72 \log (r)) \sqrt{\alpha (r) \left(\sigma (r)+x'(r)^2\right)}}{r^4 \alpha _{\theta _1} \alpha _{\theta _2}^6
   (\log (N)-9 \log (r))^3 (\log (N)-3 \log (r))^4 x'(r)^2} \Biggr),
\nonumber\\
& & \hskip -0.3in (vii)\int d{\cal V}_9 \sqrt{-g} \left(G^{xx}\right)^2\left(G^{im}\right)\left(G^{ll}\right) R^{xMNj} R_{lMNx} K_{xij}K_{xml} \sim  0,\nonumber\\
& & \hskip -0.3in (viii)\int d{\cal V}_9 \sqrt{-g}\left(G^{xx}\right)^2 \left(G^{jj}\right)\left(G^{im}\right) R^{xMNl} R_{jMNx}K_{xij}K_{xml}\sim  0, \nonumber\\
& & \hskip -0.3in (ix)\int d{\cal V}_9 \sqrt{-g} \left(G^{xx}\right)^3 \left(G^{ii}\right)\left(G^{jj}\right)\left(G^{ll}\right)\left(G^{mm}\right) R^Q_{~~mxj} R_{lixQ}K_{xij}K_{xml} \sim  0, \nonumber\\
& & \hskip -0.3in (x)\int d{\cal V}_9 \sqrt{-g}\frac{1}{2} \left(G^{xx}\right)^2 \left(G^{im}\right)\left(G^{ll}\right) R^{xjMN} R_{lxMN}K_{xij}K_{xml}\nonumber\\
& & \sim  \int d{\cal V}_9 \sqrt{-g} \Biggl[\frac{M N^{7/10} r^2 N_f^{8/3} g_s^{3/2} \log (r) (\log (N)-9 \log (r)) (\log (N)-3 \log (r))^{5/3} \sqrt{\alpha (r) \left(\sigma (r)+x'(r)^2\right)}}{ \alpha _{\theta _1}^3
   \alpha _{\theta _2}^2 x'(r)^2} \nonumber\\
   & & \times \left(\frac{ M^4 g_s^4 \log ^2(r) (\log (N)-12 \log (r))^2 \left(\sigma (r) \alpha '(r)+\alpha '(r) x'(r)^2+\alpha (r) \left(\sigma '(r)+2 x'(r) x''(r)\right)\right)^2}{64\ 6^{2/3} \pi ^{4/3} r^4 \alpha
   (r)^2 N_f^{2/3} (\log (N)-3 \log (r))^{8/3} \left(\sigma (r)+x'(r)^2\right)^2}-200 {G_5}(r)\right) \Biggr],\nonumber\\
& & \hskip -0.3in (xi)\int d{\cal V}_9 \sqrt{-g}\frac{1}{2} \left(G^{xx}\right)^2 \left(G^{im}\right)\left(G^{jj}\right) R^{xlMN} R_{jxMN}K_{xij}K_{xml}\nonumber\\
& & \sim  \int d{\cal V}_9 \sqrt{-g} \Biggl[ \frac{M N^{7/10} r^2 N_f^{8/3} g_s^{3/2} \log (r) (\log (N)-9 \log (r)) (\log (N)-3 \log (r))^{5/3} \sqrt{\alpha (r) \left(\sigma (r)+x'(r)^2\right)}}{14400\ 6^{5/6} \pi ^{19/6} \alpha _{\theta _1}^3\alpha _{\theta _2}^2 x'(r)^2} \nonumber\\
& & \times \left(\frac{29 M^4 g_s^4 \log ^2(r) (\log (N)-12 \log (r))^2 \left(\sigma (r) \alpha '(r)+\alpha '(r) x'(r)^2+\alpha (r) \left(\sigma '(r)+2 x'(r) x''(r)\right)\right)^2}{64\ 6^{2/3} \pi ^{4/3} r^4 \alpha
   (r)^2 N_f^{2/3} (\log (N)-3 \log (r))^{8/3} \left(\sigma (r)+x'(r)^2\right)^2}-200 {G_6}(r)\right)\Biggr],\nonumber\\
& & \hskip -0.3in (xii)\int d{\cal V}_9 \sqrt{-g}\frac{1}{2}\left(G^{xx}\right)^3 \left(G^{ii}\right)\left(G^{jj}\right)\left(G^{ll}\right)\left(G^{mm}\right)R^Q_{~~mxi} R_{lQxj}K_{xij}K_{xml} \sim  \int d{\cal V}_9 \left(\frac{Z_1(r)}{2}{\cal L}_1\right),\nonumber\\
& & \hskip -0.3in (xiii)\int d{\cal V}_9 \sqrt{-g}\frac{1}{2}\left(G^{xx}\right)^3 \left(G^{jj}\right)\left(G^{il}\right)R_x^{~mxP} R_{Pxxj}K_{xij}K_{xml}\nonumber\\
& & \sim  \int d{\cal V}_9 \sqrt{-g} \Biggl(-\frac{M^2 N N_f g_s^3 \log ^3(N) (5 \log (N)-196 \log (r)) (23 \log (N)-72 \log (r)) \sqrt{\alpha (r) \left(\sigma (r)+x'(r)^2\right)}}{ r^4 \alpha _{\theta _1} \alpha _{\theta _2}^6
   (\log (N)-9 \log (r))^3 (\log (N)-3 \log (r))^4 x'(r)^2} \Biggr),
\end{eqnarray}
}
where $G_5(r)$ and $G_6(r)$ are $N$ independent functions. We find that (v) and (xii) terms are the most dominant terms in the large $N$ limit. There are thirteen terms that obtain from $\frac{\partial^2 J_0}{\partial R_{titj}\partial R_{tmtl}} K_{xij}K_{xml}$:
{\scriptsize
\begin{eqnarray}
& & \hskip -0.3in (i)\int d{\cal V}_9 \sqrt{-g}\left(G^{tt}\right)^2 \left(G^{mm}\right)\left(G^{ll}\right) \delta^i_m R_t^{~TSt} R^j_{~TSl}K_{xij}K_{xml}\nonumber\\
& & \sim  \int d{\cal V}_9 \sqrt{-g} \Biggl(-\frac{M^2 N N_f g_s^3 \log ^4(N) \log ^2(r) (23 \log (N)-72 \log (r))^2 (\log (N)-27 \log (r)) \sqrt{\alpha (r) \left(\sigma (r)+x'(r)^2\right)}}{ r^4 \alpha _{\theta _1} \alpha
   _{\theta _2}^6 (\log (N)-9 \log (r))^4 (\log (N)-3 \log (r))^6 x'(r)^2}\Biggr),\nonumber\\
& & \hskip -0.3in (ii)\int d{\cal V}_9 \sqrt{-g} \left(G^{tt}\right)^3 \left(G^{ii}\right)\left(G^{jj}\right)\left(G^{mm}\right)\left(G^{ll}\right) R_{jmtQ} R^Q_{~~itl}K_{xij}K_{xml} \sim  0, \nonumber\\
& & \hskip -0.3in (iii)\int d{\cal V}_9 \sqrt{-g}\frac{1}{2}\left(G^{tt}\right)^2 \left(G^{mm}\right)\left(G^{jl}\right)R_t^{~TSt} R^i_{~TSt}K_{xij}K_{xml}\sim  0, \nonumber\\
& & \hskip -0.3in (iv)\int d{\cal V}_9 \sqrt{-g}\frac{1}{2}\left(G^{tt}\right)^2\left(G^{ii}\right)\left(G^{jl}\right) R_t^{~TSt} R^m_{~~TSi} K_{xij}K_{xml}\nonumber\\
& & \sim  \int d{\cal V}_9 \sqrt{-g} \Biggl(-\frac{M^3 N^{13/10} \alpha _{\theta _1} N_f^2 g_s^{9/2} \log (r) (23 \log (N)-72 \log (r))^3 \left(\log (N)+18 \log ^2(r)\right) \sqrt{\alpha (r) \left(\sigma (r)+x'(r)^2\right)}}{ r^4 \alpha _{\theta _2}^{10} (\log (N)-9 \log (r))^3 (\log (N)-3 \log (r))^5 x'(r)^2}\Biggr),\nonumber\\
& & \hskip -0.3in (v)\int d{\cal V}_9 \sqrt{-g}\frac{1}{2} \left(G^{tt}\right)^3\left(G^{ii}\right)\left(G^{ll}\right)\left(G^{mm}\right)\left(G^{jj}\right) R_{jQtl} R^Q_{~~itm}K_{xij}K_{xml}\sim  \int d{\cal V}_9 \left(\frac{W_1(r)}{2}{\cal L}_2\right),\nonumber\\
& & \hskip -0.3in (vi)\int d{\cal V}_9 \sqrt{-g}\frac{1}{2}  \left(G^{tt}\right)^3\left(G^{jm}\right)\left(G^{ll}\right) R_{Pttl} R_t^{~itP} K_{xij}K_{xml}\nonumber\\
& & \hskip -0.2in \sim  \int d{\cal V}_9 \sqrt{-g} \Biggl(\frac{M^2 N N_f g_s^3 \log ^3(N) (5 \log (N)-196 \log (r)) (23 \log (N)-72 \log (r)) (\log (N)-3 \log (r)-1)^2 \sqrt{\alpha (r) \left(\sigma (r)+x'(r)^2\right)}}{ r^4 \alpha _{\theta
   _1} \alpha _{\theta _2}^6 (\log (N)-9 \log (r))^3 (\log (N)-3 \log (r))^6 x'(r)^2} \Biggr),
\nonumber\\
& & \hskip -0.3in (vii) \int d{\cal V}_9 \sqrt{-g}\left(G^{tt}\right)^2\left(G^{im}\right)\left(G^{ll}\right) R^{tMNj} R_{lMNt}K_{xij}K_{xml} \sim 0, \nonumber\\
& & \hskip -0.3in (viii)\int d{\cal V}_9 \sqrt{-g}\left(G^{tt}\right)^2 \left(G^{jj}\right)\left(G^{im}\right) R^{tMNl} R_{jMNt}K_{xij}K_{xml}\sim  0, \nonumber\\
& & \hskip -0.3in (ix)\int d{\cal V}_9 \sqrt{-g} \left(G^{tt}\right)^3 \left(G^{ii}\right)\left(G^{jj}\right)\left(G^{ll}\right)\left(G^{mm}\right) R^Q_{~~mtj} R_{litQ}K_{xij}K_{xml} \sim  0, \nonumber\\
& & \hskip -0.3in (x)\int d{\cal V}_9 \sqrt{-g}\frac{1}{2} \left(G^{tt}\right)^2 \left(G^{im}\right)\left(G^{ll}\right) R^{tjMN} R_{ltMN}K_{xij}K_{xml}\nonumber\\
& & \sim  \int d{\cal V}_9 \sqrt{-g} \Biggl[\frac{M N^{7/10} r^2 N_f^{8/3} g_s^{3/2} \log (r) (\log (N)-9 \log (r)) (\log (N)-3 \log (r))^{5/3} \sqrt{\alpha (r) \left(\sigma (r)+x'(r)^2\right)}}{ \alpha _{\theta _1}^3
   \alpha _{\theta _2}^2 x'(r)^2} \nonumber\\
   & & \times \left(\frac{13 M^4 g_s^4 \log ^2(r) (\log (N)-12 \log (r))^2 \left(\sigma (r) \alpha '(r)+\alpha '(r) x'(r)^2+\alpha (r) \left(\sigma '(r)+2 x'(r) x''(r)\right)\right)^2}{64\ 6^{2/3} \pi ^{4/3} r^4 \alpha
   (r)^2 (\log (N)-3 \log (r))^2 \left(N_f (\log (N)-3 \log (r))\right){}^{2/3} \left(\sigma (r)+x'(r)^2\right)^2}-25 {G_7}(r)\right)\Biggr],\nonumber\\
& & \hskip -0.3in (xi)\int d{\cal V}_9 \sqrt{-g}\frac{1}{2} \left(G^{tt}\right)^2 \left(G^{im}\right)\left(G^{jj}\right) R^{tlMN} R_{jtMN}K_{xij}K_{xml}\nonumber\\
& & \sim  \int d{\cal V}_9 \sqrt{-g} \Biggl[\frac{M N^{7/10} r^2 N_f^{8/3} g_s^{3/2} \log (r) (\log (N)-9 \log (r)) (\log (N)-3 \log (r))^{5/3} \sqrt{\alpha (r) \left(\sigma (r)+x'(r)^2\right)}}{\alpha _{\theta _1}^3
   \alpha _{\theta _2}^2 x'(r)^2} \nonumber\\
   & & \times \left(\frac{13 M^4 g_s^4 \log ^2(r) (\log (N)-12 \log (r))^2 \left(\sigma (r) \alpha '(r)+\alpha '(r) x'(r)^2+\alpha (r) \left(\sigma '(r)+2 x'(r) x''(r)\right)\right)^2}{64\ 6^{2/3} \pi ^{4/3} r^4 \alpha
   (r)^2 (\log (N)-3 \log (r))^2 \left(N_f (\log (N)-3 \log (r))\right){}^{2/3} \left(\sigma (r)+x'(r)^2\right)^2}-25 {G_8}(r)\right)\Biggr],\nonumber\\
& & \hskip -0.3in (xii)\int d{\cal V}_9 \sqrt{-g}\frac{1}{2}\left(G^{tt}\right)^3 \left(G^{ii}\right)\left(G^{jj}\right)\left(G^{ll}\right)\left(G^{mm}\right)R^Q_{~~mti} R_{lQtj}K_{xij}K_{xml} \sim  \int d{\cal V}_9  \left(\frac{W_1(r)}{2}{\cal L}_2\right),\nonumber\\
& & \hskip -0.3in (xiii)\int d{\cal V}_9 \sqrt{-g}\frac{1}{2}\left(G^{tt}\right)^3 \left(G^{jj}\right)\left(G^{il}\right)R_t^{~mtP} R_{Pttj}K_{xij}K_{xml}\nonumber\\
& & \sim  \int d{\cal V}_9 \sqrt{-g} \Biggl(\frac{M^2 N N_f g_s^3 \log ^3(N) (5 \log (N)-196 \log (r)) (23 \log (N)-72 \log (r)) \sqrt{\alpha (r) \left(\sigma (r)+x'(r)^2\right)}}{ r^4 \alpha _{\theta _1} \alpha _{\theta
   _2}^6 (\log (N)-9 \log (r))^3 (\log (N)-3 \log (r))^4 x'(r)^2} \Biggr).
\end{eqnarray}
}
where $G_7(r)$ and $G_8(r)$ are $N$ independent functions and (v) and (xii) terms are the most dominant terms in the larg-$N$ limit. Similar to Hartman-Maldacena-like surface, there are six terms that we obtain from $\frac{\partial^2 J_0}{\partial R_{titj}\partial R_{xmxl}}K_{xij}K_{xml}$ and only first term is non-zero.
{\footnotesize
\begin{eqnarray}
& & \hskip -0.3in (i)\int d{\cal V}_9 \sqrt{-g}\left(G^{xx}\right)^2 \left(G^{mm}\right)\left(G^{ll}\right) \delta^i_m \delta^t_P \delta^t_R \delta^j_Q R_x^{~TSP} R^Q_{~TSl}K_{xij}K_{xml} \sim  \int d{\cal V}_9 \left(U_1(r) {\cal L}_3\right),\nonumber\\
& & \hskip -0.3in (ii)\int d{\cal V}_9 \sqrt{-g} \left(G^{xx}\right)^2 \left(G^{ii}\right)\left(G^{jj}\right)\left(G^{mm}\right)\left(G^{ll}\right)\left(G^{tt}\right)\delta^t_x R_{jmxQ} R^Q_{~~itl}K_{xij}K_{xml}\sim 0,\nonumber\\
& & \hskip -0.3in (iii)\int d{\cal V}_9 \sqrt{-g}\left(G^{xx}\right)^2 \left(G^{mm}\right)\left(G^{ll}\right)\left(G^{tt}\right) \delta^i_T \delta^t_S \delta^j_l  R_{Pmxt} R_x^{~TSP}K_{xij}K_{xml}\sim 0,\nonumber\\
& & \hskip -0.3in (iv)\int d{\cal V}_9 \sqrt{-g}\frac{1}{2}\left(G^{xx}\right)^2\left(G^{mm}\right)\left(G^{ll}\right)\left(G^{tt}\right)\delta^i_T \delta^t_S \delta^j_m  R_{Ptxl} R_x^{~~TSP}K_{xij}K_{xml}\sim 0, \nonumber\\
& & \hskip -0.3in (v)\int d{\cal V}_9 \sqrt{-g} \left(G^{xx}\right)\left(G^{tt}\right)^2\left(G^{ii}\right)\left(G^{jl}\right) R_t^{mxP} R_{Pitx}K_{xij}K_{xml}\sim 0,\nonumber\\
& & \hskip -0.3in (vi)\int d{\cal V}_9 \sqrt{-g}\frac{1}{2}  \left(G^{tt}\right)^2\left(G^{xx}\right)\left(G^{jj}\right)\left(G^{iK}\right)\delta^l_K R_t^{mxP} R_{Pxtj}K_{xij}K_{xml}\sim 0. 
\end{eqnarray}
}
Similar to Hartman-Maldacena-like surface, there are four terms that we obtain from$\frac{\partial^2 J_0}{\partial R_{tixj}\partial R_{xmtl}}K_{xij}K_{xml}$ and three terms are non-zero and they scale as same power of $N$.
{\footnotesize
\begin{eqnarray}
& & \hskip -0.3in (i)\int d{\cal V}_9 \sqrt{-g}\left(G^{xx}\right)^2 \left(G^{mm}\right)\left(G^{ll}\right) \delta^i_m  R_x^{~TSx} R^j_{~TSl}K_{xij}K_{xml} \sim  \int d{\cal V}_9 \left(U_1(r) {\cal L}_4\right),\nonumber\\
& & \hskip -0.3in (ii)\int d{\cal V}_9 \sqrt{-g} \left(G^{xx}\right)^3 \left(G^{ii}\right)\left(G^{jj}\right)\left(G^{ll}\right)\left(G^{mm}\right) R_{jmxQ} R^Q_{~~ixl}K_{xij}K_{xml} \sim 0, \nonumber\\
& & \hskip -0.3in (iii)\int d{\cal V}_9 \sqrt{-g}\frac{1}{2}\left(G^{xx}\right) \left(G^{tt}\right)^2\left(G^{jm}\right)\left(G^{ll}\right)  R_x^{~ixP} R_{Pttl}K_{xij}K_{xml} \sim  \int d{\cal V}_9 \left(V_1(r) {\cal L}_4\right),\nonumber\\
& & \hskip -0.3in (iv)\int d{\cal V}_9 \sqrt{-g}\frac{1}{2}\left(G^{xx}\right)^2\left(G^{tt}\right)\left(G^{jj}\right)\left(G^{il}\right)  R_t^{~mtP} R_{Pxxj} K_{xij}K_{xml}\sim  \int d{\cal V}_9 \left(V_1(r) {\cal L}_4\right).
\end{eqnarray}
}
 We have multiplied each term by $\frac{8}{(q_{\alpha}+1)}$.
%%%%%%%%%%%%%%%%%%%%

\end{document}